\documentclass{aa}  

\usepackage{graphicx}
\usepackage{booktabs}
\usepackage{subfigure}
\usepackage{mathrsfs}
\usepackage{txfonts}
\usepackage[colorlinks=true, citecolor=blue, linkcolor = blue]{hyperref}

\newcommand{\dd}{\ensuremath{\mathrm{d}}}
\newcommand{\msafit}{{\sc msafit}}
\newcommand{\vobsi}{$V_{i,{\rm obs}}$}
\newcommand{\sigmaobsi}{$\sigma_{v,i,{\rm obs}}$}
\newcommand{\kms}{km~s$^{-1}$}
\newcommand{\vsigma}{$(V/\sigma)_{r_e}$}
\begin{document} 

   \title{Fast rotators at cosmic noon: Stellar kinematics for 15 quiescent galaxies from JWST-SUSPENSE}
   \author{Martje Slob\inst{1}
          \and 
          Mariska Kriek\inst{1}
          \and 
          Anna de Graaff\inst{2}
          \and
          Chloe M. Cheng\inst{1}
          \and
          Aliza G. Beverage\inst{3}
          \and
          Rachel Bezanson\inst{4}
          \and
          Natascha M. F\"orster Schreiber\inst{5}
          \and
          Brian Lorenz\inst{3}
          \and
          Pavel E. Mancera Pi\~na\inst{1}
          \and
          Danilo Marchesini\inst{6}
          \and
          Adam Muzzin\inst{7}
          \and
          Andrew B. Newman\inst{8}
          \and
          Sedona H. Price\inst{9,4}
          \and
          Katherine A. Suess\inst{10}
          \and
          Jesse van de Sande\inst{11}
          \and
          Pieter van Dokkum\inst{12}
          \and
          Daniel R. Weisz\inst{3}
          }

   \institute{Leiden Observatory, Leiden University, P.O. Box 9513, 2300 RA Leiden, The Netherlands. \email{slob@strw.leidenuniv.nl} 
   \and
   Max-Planck-Institut f\"ur Astronomie, K\"onigstuhl 17, D-69117, Heidelberg, Germany
   \and
   Department of Astronomy, University of California, Berkeley, CA 94720, USA
   \and
   Department of Physics and Astronomy and PITT PACC, University of Pittsburgh, Pittsburgh, PA 15260, USA
   \and
   Max-Planck-Institut f\"ur extraterrestrische Physik, Giessenbachstrasse 1, D-85748 Garching, Germany
   \and
   Department of Physics \& Astronomy, Tufts University, MA 02155, USA
   \and
   Department of Physics and Astronomy, York University, 4700 Keele Street, Toronto, Ontario, ON MJ3 1P3, Canada
   \and
   Observatories of the Carnegie Institution for Science, 813 Santa Barbara St., Pasadena, CA 91101, USA
   \and
   Space Telescope Science Institute (STScI), 3700 San Martin Drive, Baltimore, MD 21218, USA
   \and
   Department for Astrophysical \& Planetary Science, University of Colorado, Boulder, CO 80309, USA
   \and
   School of Physics, University of New South Wales, NSW 2052, Australia
   \and
   Astronomy Department, Yale University, 52 Hillhouse Ave, New Haven, CT 06511, USA
   }

   \date{Received XX; accepted XX}
 
  \abstract{
  We present spatially resolved stellar kinematics of 15 massive ($M_*=10^{10.5-11.5}M_{\odot}$) quiescent galaxies at $z\sim1.2-2.3$ from the JWST-SUSPENSE program. This is the largest sample of spatially resolved kinematic measurements of quiescent galaxies at cosmic noon to date. Our measurements are derived from ultra-deep NIRSpec/MSA stellar absorption line spectra using a forward-modelling approach that accounts for optics, source morphology, positioning, and data reduction effects.
  Ten out of 15 galaxies are orientated such that we can measure rotational support. Remarkably, all 10 galaxies show significant rotation ($V_{r_e}=117-345$~km\,s$^{-1}$, $\sigma_0 = 180-387$~km\,s$^{-1}$) and are classified as fast rotators from their spin parameter. The remaining galaxies are too misaligned with respect to the slitlet to constrain their rotational velocities.
  The widespread rotational support in our sample indicates that the process responsible for quenching star formation in early massive galaxies did not destroy rotating disc structures.
  When combined with other quiescent galaxy samples at $z\sim0.5-2.5$, we find a trend between rotational support and age, where younger quiescent galaxies are more rotationally supported. This age trend was also found at $z\sim0$, and it likely explains why our high-redshift galaxies have more rotational support than massive early-type galaxies at $z\sim0$, which are older on average.
  Our kinematic modelling also enabled us to calculate dynamical masses. These dynamical masses greatly exceed the stellar masses for our sample (median $M_{\text{dyn}}/M_*=2.73$); they even allow for the bottom-heavy initial mass function found in the cores of low-$z$ massive ellipticals.
  Altogether, our results support a scenario in which distant quiescent galaxies evolve into nearby massive early-type galaxies, by gradually building up their outskirts, and simultaneously losing rotation through a series of (mostly minor) mergers.}


   \maketitle
\section{Introduction}
The most massive and oldest quiescent galaxies that exist in the Universe today are predominantly slow-rotating ellipticals \citep{JvandeSande2017, MVeale2017, EEmsellem2011}.
To explain their slow rotation and morphologies, a picture emerged in which the mechanism responsible for the quenching of star formation may also cause significant structural changes, as suggested by the classic major merger scenario \citep[e.g.][]{JBarnes1988, LHernquist1993, CStruck1999, PHopkins2008}.
This classic scenario of massive galaxy evolution thus posits that the rotating-disc structure of massive galaxies may be destroyed during the initial quenching event, and that the structures of distant, quiescent massive galaxies resemble those of nearby massive early-type galaxies.

However, over the past few decades this classic picture has been challenged by detailed studies of distant quiescent galaxies. First, distant quiescent galaxies are more compact than low-$z$ early-type galaxies, which indicates that significant size growth occurs after quenching \citep[e.g.][]{EDaddi2005, PvanDokkum2008, AvanderWel2014, KSuess2021}. This picture is supported by colour-gradient studies, which show that the outskirts of old quiescent galaxies are bluer than their cores, and that they thus may have been built up by the accretion of low-mass low-metallicity galaxies \citep[e.g.][]{JGreene2015, IMartinNavarro2018, KSuess2019, Cheng_2024}. Moreover, photometric studies showed that a significant fraction of distant massive quiescent galaxies are disc-like, with lower S\'ersic indices and flatter shapes than nearby early-type galaxies \citep{EMcGrath2008, PvanDokkum2008, AvanderWel2011, RMcClure2013, VBruce2012, VBruce2014, FBuitrago2013, YChang2013a, YChang2013b}. Detailed studies of nearby early-type galaxies also challenge this classic picture. While the most massive oldest low-$z$ early-type galaxies show little rotational support, younger low-$z$ quenched galaxies still have significant rotation \citep[e.g.][]{JvandeSande2017, JVandeSande2018, SCroom2024}.

To confirm whether quiescent galaxies beyond the low-$z$ universe are rotationally supported discs, direct measurements of their dynamical properties from resolved stellar kinematic studies are needed. Indeed, the results from the LEGA-C\footnote{Large Early Galaxy Astrophysics Census} survey revealed that a significant fraction of massive quiescent galaxies at $z\sim0.7$ are rotationally supported \citep{RBezanson2018, JvanHoudt2021}. 
Beyond $z>1$, studies like this were only possible for lensed galaxies with large ground-based telescopes until recently because these measurements rely on spatially resolved faint absorption lines. In particular, this was done for four strongly lensed quiescent galaxies at $z\sim2$, all of which show significant rotation \citep{ANewman2015, ANewman2018b, SToft2017}.

If they hold true for the general population of distant quiescent galaxies, these results have important implications for massive galaxy evolution theory. They imply that rotating discs are not destroyed during the quenching of star formation. Other physical processes, including the accretion of low-mass satellites after the initial quenching event, might instead perturb the galaxy structures and form the low-$z$ population of slow-rotating ellipticals. This picture is also supported by simulations, which show that dry mergers can effectively destroy ordered rotation in galaxies from cosmic noon to $z\sim0$ \citep[e.g.][]{FBournaud2007,TNaab2009,FSchulze2018,YDubois2016,CLagos2018a,CLagos2018b,CLagos2022}.

With the launch of the James Webb Space Telescope (JWST), the high sensitivity and spatial resolution of JWST/NIRSpec \citep{PFerruit2022} now enable us to study the detailed kinematic properties of statistically significant samples of distant quiescent galaxies. Indeed, studies based on NIRSpec/IFU have identified quiescent galaxies at $z>3$ with strong rotational support \citep{FdEugenio2024,RPascalau2025}. Moreover, \citet{ANewman2025} used the NIRSpec/IFU to confirm the rotational support of one of the three lensed galaxies presented by \citet{ANewman2018b}. Although these results confirmed that NIRSpec/IFU is a powerful tool for measuring accurate resolved stellar kinematics of individual galaxies, its field of view and sensitivity limit the sample size of distant quiescent galaxies for which kinematic parameters can be measured.

Instead, the micro-shutter array (MSA) on NIRSpec offers a powerful and more efficient strategy for measuring stellar kinematics for large samples of galaxies at $z>1$. NIRSpec/MSA can simultaneously observe $\sim$100 objects, which makes it highly efficient for studying large samples of high-redshift galaxies. By leveraging spatial information from high-resolution imaging, we can mitigate the fact that the slit-based observations from NIRSpec/MSA only cover one spatial dimension. Combined with detailed models of the instrument and detector optics, we can thus measure the spatial and kinematic properties from NIRSpec/MSA spectra with a forward-modelling approach (\citealt{AdeGraaff2024_kinematics}, see also \citealt{SPrice2016,SPrice2020, JvanHoudt2021,CStraatman2022}). 

In this paper, we present the kinematic properties of 15 massive ($\log M_*/M_{\odot} \sim 10.2-11.5$) quiescent galaxies at $z = 1 -2.5$ from the JWST-SUSPENSE programme \citep{MSlob2024}. These galaxies represent the first sample of kinematically modelled quiescent galaxies at $z>1$. The galaxies are all observed in a single NIRSpec/MSA pointing and have deep JWST/NIRCam or \textit{Hubble} Space Telescope (HST) imaging, which provide spatially resolved spectroscopic and imaging data. From these data, we derive the intrinsic stellar kinematics using a forward-modelling technique based on \msafit\ \citep{AdeGraaff2024_kinematics}, and we accordingly constrain the kinematic properties of these galaxies. We use these measurements to assess the prevalence of quiescent rotating discs in our sample and discuss the implications for galaxy evolution models.

The paper is organised as follows. In Section \ref{sec:data} we present the data. Our measurements and forward-modelling technique are outlined in Section \ref{sec:modelling}. In Section \ref{sec:results} we present the results of our modelling, the inferred measures of rotational support for our sample, and the dynamical masses. We discuss caveats and the implications for galaxy evolution, the initial mass function (IMF), and virial mass estimates in Section \ref{sec:discussion}, and we finally summarise our findings in Section \ref{sec:summary}. Throughout this work, we assume a $\Lambda$ cold dark matter cosmology with $\Omega_{\rm m}= 0.3$, $\Omega_\Lambda=0.7$, and $H_0 =70\rm \, km s^{-1} \,Mpc^{-1}$.

\section{Data}\label{sec:data}
\subsection{NIRSpec MSA spectroscopy}
We used NIRSpec/MSA observations from the JWST-SUSPENSE programme (programme ID 2110, \citealt{MSlob2024}), which observed 20 massive ($\log M_*/M_{\odot} \sim 10.2-11.5$) quiescent galaxies at $z = 1-3$. These galaxies were selected to be quiescent based on their location in the UVJ diagram (\citealt{AMuzzin2013}; see \citealt{MSlob2024} for details of the selection procedure). Below, we provide a summary of the observing strategy and reduction of the data. For a full description, we refer to \citet{MSlob2024}.

We used the NIRSpec/MSA with the medium resolution ($R\sim1000$) G140M-F100LP disperser and filter combination, which covers the rest-frame optical wavelengths for our sample. The galaxies were observed using a custom two-point nodding pattern with a nod size of two shutters, which ensured that self-subtraction for the extended galaxies in the sample was minimised. The data were reduced using a modified version of the JWST Science Calibration Pipeline \citep{HBushouse2023}, with a custom outlier-detection algorithm described by \citet{MSlob2024}. With a total on-source integration time of 16.4 hr, the resulting 2D spectra have median signal-to-noise ratios between rest-frame $4600-4800$~\AA\ of $S/N=10-58$~\AA$^{-1}$, and extend over multiple effective radii for the majority of sources in the sample.

To study the resolved kinematics for the galaxies in our sample, we extracted spectra for each individual row along the spatial direction of the rectified 2D spectra. The data were taken at two dither positions, and we combined the extracted 1D spectra row by row for each dither separately. Due to the undersampling of the NIRSpec point spread function (PSF) along the spatial axis, these row-by-row spectra are affected by sinusoidal patterns (wiggles) in their fluxes (see \citealt{MPerna2023} for an example with NIRSpec/IFU data). We corrected for these wiggles by fitting a sinusoidal profile to each spectral row and divided our data by this fit, as described by C. Cheng et al. (in prep.). 

Throughout this work, we adopt the stellar masses derived from full spectral energy distribution (SED) fitting, including the NIRSpec spectrometry, reported by \citet{MSlob2024}. These masses were derived using a \citet{GChabrier2003} IMF, with a non-parametric star formation history.

\subsection{Imaging}
Fifteen out of 20 galaxies in our sample fall within the imaging footprint from the COSMOS-Web programme \citep[Programme ID: 1727][]{CCasey2023}. These galaxies were observed with four different NIRCam filters (F115W, F150W, F277W, and F444W). For our analysis, we used the images that are available in MAST\footnote{\url{https://mast.stsci.edu/}}, which were reduced using the JWST Calibration Pipeline v11.17.14 with the CRDS pipeline mapping (pmap) context 1230. For the 5 galaxies in the JWST-SUSPENSE sample that are not covered by COSMOS-Web, we used HST/ACS imaging in the F814W filter \citep{NScoville2007}. Although this filter is bluer than COSMOS-Web imaging, the galaxies for which we used F814W are at $z\sim1.2$. At these redshifts, the F814W filter still covers rest-frame optical wavelengths.

\section{Kinematic modelling}\label{sec:modelling}
\begin{figure*}[t]
    \sidecaption
    \includegraphics[width=12cm]{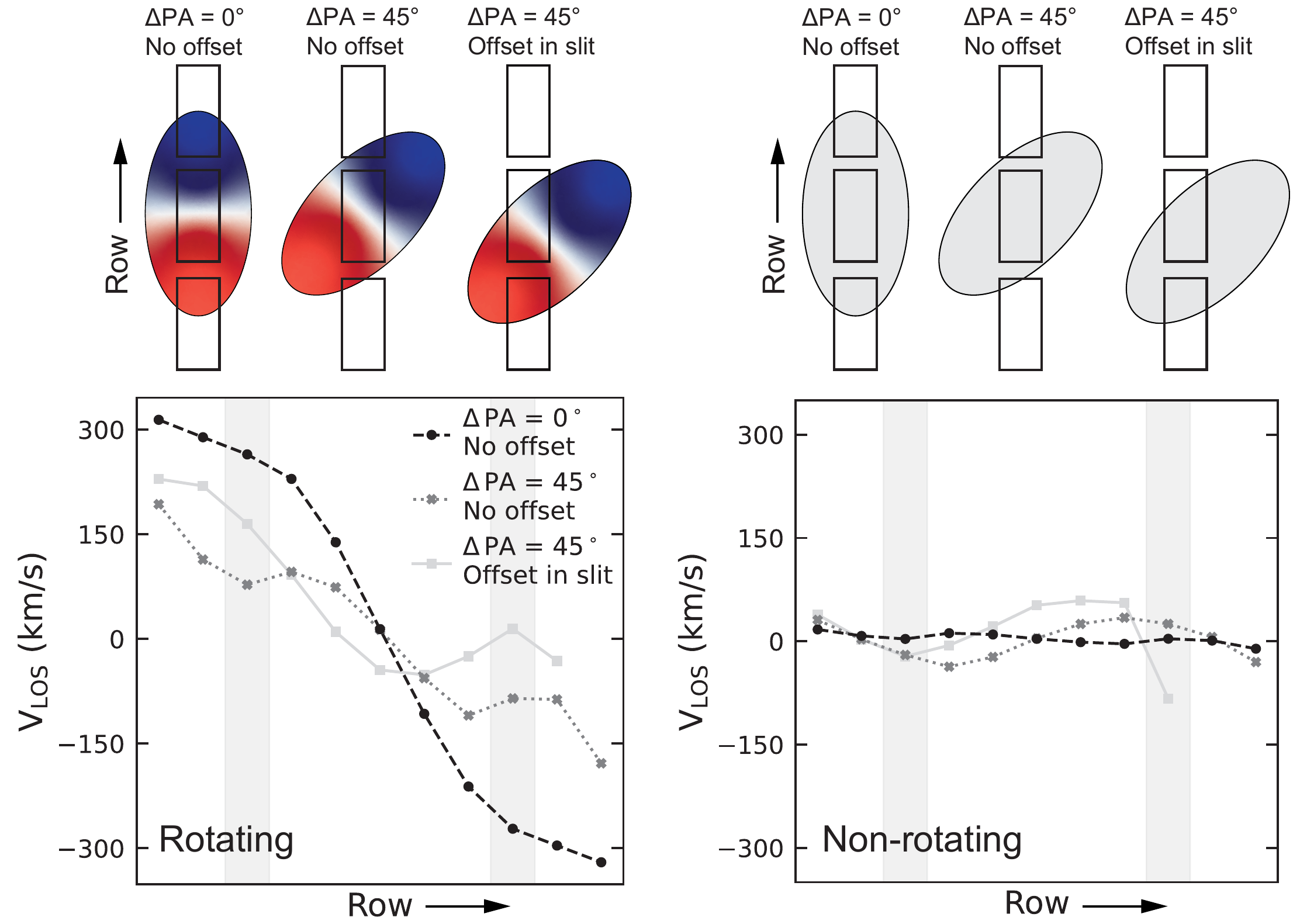}
    \caption{LOS velocity profiles for a model galaxy that is perfectly aligned and centred (left, black circles), a model galaxy that is perfectly centred but misaligned (middle, dark grey crosses), and a model galaxy that is offset and misaligned (right, light grey squares) with respect to the central micro-shutter. The left panels show the models for a galaxy with an intrinsic velocity field with a maximum rotational velocity of 300~\kms. The right panels show the LOS velocity profiles for a galaxy without intrinsic rotational velocity, representing the effects from optics, the PSF, and data reduction steps. These effects can lead to observed LOS velocities of up to $\sim50$~\kms. The modelled galaxies have $r_e = 0.13$~\arcsec (1.1~kpc at $z\sim1.5$), a S\'ersic index of 1.5, and an axis ratio $q$ of $0.3$. The grey bands indicate the areas of the velocity profile that are affected by bar shadows from the MSA.}
    \label{fig:extreme_cases}
\end{figure*}

It is challenging to obtain intrinsic stellar kinematics from NIRSpec/MSA data because these slit-based spectra cover only one randomly aligned spatial dimension. In Figure \ref{fig:extreme_cases} we demonstrate the significant change in the observed line-of-sight (LOS) velocity profile when a source is misaligned and/or positioned off-centre within an MSA shutter. However, based on the detailed imaging and measurements of structural parameters of the galaxies in our sample, we know the position, photometric orientation, and inclination of the galaxies with respect to the MSA shutters. When we assume that the kinematic and photometric axes are aligned and that the galaxies are axisymmetric, the imaging and structural parameters can be used to measure the intrinsic kinematic properties of our galaxies using a forward-modelling approach based on \msafit\ \citep{AdeGraaff2024_kinematics}. This software package takes the source position and morphology and the complicated instrumental effects of NIRSpec/MSA into account.

To obtain stellar kinematics from absorption line spectra, several adaptions to \msafit\ were made. \msafit\ is designed to model emission lines in unrectified spectra. However, stellar kinematics of distant quiescent galaxies are based on many absorption lines with low equivalent widths, which are most readily measured in a combined rectified spectrum. Our approach thus requires comparing the model and the data in reduced space, specifically, by directly matching their 1D velocity profiles. We therefore used \msafit\ to generate 2D model spectra, and then applied our observing strategy and reduction procedure. The resulting 2D model spectra can be directly compared to our reduced observations. Below, we describe our full procedure, starting with the velocity and structural measurements in Section \ref{sec:vel_measurements} and \ref{sec:galfit}, respectively. In Sections \ref{sec:forward_model} and \ref{sec:sampling}, we describe the forward model and our sampling method.

\subsection{Velocity measurements}\label{sec:vel_measurements}
We measured velocity and dispersion profiles for each galaxy in our sample by determining the velocity shift (\vobsi) and velocity dispersion (\sigmaobsi) for each row $i$ in the 2D spectrum using \textsc{ppxf} \citep{MCappellari2004, MCappellari2017,MCappellari2023}. To obtain reliable \vobsi\ and \sigmaobsi\ measurements, even in galaxy outskirts where the $S/N$ is low ($(S/N)_i \sim 3-5$\,\AA$^{-1}$, rest frame), we restricted the stellar population templates used in {\sc ppxf} fitting \citep{JvandeSande2017, FdEugenio2024}. We selected the template for each galaxy by first fitting the integrated spectrum with \textsc{ppxf}. We also included a 12th-order polynomial in the fit to correct for flux-calibration issues. We used the resulting stellar population template for each row. Thus, we assumed that the stellar populations do not vary significantly across the spatial profile of the galaxy. While spatial gradients in stellar population parameters have been observed in distant quiescent galaxies \citep[e.g.][]{Jafariyazani_2020,Cheng_2024}, they are generally mild and thus are unlikely to affect the measured kinematic information.

We used the full wavelength range of our spectra in the fitting, but to ensure that our results were not biased by nebular emission, we masked out any emission lines. Furthermore, we masked the Na {\sc i} absorption doublet because this line is a tracer of neutral gas in the interstellar medium (ISM) as well as active galactic nucleus (AGN) outflows, and can have a significant kinematic offset from stellar absorption features \citep[e.g.][]{SBelli2024, RDavies2024}.

To measure \sigmaobsi\ with {\sc ppxf}, we must take the line spread function (LSF) of NIRSpec/MSA into account. For the G140M grating, the LSF of NIRSpec/MSA changes by a wavelength-independent factor that depends on source position and morphology \citep{AdeGraaff2024_kinematics, MSlob2024} and is thus different for each galaxy in our sample. We determined the exact slit position with our forward-modelling method and therefore initially did not have the exact LSF for each source. For the first iteration, we thus assumed an LSF that was narrower by a factor of 10 (rather than a typical factor of $\sim 1.3$) than the NIRSpec/MSA LSF reported in the JDox user documentation\footnote{\url{https://jwst-docs.stsci.edu/jwst-near-infrared-spectrograph/nirspec-instrumentation/nirspec-dispersers-and-filters}} for the model and the data. This high factor ensured that we underestimated the true LSF.

After the initial model fit, we had the exact position of the source in the slit, and we modelled the true LSF for each galaxy following the method of \citet{MSlob2024}. To obtain the LSF, we used our best-fit galaxy position in the slit and generated 50 emission lines without intrinsic velocity or dispersion, applied all relevant data reduction steps, and fitted a Gaussian profile to each emission line. We then used the width of the Gaussians over the wavelength range of the detector to determine the rescaling factor for the JDox NIRSpec/MSA LSF for each galaxy. We refitted the dispersion and velocity of each row in the data and the model after convolving the stellar templates with the rescaled LSF of each galaxy.

For each galaxy in our sample, we fitted all rows with a $(S/N)_i > 4$~pixel$^{-1}$. To ensure that the \vobsi\ and \sigmaobsi\ measurements were of good quality, we removed rows from our analysis for which the error on \vobsi\ and \sigmaobsi\ from {\sc ppxf} was larger than 70~km/s. Visual inspection showed that the fits become unreliable for larger errors. We set the systemic velocity of the galaxy to the best-fit velocity of the integrated spectrum.

Our resolved fitting procedure resulted in \vobsi\ and \sigmaobsi\ measurements for up to 11 rows for the galaxies in our sample. In Figure \ref{fig:2d_line_prof} we compare the velocity and velocity dispersion profiles from {\sc ppxf} to example absorption lines from our observed spectra.
In order to constrain the kinematic profiles, we required \vobsi\ and \sigmaobsi\ measurements in at least 4 rows and excluded the three galaxies from our sample that did not meet this criterion. We thus have resolved velocity curves for 17 galaxies.

\subsection{Structural parameters}\label{sec:galfit}
We used \textsc{GALFIT} \citep{CPeng2002,CPeng2010} to measure the morphological parameters of our sample using NIRCam imaging in the F115W, F150W, F277W, and F444W filters from the COSMOS-Web programme (Programme ID: 1727, \citealt{CCasey2023}; see Appendix \ref{ap:galfit} for more details on our fitting procedure). The morphological parameters were required to account for the source morphology in our kinematic modelling. 

The structural parameters measured using the different NIRCam filters agree well, without systematic offsets between different filters. For our modelling, we used the median structural parameters of the NIRCam filters, which cover rest-frame optical wavelengths for our galaxies. In practice, this meant that we excluded the F115W filter for galaxies with $z>1.3$, and for galaxies with $z>2$, we also excluded the F150W filter.
We note that our kinematic modelling results do not change significantly for the structural parameters of one specific filter instead of the median values. For sources that were not covered by the COSMOS-Web footprint, we used the morphological parameters from HST/ACS F814W observations \citep{RGriffith2012}. The morphological parameters from COSMOS-Web and HST/ACS F814W agree within 1$\sigma$ for galaxies with $z<1.3$, and thus, we expect no significant bias in the structural parameters of the galaxies for which we used the F814W parameters.
Two out of 17 galaxies for which we measured resolved velocity curves were excluded from our sample because they are not covered by COSMOS-Web and are too faint to be included in the \citet{RGriffith2012} morphological catalogue. Our final sample thus consists of 15 galaxies, whose measured morphological parameters we report in Table \ref{tab:fit_properties}.

\begin{table*}[h]
\renewcommand{\arraystretch}{1.35}
    \flushleft
    \caption{GALFIT measurements and dynamic modelling results.}\vspace{-0.1in}
\resizebox{1.0\textwidth}{!}{
    \begin{tabular}{l l l l l l l l l l l l}
        \hline\hline 
        \multicolumn{2}{c}{} & \multicolumn{4}{c}{{Structural parameters}} & \multicolumn{4}{c}{{Inferred velocity parameters}}  & \multicolumn{2}{c}{} \\
        \cmidrule(lr){3-6}\cmidrule(lr){7-10}
        ID & $z_{\text{spec}}$& $r_e$ & $n$ & $q$ & $\Delta$PA\tablefootmark{a} & $V_{r_e}$ & $\sigma_0$ & $V_{r_e}/\sigma_0$ & $\lambda_{r_e}$  & $\log_{10}(M_{\text{dyn}})$ & $\log_{10}(M_*)$\tablefootmark{b}\\
        & & (kpc) & & & (\textdegree) & (km s$^{-1}$) & (km s$^{-1}$) & & & $(M_{\odot})$ & $(M_{\odot})$\\
        \hline
        127108 & 1.335 & $1.51 \pm 0.17$ & $1.9 \pm 0.1$ & $0.82 \pm 0.01$ & $52 \pm 1$ & $114_{-42}^{+24} $ & $210_{-10}^{+5}$ & $0.54_{-0.20}^{+0.12}$ & $0.18_{-0.08}^{+0.03}$ &$11.0_{-0.1}^{+0.1}$& $10.24_{-0.03}^{+0.02}$ \\
127154\tablefootmark{d} & 1.205 & $2.06 \pm 0.25$ & $2.3 \pm 0.1$ & $0.41 \pm 0.01$ & $66 \pm 1$ & $>1$ & $242_{-6}^{+4}$ & $>0.00$ & $>0.00$ & $11.2_{-0.1}^{+0.1}$ & $10.75_{-0.01}^{+0.01}$ \\
127700\tablefootmark{d} & 2.013 & $1.49 \pm 0.17$ & $4.0 \pm 0.1$ & $0.96 \pm 0.01$ & $79 \pm 11$ & $>1$ & $241_{-4}^{+1}$ & $>0.00$ & $>0.00$ & $11.1_{-0.1}^{+0.1}$ & $10.92_{-0.05}^{+0.03}$ \\
127941\tablefootmark{d} & 2.141 & $1.84 \pm 0.83$ & $4.9 \pm 0.2$ & $0.54 \pm 0.01$ & $82 \pm 1$ & $>131$ & $259_{-16}^{+9}$ & $>0.49$ & $>0.18$ & $11.4_{-0.3}^{+0.2}$ & $10.80_{-0.02}^{+0.02}$ \\
128036 & 2.196 & $1.02 \pm 0.08$ & $1.5 \pm 0.1$ & $0.67 \pm 0.01$ & $26 \pm 1$ & $248_{-15}^{+17} $ & $210_{-6}^{+3}$ & $1.18_{-0.08}^{+0.09}$ & $0.42_{-0.04}^{+0.03}$ &$11.0_{-0.0}^{+0.0}$& $10.92_{-0.03}^{+0.04}$ \\
128041 & 1.760 & $1.50 \pm 0.08$ & $1.6 \pm 0.1$ & $0.45 \pm 0.01$ & $9 \pm 1$ & $336_{-14}^{+14} $ & $226_{-1}^{+1}$ & $1.48_{-0.06}^{+0.07}$ & $0.48_{-0.00}^{+0.04}$ &$11.3_{-0.0}^{+0.0}$& $10.71_{-0.01}^{+0.01}$ \\
128452 & 1.205 & $1.59 \pm 0.21$ & $4.0 \pm 0.1$ & $0.72 \pm 0.01$ & $66 \pm 1$ & $154_{-79}^{+59} $ & $275_{-9}^{+9}$ & $0.56_{-0.29}^{+0.22}$ & $0.18_{-0.09}^{+0.09}$ &$11.3_{-0.1}^{+0.1}$& $10.99_{-0.00}^{+0.01}$ \\
128913 & 2.285 & $2.32 \pm 0.74$ & $3.7 \pm 0.1$ & $0.77 \pm 0.01$ & $42 \pm 1$ & $152_{-49}^{+56} $ & $180_{-1}^{+3}$ & $0.84_{-0.27}^{+0.32}$ & $0.22_{-0.07}^{+0.08}$ &$11.1_{-0.2}^{+0.1}$& $10.91_{-0.03}^{+0.03}$ \\
129133 & 2.139 & $1.04 \pm 0.08$ & $2.3 \pm 0.1$ & $0.42 \pm 0.01$ & $45 \pm 1$ & $305_{-15}^{+17} $ & $257_{-1}^{+2}$ & $1.19_{-0.07}^{+0.07}$ & $0.52_{-0.04}^{+0.03}$ &$11.1_{-0.0}^{+0.0}$& $11.09_{-0.02}^{+0.02}$ \\
 129149 & 1.579 & $1.01 \pm 0.08$ & $1.9 \pm 0.1$ & $0.36 \pm 0.01$ & $64 \pm 1$ & $191_{-41}^{+21} $ & $387_{-34}^{+4}$ & $0.49_{-0.12}^{+0.07}$ & $0.27_{-0.07}^{+0.05}$ &$11.4_{-0.1}^{+0.1}$& $11.02_{-0.01}^{+0.01}$ \\
129197\tablefootmark{d} & 1.474 & $1.20 \pm 0.25$ & $4.4 \pm 0.1$ & $0.83 \pm 0.01$ & $87 \pm 1$ & $>100$ & $197_{-6}^{+12}$ & $>0.48$ & $>0.11$ & $10.9_{-0.1}^{+0.1}$ & $10.52_{-0.02}^{+0.02}$ \\
129982 & 1.249 & $6.48 \pm 1.13$ & $3.6 \pm 0.1$ & $0.39 \pm 0.01$ & $30 \pm 1$ & $334_{-46}^{+52} $ & $213_{-2}^{+11}$ & $1.57_{-0.23}^{+0.26}$ & $0.52_{-0.06}^{+0.07}$ &$11.9_{-0.1}^{+0.1}$& $11.22_{-0.01}^{+0.01}$ \\
130040\tablefootmark{c} & 1.170 & $5.15 \pm 0.50$ & $3.9 \pm 0.2$ & $0.53 \pm 0.02$ & $67 \pm 2$ & $180_{-19}^{+17} $ & $285_{-5}^{+5}$ & $0.63_{-0.07}^{+0.06}$ & $0.29_{-0.04}^{+0.03}$ &$11.8_{-0.1}^{+0.0}$& $11.21_{-0.01}^{+0.01}$  \\
130208\tablefootmark{c} & 1.231 & $2.41 \pm 0.07$ & $2.3 \pm 0.1$ & $0.51 \pm 0.01$ & $60 \pm 1$ & $216_{-28}^{+28} $ & $291_{-16}^{+18}$ & $0.74_{-0.11}^{+0.11}$ & $0.28_{-0.04}^{+0.04}$ &$11.5_{-0.1}^{+0.1}$& $10.95_{-0.00}^{+0.00}$  \\
130647\tablefootmark{d} & 1.508 & $4.15 \pm 0.85$ & $3.8 \pm 0.1$ & $0.60 \pm 0.01$ & $89 \pm 1$ & $>80$ & $371_{-21}^{+15}$ & $>0.21$ & $>0.08$ & $11.9_{-0.1}^{+0.1}$ & $11.48_{-0.00}^{+0.00}$\\
        \hline\hline
    \end{tabular}}
    \tablefoottext{a}{PA offset with respect to the MSA shutter. 0\textdegree\ corresponds to the major-axis alignment, 90\textdegree\ corresponds to the minor-axis alignment.}\newline
    \tablefoottext{b}{Stellar masses for a \citet{GChabrier2003} IMF from \citet{MSlob2024}.}\newline
    \tablefoottext{c}{Galaxies without COSMOS-Web coverage. The structural parameters are taken from HST/ACS F814W imaging \citep{NScoville2007,RGriffith2012}.}\newline
    \tablefoottext{d}{Galaxies for which we cannot constrain the rotational velocity. We report 2$\sigma$ lower limits for $V_{r_e}, V_{r_e}/\sigma_0$, and $\lambda_{r_e}$.}
    \label{tab:fit_properties}
    \vspace{-0.15in}
\end{table*}

\subsection{Forward modelling}\label{sec:forward_model}
We used a forward-modelling procedure to derive the intrinsic kinematics from the observed velocity and the velocity dispersion profiles for each galaxy. The first step in this procedure is to generate model line profiles for a set of morphological and kinematic parameters using \msafit\ \citep{AdeGraaff2024_kinematics}. We refer to \citet{AdeGraaff2024_kinematics} for a detailed description of the modelling set-up, but provide a summary of the main steps and key differences between our methods and \msafit\ below.

Our kinematic model depends on nine parameters, which describe the surface brightness profile, location, and orientation with respect to the micro-shutters and intrinsic stellar kinematics of the source. For the surface brightness profile, we used a S\'ersic model \citep{JSersic1968}, parametrised by the half-light radius $r_e$, the minor-to-major axis ratio $q$, and the S\'ersic index $n$. The position in the central shutter ($\dd x, \dd y$), and the position angle (PA) with respect to the shutter together with the surface brightness profile define which part of the galaxy falls within the MSA shutters. We are only interested in $v$ and $\sigma_v$ and not in the intensity of the modelled line profiles. We therefore set the total flux $F$ of the S\'ersic profiles to a fixed number for each generated model. 

We used a thin-disc model for the velocity field, which is defined using an arctangent rotation curve \citep{SCourteau1997},
\begin{equation}\label{eq:arctan_v}
    v(r) = \frac{2}{\pi} v_a \arctan\left(\frac{r}{r_t}\right),
\end{equation}
where $v_a$ is the maximum velocity with respect to the systemic velocity of the galaxy, and $r_t$ is the turnover radius. To allow for a dynamically warm disc, we included a constant velocity dispersion profile $\sigma(r) = \sigma_0$. We note that a thin-disc model is likely too simplistic a description of the velocity profile of most galaxies in our sample, and we discuss the effects of this assumption in Section \ref{sec:discussion}.\\

The model was initialised using the nine parameters described above by generating flux and velocity profiles to obtain a model flux cube ($I(x,y,\lambda)$) in the MSA plane coordinates ($x, y$). To evaluate the steep S\'ersic profile in the centre, we over-sampled the innermost region of the profile ($< 0.05 r_e$) by a factor of 500 and the outer region ($> 0.2 r_e$) by a factor of 10. The profile was then integrated onto a coarser grid with a sampling equal to the NIRSpec pixel size (0.01\arcsec) or to the Nyquist frequency of the PSF if that was lower.

To mitigate PSF under-sampling effects and ensure that we took the spectral information into account across the full wavelength range of the spectrum, we generated three emission lines for each model galaxy, spaced evenly in wavelength. This number was a compromise between model accuracy and computation time. Although including more lines would slightly decrease the uncertainties, this would not affect our overall results. 

To model the NIRSpec/MSA PSF, we used a custom version of the PSF model described by \citet{AdeGraaff2024_kinematics}, with an oversampling greater by a factor of 2. This higher oversampling was necessary to ensure that the PSF was over-sampled compared to the Nyquist sampling.
Using \msafit, we convolved the model flux cube with the PSF following \citet{AdeGraaff2024_kinematics}, and this PSF-convolved model was then projected onto the two $2048\times2048$ pixel NIRSpec detectors, following the trace for the shutter ($s_{ij}$) in which the source was observed. The resulting model spectrum, showing three emission lines, encapsulates slit losses, bar shadows, interpixel capacitance, and spatial under-sampling. This model is equivalent to a noiseless un-rectified single-frame NIRSpec/MSA observation. 

To convert these single-frame model spectra into combined rectified model spectra, we generated a model for each nod (i.e. shutter position $s_{ij}$) using the above method. We then applied all reduction steps that were made on the observed data (see Section \ref{sec:data}) to reduce and combine the individual frames. This step resulted in a single rectified combined 2D model spectrum containing the three modelled emission lines. The line profiles of these lines accounted for optics and detector effects (from \msafit) and for resampling and combination effects (from the data reduction steps).

Finally, we obtained  row-by-row line-of-sight velocity ($V_{l,i,\text{mod}}$) and dispersion ($\sigma_{v,l,i,\text{mod}}$) measurements of each line $l$ by fitting a Gaussian profile to each row $i$. By taking the median value of $V_{i,\text{mod}}$ and $\sigma_{v,i,\text{mod}}$ of the three lines, we obtained line-of-sight velocity and dispersion profiles for each model. In Figure \ref{fig:extreme_cases} we illustrate the line-of-sight velocity profiles for galaxy models with different velocity profiles and slit alignments.

\subsection{Parameter inference}\label{sec:sampling}
\msafit\ was originally developed to perform Markov chain Monte Carlo (MCMC) sampling. However, the adaptations to the code described in the previous section make it significantly more expensive in both RAM and CPU usage,  and thus it is currently unfeasible to use MCMC sampling for our method. We therefore performed the fitting using {\sc scipy.optimize}, evaluated using the $\chi^2$ metric, which requires fewer models to be generated.

We fit for five parameters in our modelling ([$\dd x, \dd y, v_a, r_t, \sigma_0$]). The remaining four parameters, $n, q, r_e,$ and PA were fixed to the {\sc GALFIT} measurements because the uncertainty in these parameters was small, and the fitting times were significantly reduced with fewer free parameters. We initialised the slit position $\dd x$ and $\dd y$ as the astrometric solution in the slit, and let the values vary by 0.75 pixels in either direction to allow for pointing and astrometric inaccuracies. For each galaxy, we estimated the initial guess for $v_a$ by eye and set the initial guess for $\sigma_0$ as the integrated velocity dispersion measured from {\sc ppxf}. We allowed $v_a$ and $\sigma_0$ to vary 200\,km/s around this initial guess in the fit. The choice of initial $v_a$ and $\sigma_0$ does not affect the results of our fitting, but a reasonable initial guess speeds up the sampling process. We initialised $r_t$ as 0.75$r_e$ and allowed it to vary from 0.05$r_e$ to 2$r_e$.

We performed the fitting using two iterations of {\sc scipy.optimize}. In the first iteration, we used the Powell method, with a tolerance of $10^{-2}$. This sampling method incorporates random jumps within the bounds of the parameter space to ensure that the sampler explores the full grid and does not become stuck in a local minimum. For the second iteration, the model was initialised at the best-fit parameters of the first iteration, and we then ran {\sc scipy.optimize} with the Nelder-Mead method and a tolerance of $10^{-4}$ to explore a finer grid around these best-fit values. 

We obtained errors on the best-fit model and inferred velocity parameters by creating 2000 simulations of the data, randomly perturbed around the uncertainties of the data points and GALFIT parameters. For each perturbed sample, we determined the best-fit model and parameters from the models that were generated in the fitting routine. The uncertainties on the model and best-fit parameters were calculated from the standard deviation of all perturbed samples.

\section{Results}\label{sec:results}
\subsection{Kinematic properties}
\begin{figure}
    \centering
    \includegraphics[width=\linewidth]{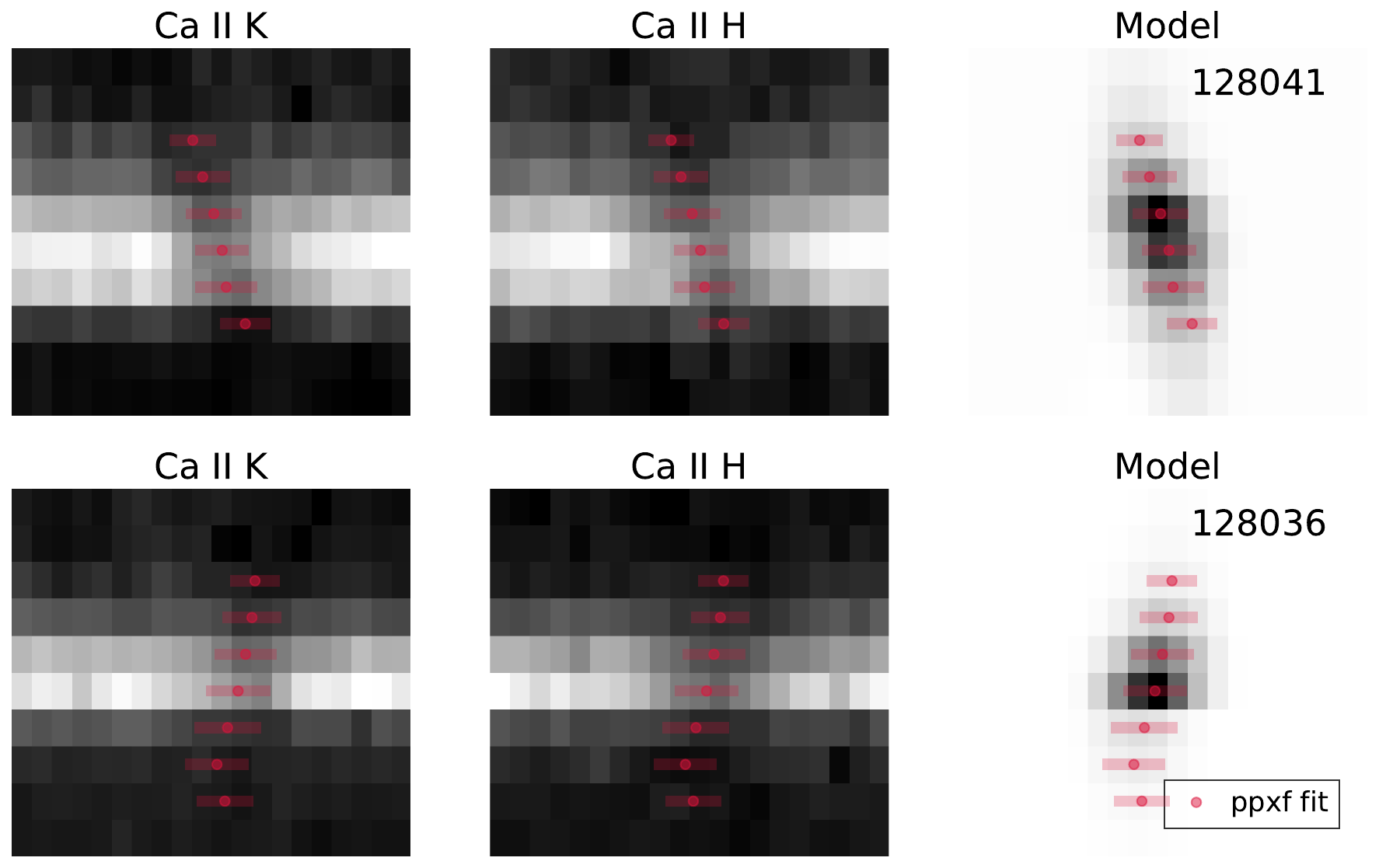}
    \caption{2D absorption line profiles of Ca{\sc ii} K (left) and H (middle), and the best-fit model (right) for two example galaxies. The red points (lines) show the best-fit velocity (velocity dispersion) profile from {\sc ppxf} that was used as input for our forward modelling in Section \ref{sec:forward_model}. We note that the {\sc ppxf} fits were obtained from the entire wavelength range of the spectra, and the two absorption lines illustrated in this figure serve as an example.}
    \label{fig:2d_line_prof}
    \vspace{-0.15in}
\end{figure}

We present the results of our kinematic forward-modelling procedure for all 15 distant quiescent galaxies in Table \ref{tab:fit_properties}. In Figure \ref{fig:2d_line_prof} we show example absorption line profiles in the 2D spectra and our inferred best-fit model, with the best-fit {\sc ppxf} line profiles overlaid. In Figure \ref{fig:v_curves} we show the best-fit line-of-sight velocity and dispersion models in 1D, and inferred 2D space for the ten galaxies for which we can constrain the rotational velocity. We show the remaining five galaxies in Appendix \ref{ap:unconstrained_fits}. We also show the NIRCam or HST/F814W imaging for all galaxies. We overplot the MSA slit positions for the imaging and the inferred 2D velocity fields.

In total, we can constrain the rotational velocity for 10 out of 15 targets in our sample. Four out of five of the remaining galaxies in our sample are aligned along the minor axis with respect to the MSA shutter ($\Delta$PA$>70$\textdegree), and galaxy 127154 has posterior with a strong degeneracy between the source position and velocity. We obtained only lower limits on the rotational velocity of these 5 galaxies. The remaining 10 galaxies for which we could, in principle, measure rotational velocities rotate substantially, with the model deprojected $V_{r_e}$ ranging from $114 - 336$~\kms, with a median value of 203~\kms. The intrinsic velocity dispersions range from $197-387$~\kms, with a median value of 241~\kms.

We note that for two galaxies (129197 and 130647), the stellar velocity models provide a poor fit to the kinematic data in Figure \ref{fig:v_models_unconstrained}. For 129197 the discrepancy between model and data likely arises because the galaxy is misaligned with respect to the MSA shutter, which prevents us from constraining the velocity curve in the outskirts. While 130647 is also aligned along the minor axis with respect to the shutter, we observe a velocity excess in centre of the galaxy, rather than in the outskirts. This velocity excess may be due to a massive central black hole \citep{ANewman2025}. In this context, it is interesting to note that the NIRSpec spectrum of this galaxy exhibits strong emission lines that likely originate from an AGN \citep{MSlob2024}. Spatially resolved IFU observations are needed to fully constrain the stellar kinematics in the central region of this galaxy and assess the origin of the central velocity excess.

\begin{figure*}[p]
    \centering
    \vspace{-.55in}
    \subfigure{\includegraphics[width=0.473\linewidth]{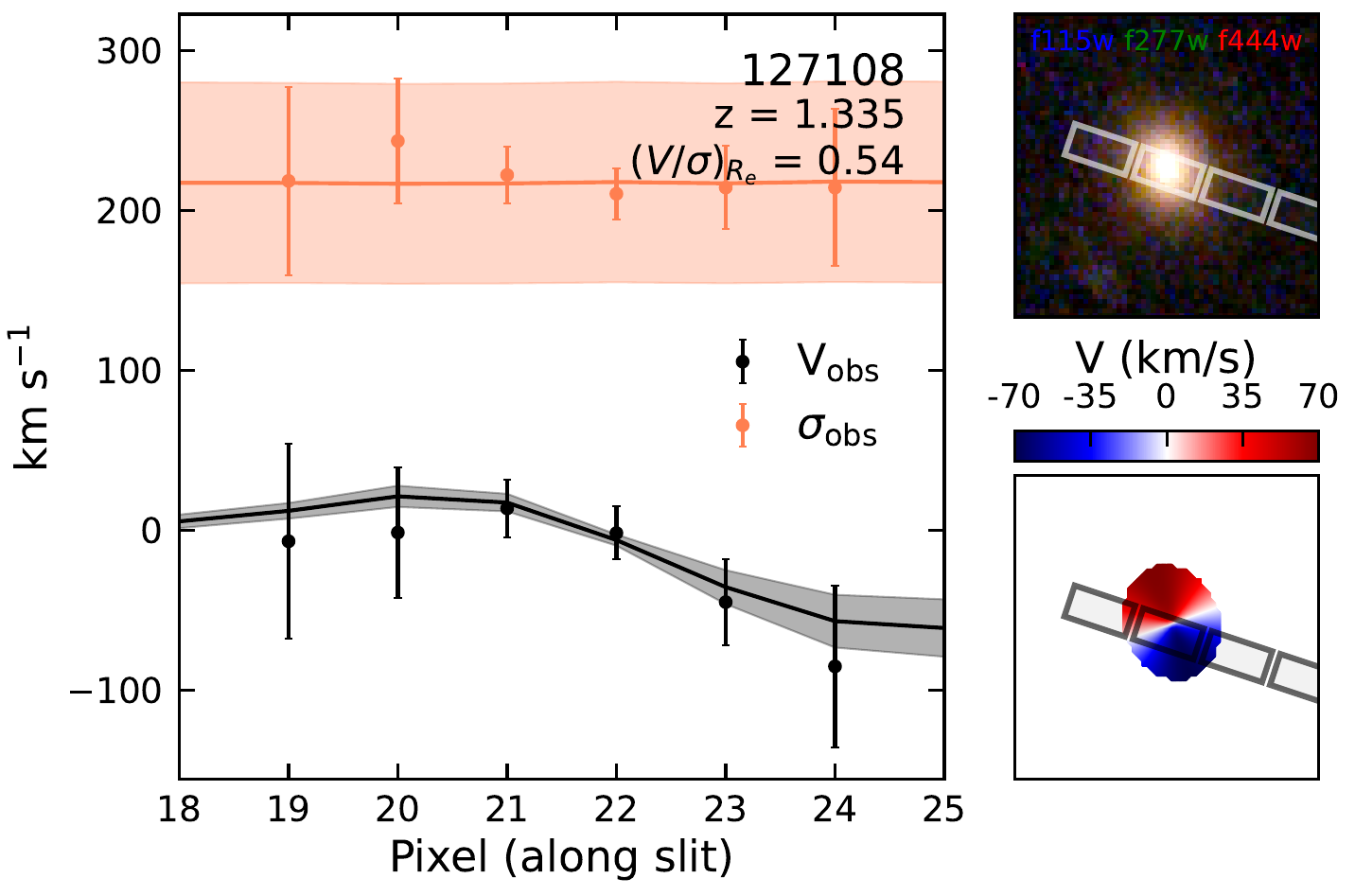}}%
    \subfigure{\includegraphics[width=0.473\linewidth]{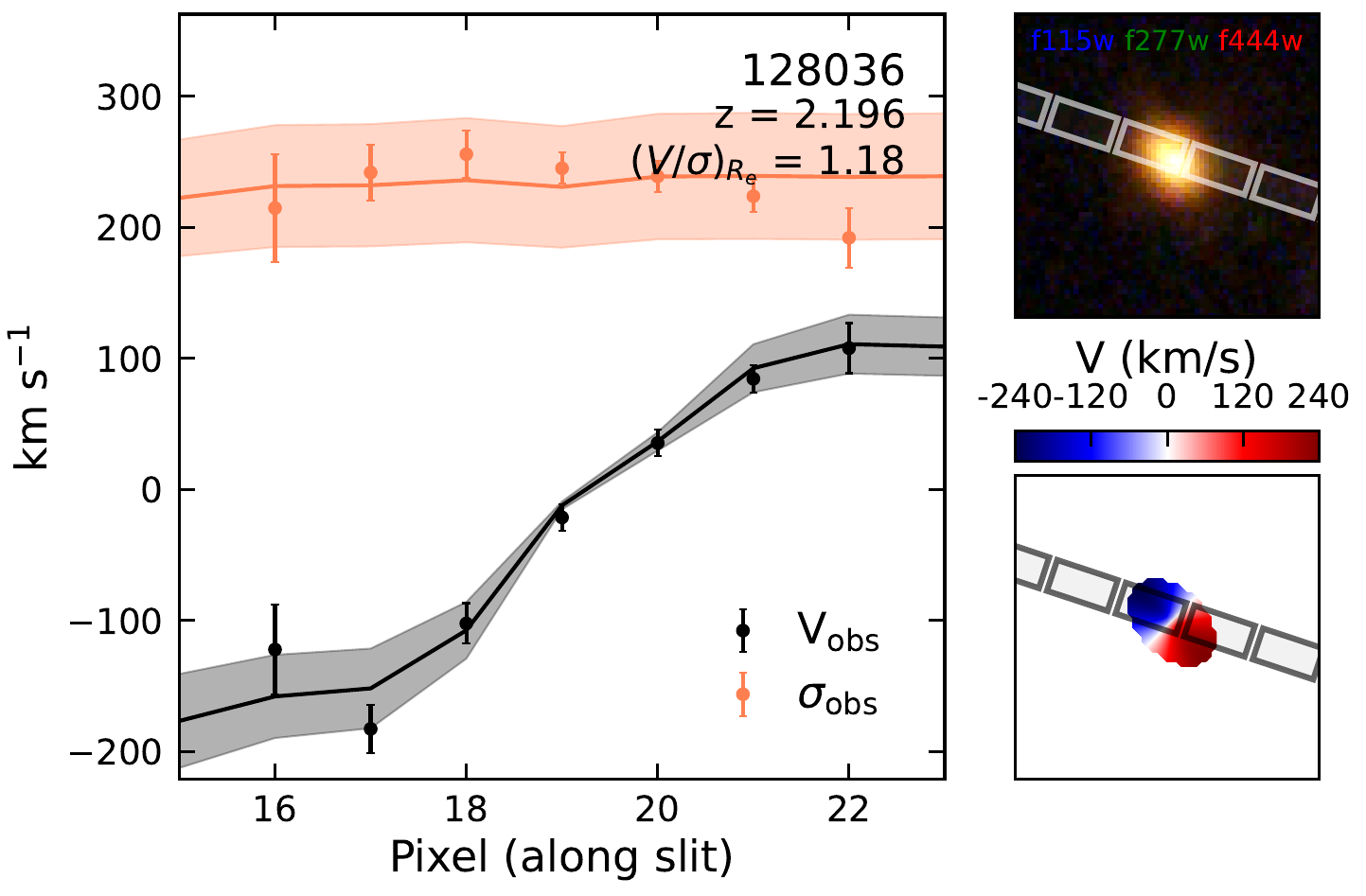}}    \subfigure{\includegraphics[width=0.473\linewidth]{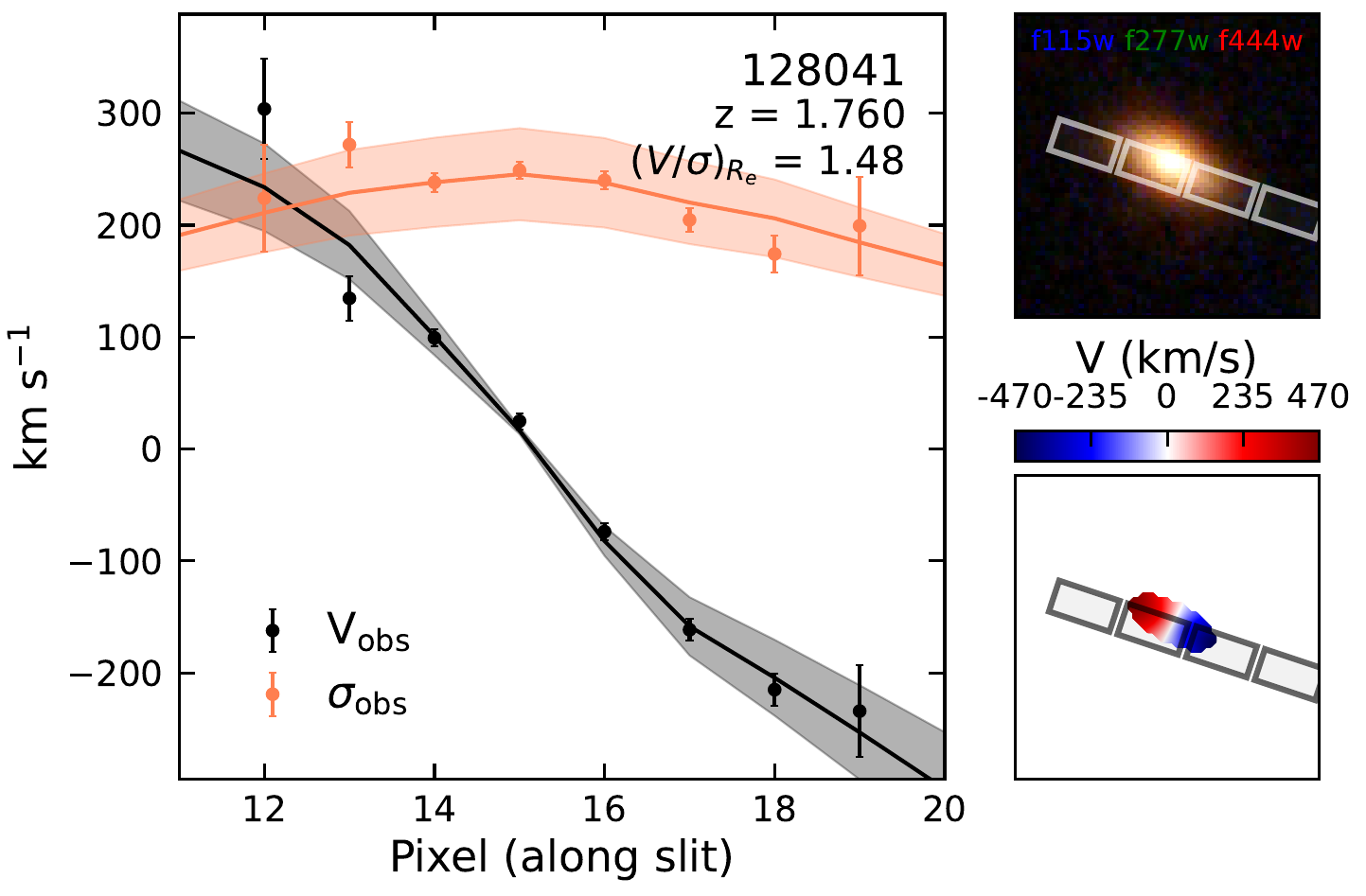}}%
    \subfigure{\includegraphics[width=0.473\linewidth]{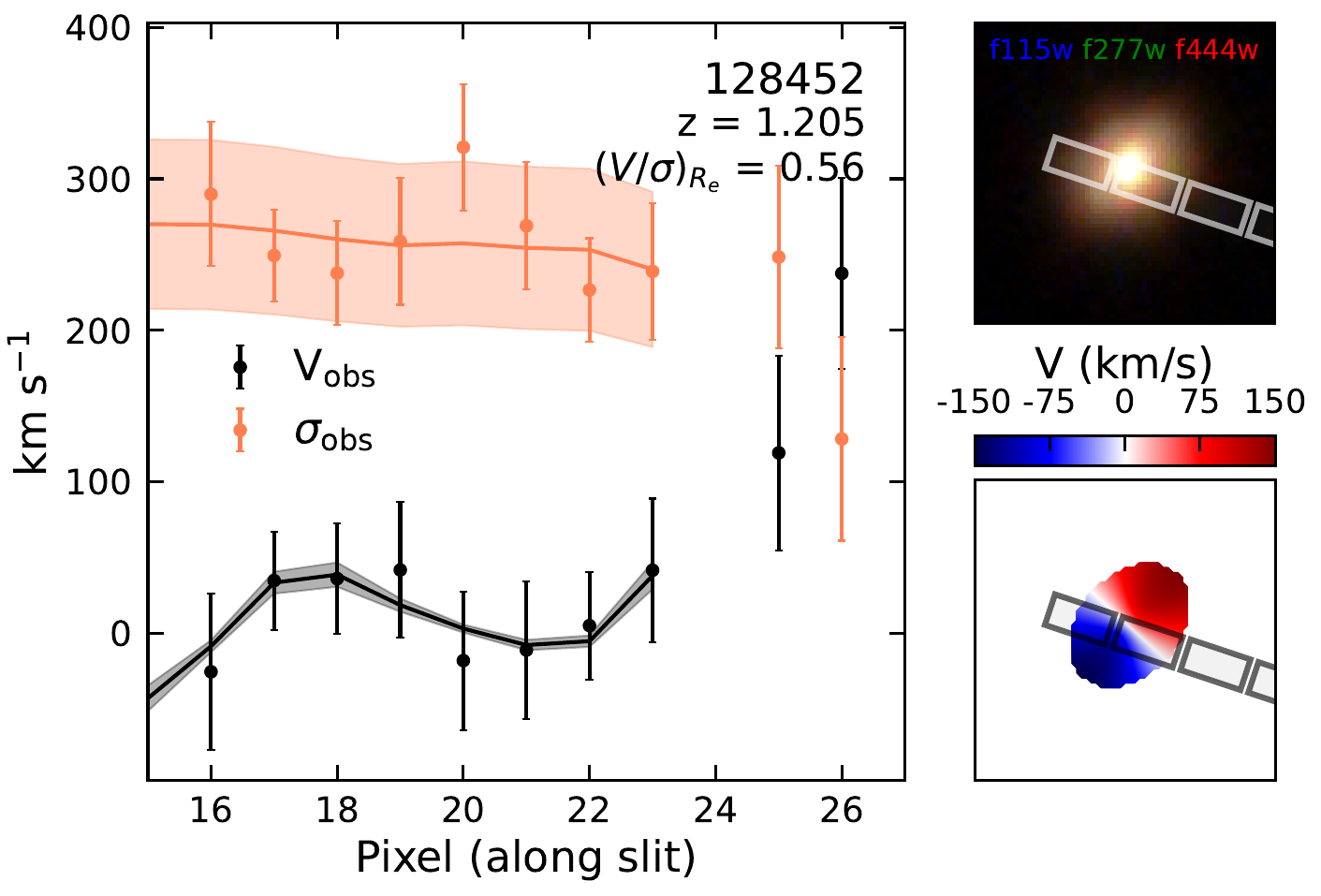}}
    \subfigure{\includegraphics[width=0.473\linewidth]{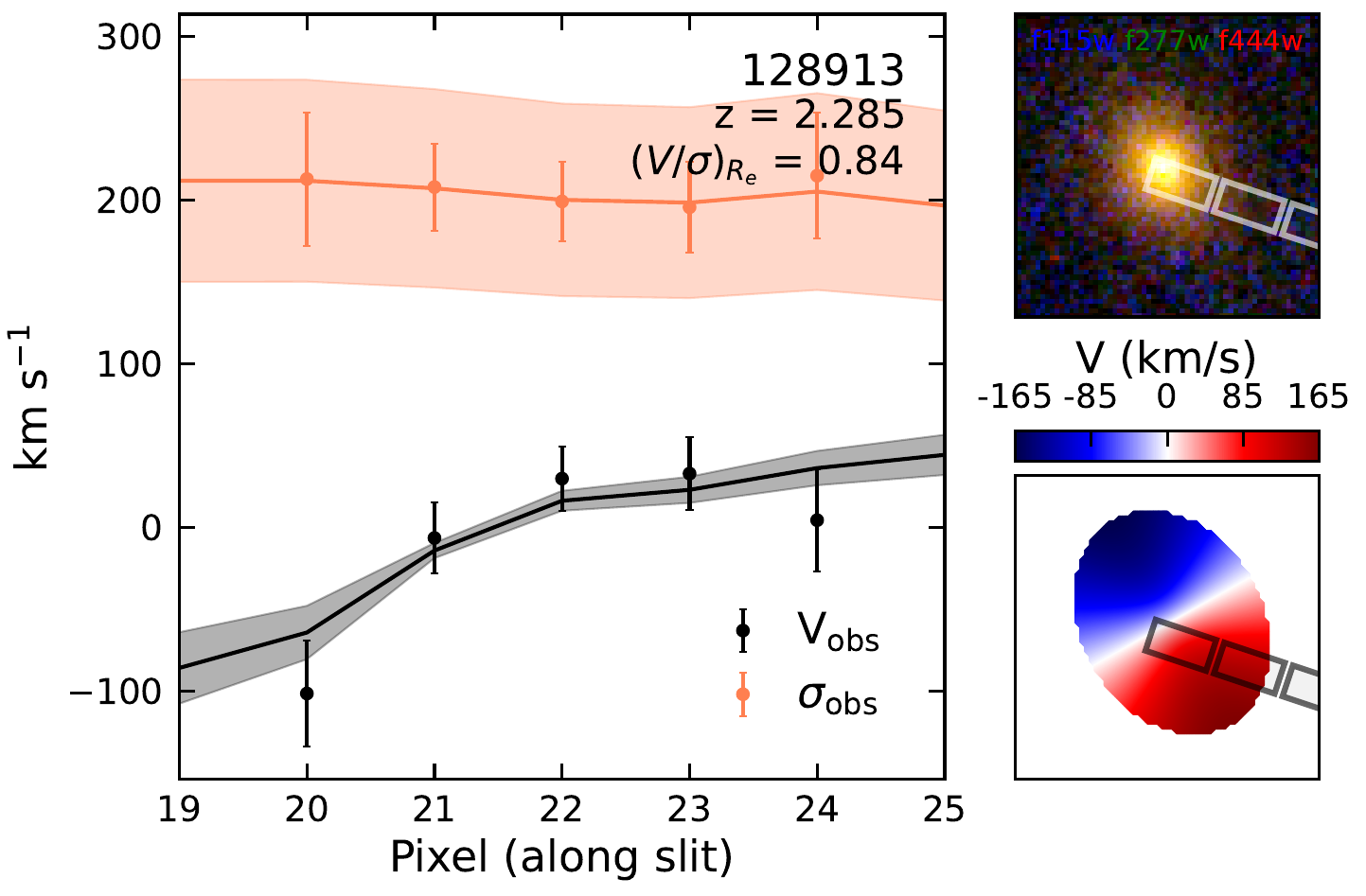}}%
    \subfigure{\includegraphics[width=0.473\linewidth]{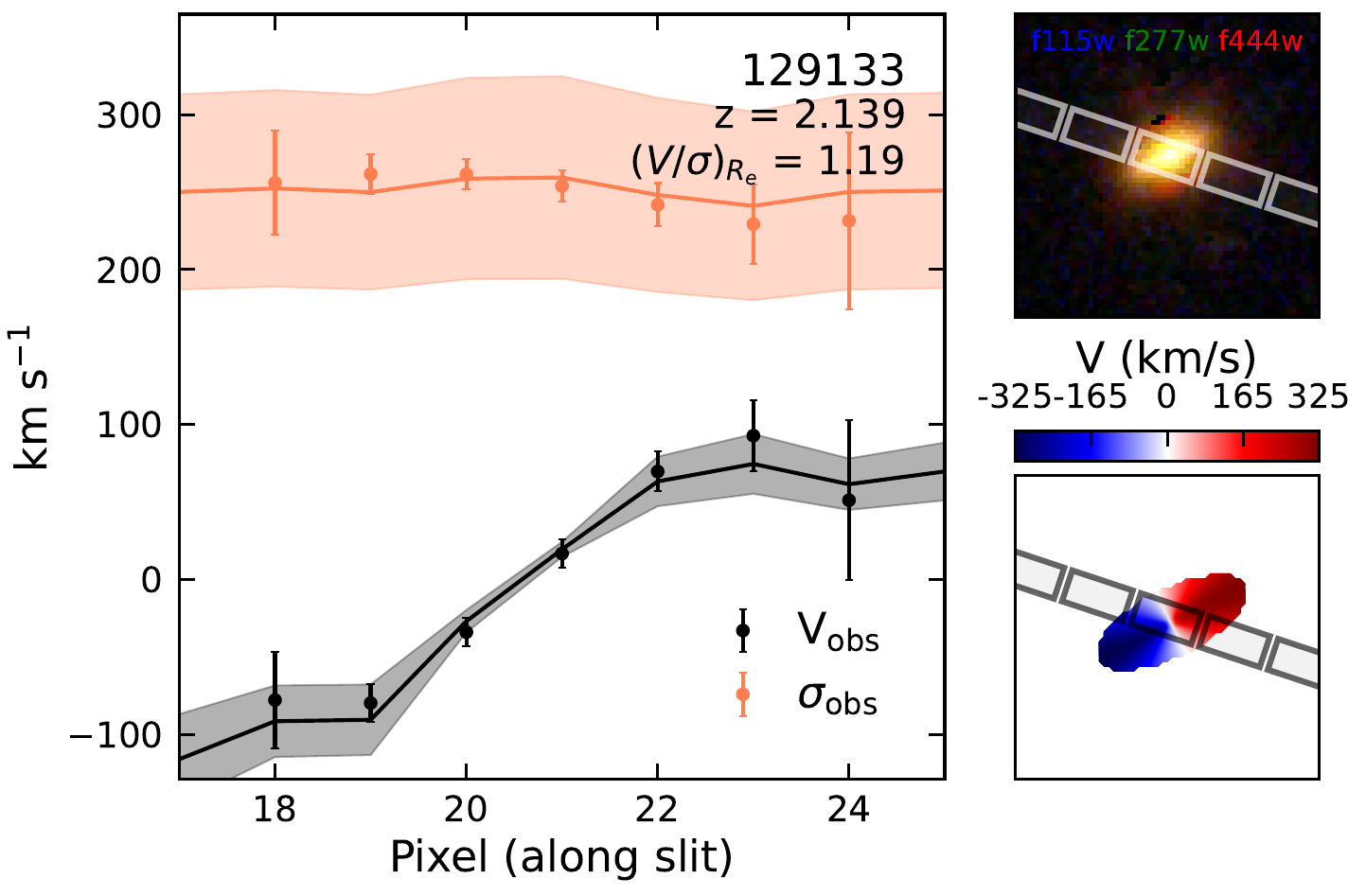}}
    \subfigure{\includegraphics[width=0.473\linewidth]{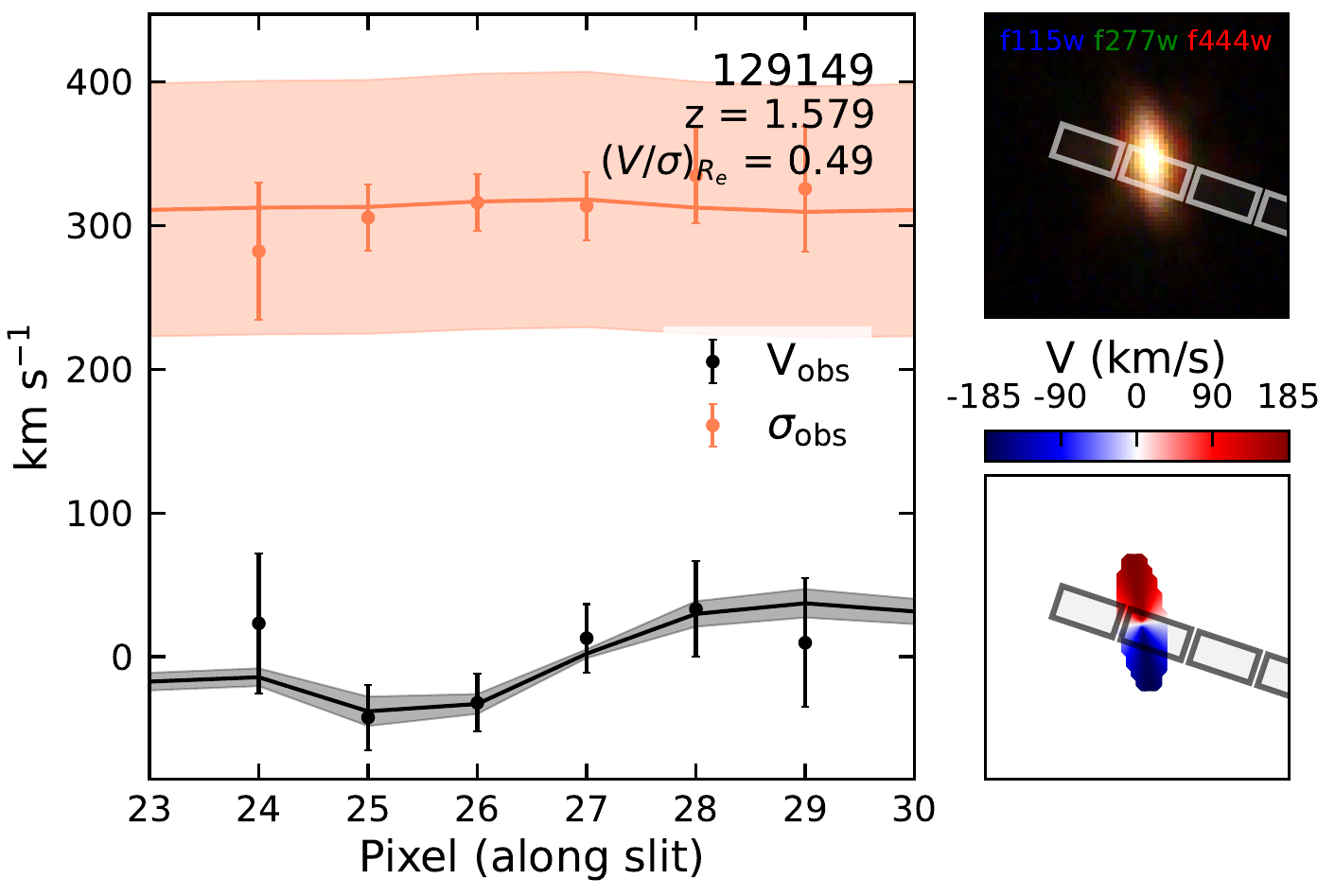}}%
    \subfigure{\includegraphics[width=0.473\linewidth]{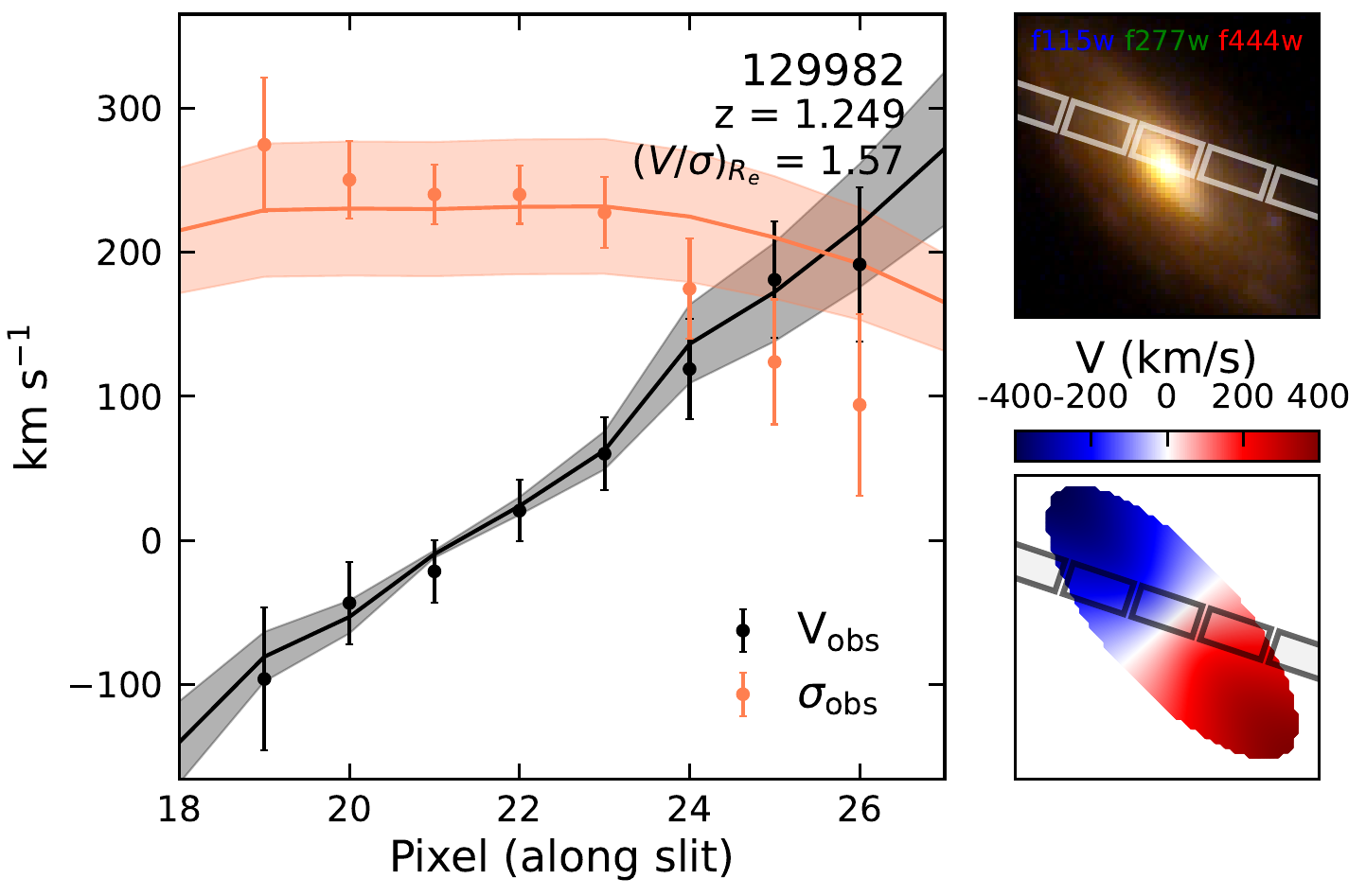}}
    \caption{Observed kinematics and best-fit models for the ten distant quiescent galaxies for which we can constrain rotational velocities. In the left panels, we show the observed velocities (dispersions) as the black (orange) data points, and the best-fit models are represented by the corresponding lines. In the top right panels we show  NIRCam RGB imaging \citep{CCasey2023} where available, or HST F814W imaging otherwise \citep{NScoville2007}. In the bottom right panels we show the inferred 2D line of sight velocity fields. We overlay the MSA microshutter positions in both panels on the right.}
    \label{fig:v_curves}
\end{figure*}

\begin{figure*}[t]
    \setcounter{figure}{2}
    \centering
    \subfigure{\includegraphics[width=0.473\linewidth]{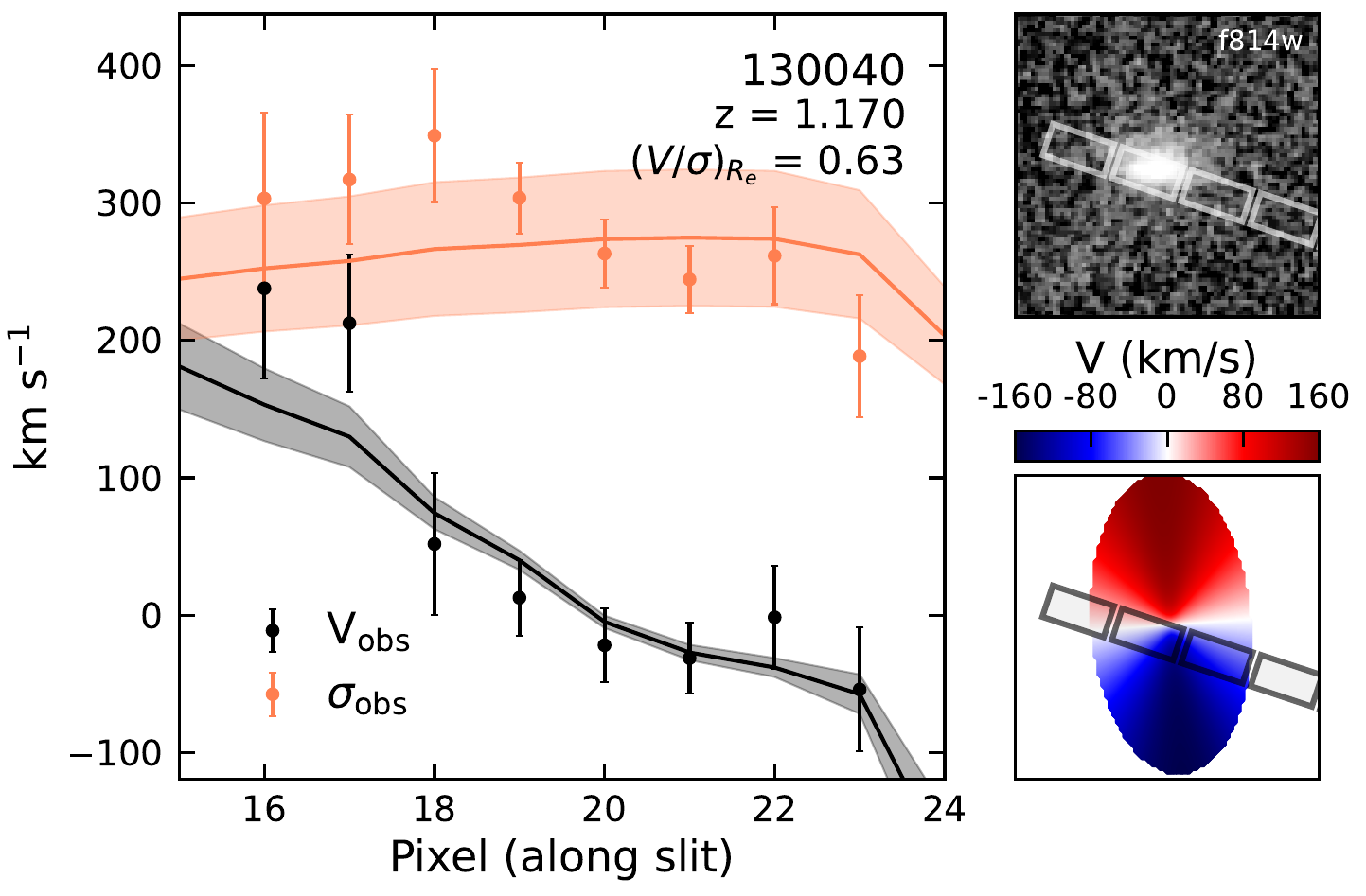}}%
    \subfigure{\includegraphics[width=0.473\linewidth]{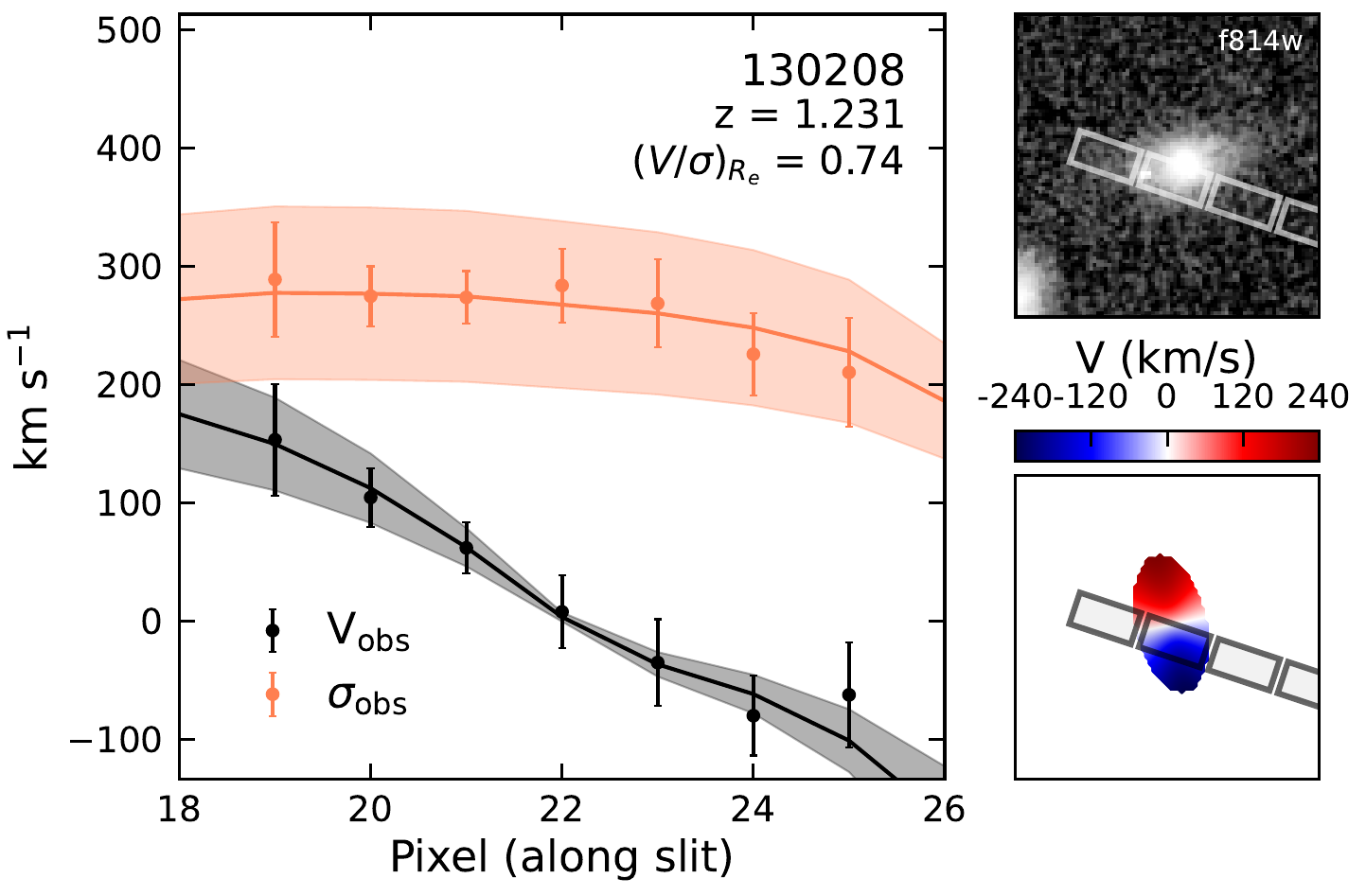}}
    \caption{Continued.}
    \vspace{-0.15in}
\end{figure*}

To quantify the degree of rotational support in our sample, we computed the deprojected $(V/\sigma)_{r_e} \equiv V(r_e)/\sigma_0$, which is a direct measure of the intrinsic rotational support in disc galaxies. For the ten galaxies in our sample for which we can constrain the rotational velocity, $(V/\sigma)_{r_e}$ ranges from 0.54 to 1.57, with a median of 0.79. Four galaxies have \vsigma $>1$, corresponding to a rotationally supported disc.

\begin{figure}[t]
    \centering
    \includegraphics[width=0.97\linewidth]{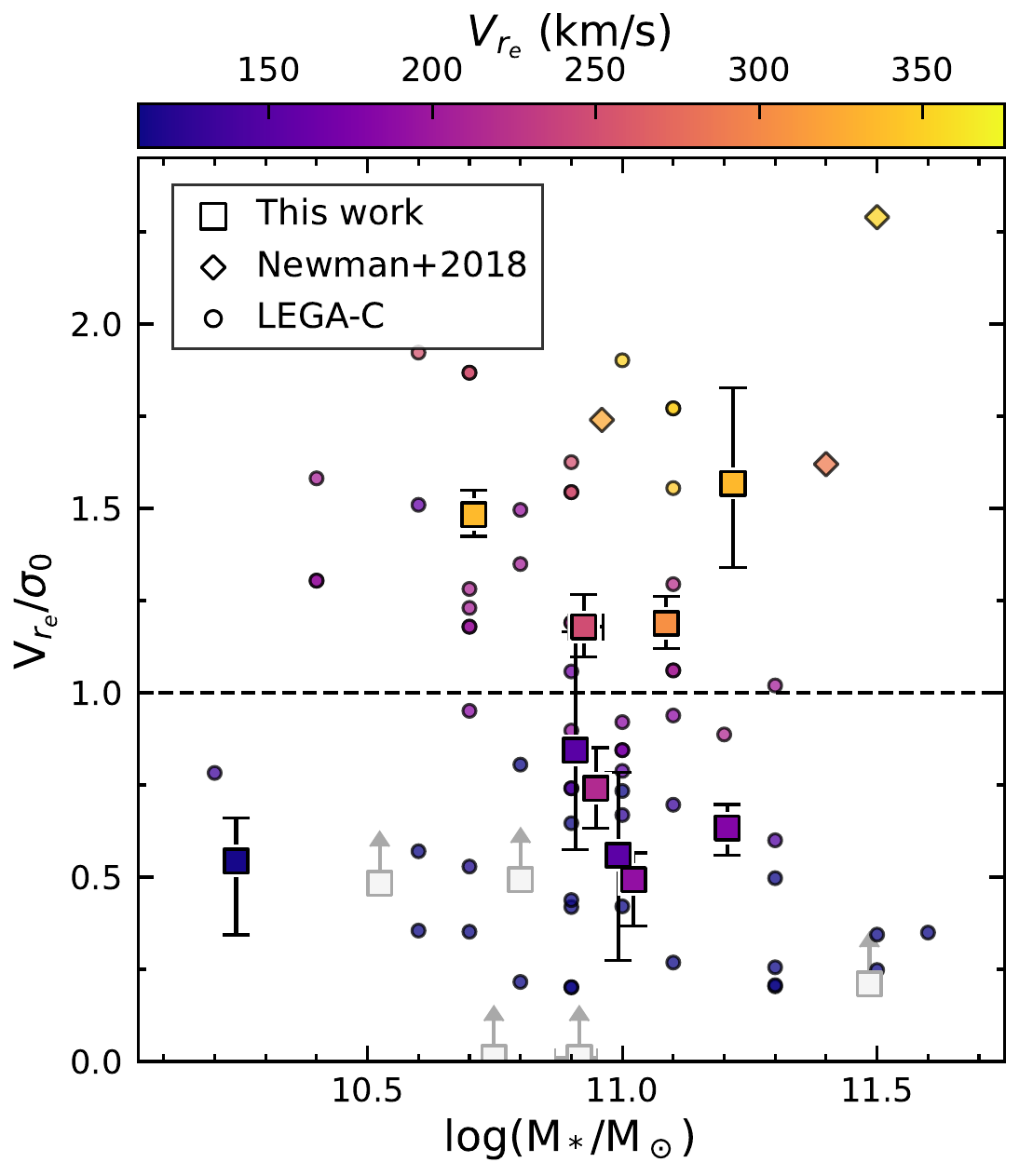}
    \caption{$V_{r_e}/\sigma_0$ as a function of stellar mass for the 15 distant quiescent galaxies in our sample (squares). The grey squares indicate galaxies for which we can only obtain lower limits to $V_{r_e}/\sigma_0$. We also show quiescent galaxies at $z\sim0.8$ from LEGA-C \citep[circles,][]{JvanHoudt2021} and three lensed quiescent galaxies at $z\sim2$ \citep[diamonds,][]{ANewman2018b}. The dashed line indicates a ratio of 1, corresponding to the definition of rotational support. The symbols are coloured by $V_{r_e}$.}
    \label{fig:v_over_sigma_mass}
    \vspace{-0.15in}
\end{figure}

We plot $(V/\sigma)_{r_e}$ as a function of stellar mass in Figure \ref{fig:v_over_sigma_mass}. We also show the three quiescent galaxies from \citet{ANewman2018b} at $z>2$ in this plot. Furthermore, to compare with galaxies with similar masses at lower redshifts, we show the $V/\sigma$ ratios for quiescent galaxies in the LEGA-C survey \citep{AvanderWel2021} at $z\sim0.7$ (\citealt{JvanHoudt2021}; see also \citealt{RBezanson2018}). We find no trend between mass and \vsigma\ for the quiescent galaxies in these samples. We colour the symbols in Figure \ref{fig:v_over_sigma_mass} by $V_{r_e}$ and find that galaxies with similar degrees of rotational support (\vsigma) cover a range of $V_{r_e}$. We discuss the implications of these results further in Section \ref{sec:discussion_gal_evol}.

To further compare the rotational properties of our sample with samples of nearby early-type galaxies, we also computed the spin parameter, $\lambda_{r_e}$ \citep{EEmsellem2007,EEmsellem2011}. This parameter is a proxy for the angular momentum per unit mass, defined as 
\begin{align}\label{eq:spin_param}
    \lambda_{r_e} = \frac{\Sigma_{r < r_e}F_i r_i |V_i|}{\Sigma_{r<r_e}F_i r_i \sqrt{V_i^2 + \sigma_{v,i}^2}}.
\end{align}
Here, $r_i$ is the radius of a spatial bin, $F_i$ is the flux of the S\'ersic profile, $V_i$ is the line-of-sight stellar velocity, and $\sigma_{v,i}$ is the line-of-sight stellar dispersion. We computed $V_i$ by projecting the best-fit arctangent rotation curve (Eq. \eqref{eq:arctan_v}) to 2D space, applying the relevant inclination effects for a thin disc. We took $\sigma_{v,i} = \sigma_0$ across the galaxy.

\begin{figure}[t]
    \centering
    \includegraphics[width=0.97\linewidth]{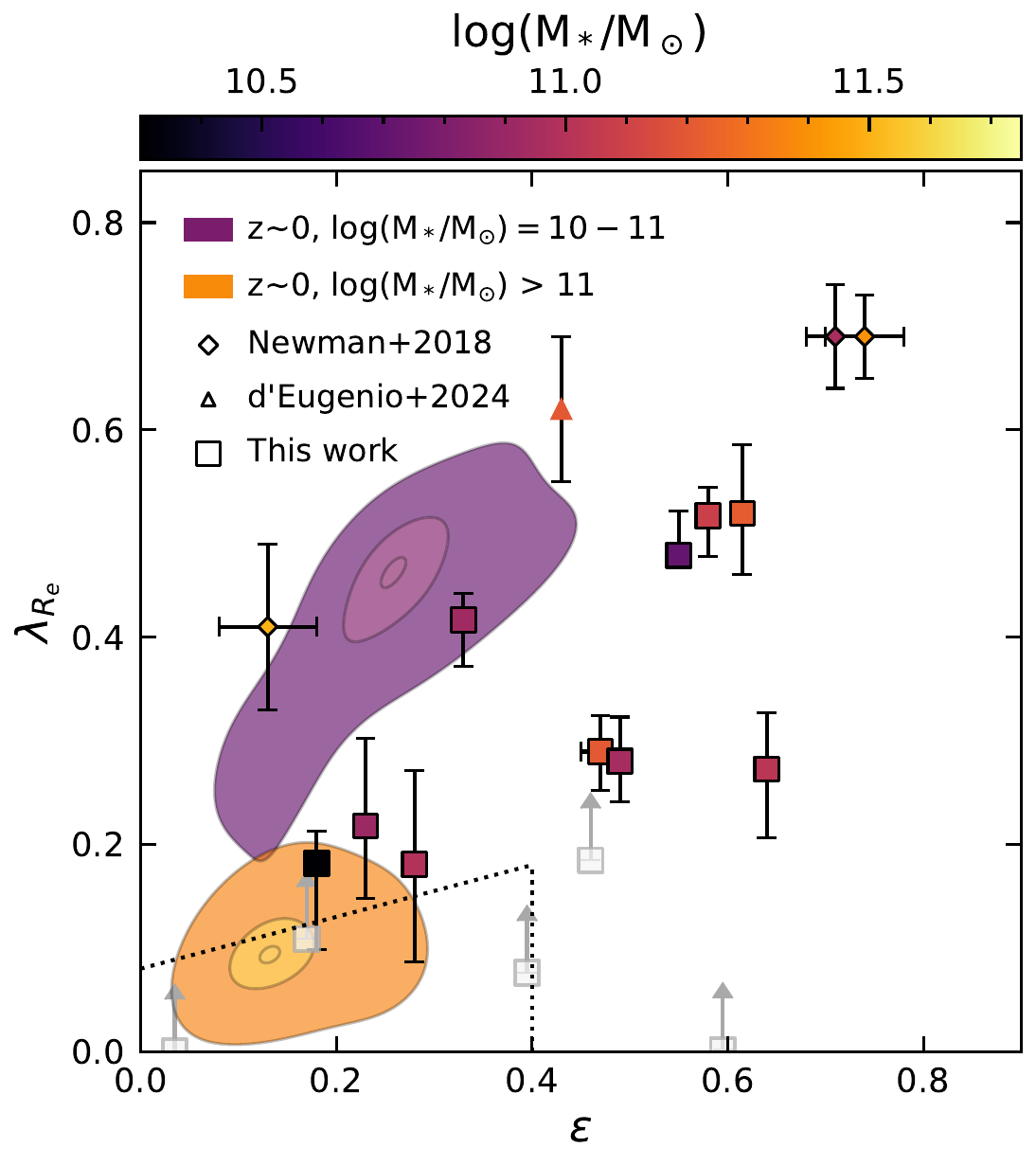}
    \caption{Spin parameter $\lambda_{r_e}$ as a function of projected ellipticity $\epsilon$ for the distant quiescent galaxies in our sample (squares). We also show four previously identified quiescent rotating discs at $z = 1.95-3.06$ from \citet{ANewman2018b} (diamonds) and \citet{FdEugenio2024} (triangle). The data points are coloured by stellar mass. We show the distribution of low-$z$ early-type galaxies from the SAMI \citep{JvandeSande2017}, ATLAS3d \citep{EEmsellem2011}, and MASSIVE surveys \citep{MVeale2017} in two  mass ranges as contours. The dotted line shows the separation between fast and slow rotators defined by \citet{MCappellari2016}; all ten galaxies in our sample for which we can constrain rotational support are classified as fast rotators. The five galaxies for which we could only obtain lower limits for the spin parameter are shown in grey.}
    \label{fig:lamda_re_vs_ell}
    \vspace{-0.15in}
\end{figure}

In Figure \ref{fig:lamda_re_vs_ell} we show $\lambda_{r_e}$ as a function of ellipticity $\epsilon$. Here, $\epsilon$ is the ellipticity from the elliptical isophote fit to the relevant imaging at one $r_e$. For the five galaxies for which we cannot constrain rotational velocities (grey crosses), we show the lower limits of $\lambda_{r_e}$. The ten galaxies for which we were able to measure the rotational support are all classified as fast rotators based on their location in the $\lambda_{r_e}-\epsilon$ plane \citep{MCappellari2016}. Compared to the four quiescent galaxies at $z>2$ for which kinematic measurements were previously available \citep{ANewman2018b, FdEugenio2024}, which are also shown in this figure, the galaxies in our sample appear to be less rotationally supported from their position in the $\lambda_{r_e}-\epsilon$ plane. We discuss possible explanations for this difference in Section \ref{sec:discussion_gal_evol}.

We note that while all ten galaxies in our sample for which we can constrain the velocity are identified as fast rotators from their position in the $\lambda_{r_e}-\epsilon$ plane, only four of these galaxies have $(V/\sigma)_{r_e} > 1$, corresponding to a rotationally supported disc. This difference in classification likely arises because the \vsigma\ criterion is too simplistic for stellar kinematics of quiescent galaxies, as they are probably not described by simple thin disc models. Furthermore, the boundary between fast and slow rotators in the $\lambda_{r_e}-\epsilon$ plane is defined empirically, so we do not expect this to directly correspond to the $(V/\sigma)_{r_e} > 1$ definition. We discuss this in more detail in Section \ref{sec:discussion_caveats}.

We also show the distributions of low-$z$ quiescent galaxies from the SAMI \citep{SCroom2012,JvandeSande2017,SCroom2021}, ATLAS3D \citep{MCappellari2011,EEmsellem2011}, and MASSIVE surveys \citep{CMa2014,MVeale2017} in the $\lambda_{r_e}-\epsilon$ plane using the catalogues described by \citet{Jvandesande2019}. Our galaxies are faster rotators than nearby early-type galaxies with $\log(M_{\odot}/M_*)>11$. However, at lower masses (10<$\log(M_{\odot}/M_*)<11$), the low-$z$ population occupies the same region in the $\lambda_{r_e}-\epsilon$ plane as our sample. We discuss this further in Section \ref{sec:discussion_gal_evol}.

\subsection{Galaxy sizes}\label{sec:sizes}
\begin{figure}
    \centering
    \includegraphics[width=0.97\linewidth]{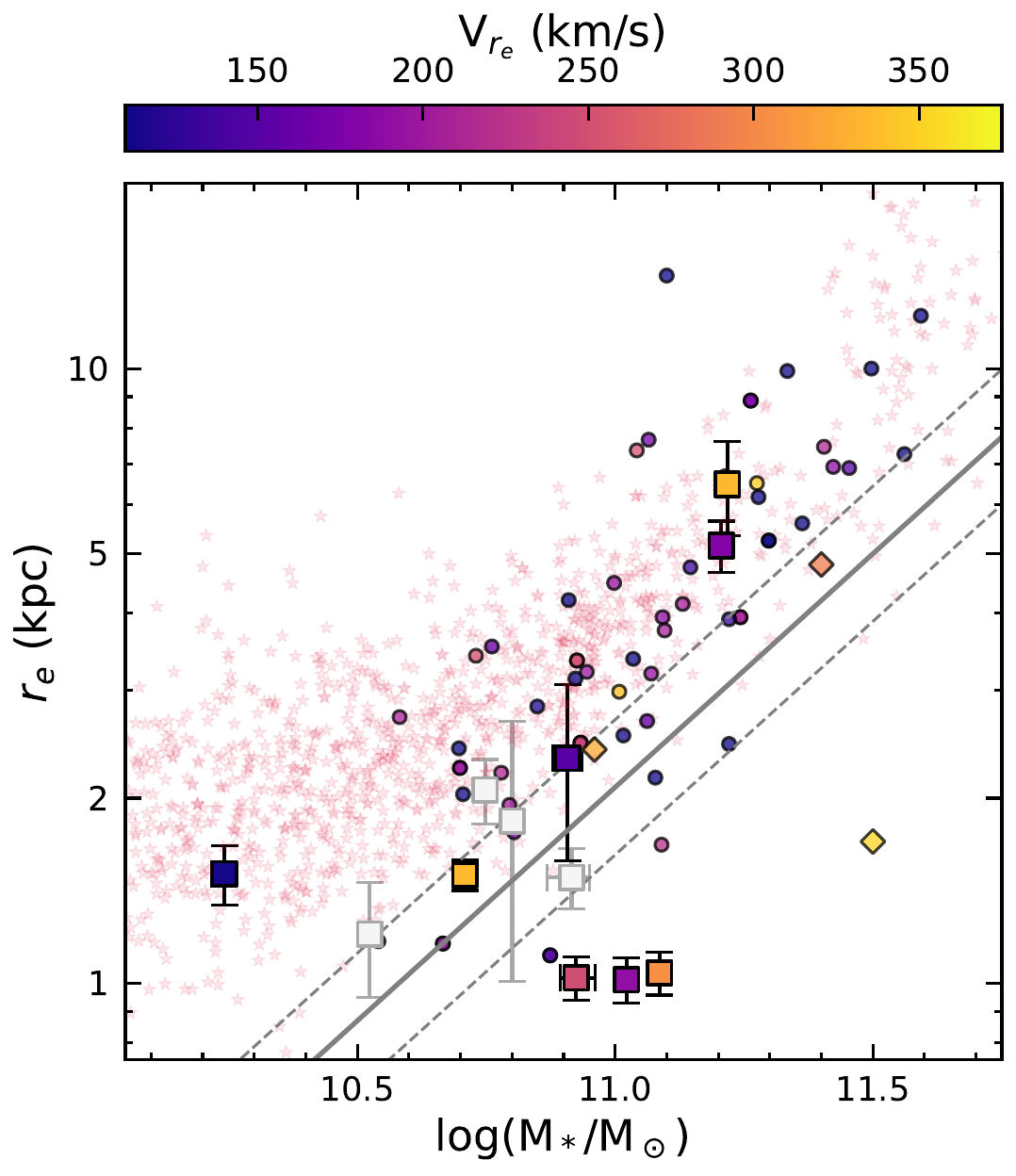}
    \caption{$r_e$ vs stellar mass for the galaxies in our sample (squares). We also show quiescent galaxies at $z\sim0.7$ from LEGA-C \citep[circles,][]{JvanHoudt2021}, three lensed quiescent galaxies at $z\sim2$ \citep[diamonds,][]{ANewman2018b}, and galaxies at $z\sim0$ from the SAMI \citep{JvanHoudt2021}, MASSIVE \citep{MVeale2017}, and ATLAS3D \citep{EEmsellem2011} surveys as red stars. The LEGA-C and SUSPENSE points are coloured by $V_{r_e}$, and the galaxies for which we could only obtain lower limits for $V_{r_e}$ are shown in grey. We show the linear fit to quiescent galaxies at $z\sim1.75$ from the 3D-HST+CANDELS survey \citep{AvanderWel2014} in grey, with the dashed lines indicating the 1$\sigma$ scatter.}
    \label{fig:mass_size}
\end{figure}

In Figure \ref{fig:mass_size} we show where our galaxies fall in the mass-size plane, using the rest-frame optical sizes derived in Section \ref{sec:galfit}. We also plot the HST/ACS F814W sizes of quiescent galaxies from the LEGA-C sample at $z\sim0.7$ \citep{JvanHoudt2021} and the HST/WFC3 F160W sizes of three lensed quiescent galaxies at $z\sim2$ \citep{ANewman2018b}. For both of these samples, the filters used to measure the sizes correspond to rest-frame optical wavelengths at their respective redshifts.
We also plot the optical sizes of $z\sim0$ quiescent galaxies from the SAMI \citep{JvandeSande2017}, MASSIVE \citep{MVeale2017}, and ATLAS3D \citep{EEmsellem2011} surveys. The majority of the galaxies in our sample are classified as ``compact" \citep[][]{AvanderWel2014} and are consistent with the mass-size relation of quiescent galaxies at $z\sim1.75$ from the 3D-HST+CANDELS survey \citep{AvanderWel2014}. We coloured the data points by $V_{r_e}$ and found no trend correlation between the amount of rotation and the offset from the coeval mass-size relation.

\subsection{Dynamical masses}\label{sec:dyn_mass}
From the best-fit kinematic and structural parameters, we can derive the total dynamical mass of the galaxies in our sample, as defined by \citet{SPrice2022},
\begin{align}\label{eq:Mdyn_Price22}
    M_{\mathrm{dyn, tot}} = k_{\mathrm{tot}} \frac{v^2_{\mathrm{circ}}(r_e) r_e}{G}.
\end{align}
Here, $v_{\mathrm{circ}}^2 = v^2(r_e) + 3.35\ \sigma_0^2$ \citep{ABurkert2010}, where the factor 3.35 accounts for an asymmetric drift correction from the velocity dispersion. We discuss this further in Section \ref{sec:discussion_caveats}. We assumed a thin-disc model in our dynamic modelling and therefore used $k_{\mathrm{tot}} = 1.8$, which is the virial coefficient for an oblate potential with an intrinsic axis ratio $q_0 = 0.2$ and $n \sim 1-4$ \citep{SPrice2022}. We note that the thin-disc model is likely a too simple description for many galaxies in our sample, and the true value of $k_{\mathrm{tot}}$ may be higher. In Section \ref{sec:discussion_caveats} we further discuss how the thin-disc assumption influences our dynamical mass estimates.

\begin{figure}[t]
    \centering
    \includegraphics[width = 0.97\linewidth]{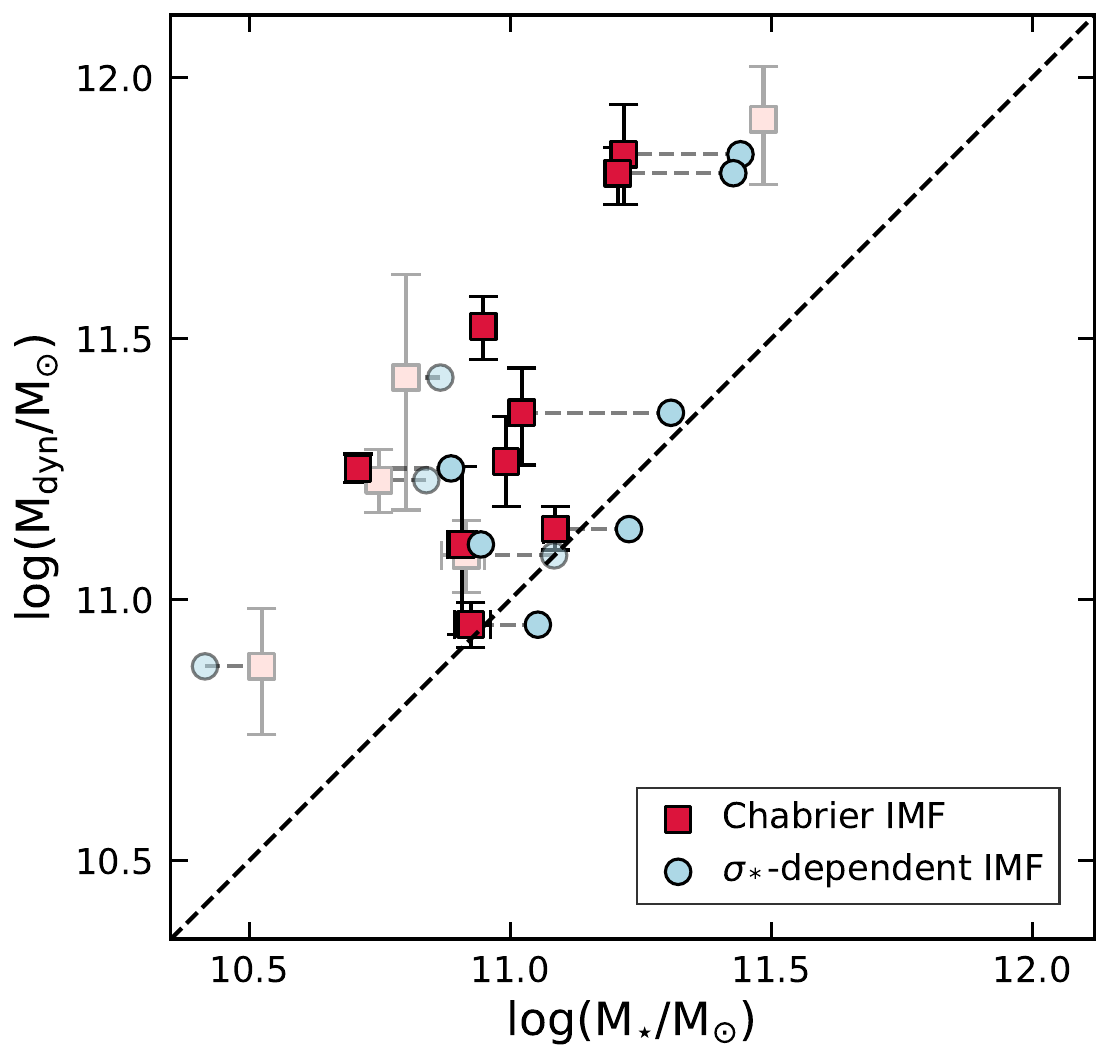}
    \caption{Dynamical against stellar masses assuming a \citet{GChabrier2003} IMF (red squares) and a $\sigma_v$-dependent IMF from \citet{TTreu2010} (blue circles) for the distant quiescent galaxies in our sample. The dashed line indicates the one-to-one ratio of dynamical vs stellar mass. Light points indicate galaxies for which only lower limits to the rotational velocity could be obtained. }
    \label{fig:Mdyn}
\end{figure}

The resulting dynamical masses are reported in Table \ref{tab:fit_properties} and are shown against stellar masses measured from SPS modelling \citep{MSlob2024} in Figure \ref{fig:Mdyn}. For all galaxies in this study, the dynamical masses are higher than the stellar masses, assuming a \citet{GChabrier2003} IMF, as is expected from the fact that the dynamical masses includes the stellar, gas, and dark matter components of a galaxy. When we assume that the gas mass of the galaxies in our sample is low \citep[e.g.][]{CWilliams2021, SBelli2021}, we find that our sample has a median dark matter fraction of 63\% within 1$r_e$, with individual galaxies having fractions up to 83\%. 

The dynamical-to-stellar mass ratios of our sample are $\sim0.2$~dex higher than those of quiescent galaxies with similar masses and redshifts for the same IMF assumption \citep[e.g.][]{JvandeSande2013, SBelli2017, BForrest2022, MKriek2024}. In Section \ref{sec:discussion_mdyn} we discuss this offset further, and we also investigate 
the implications of our dynamical-to-stellar mass ratios for the IMF.

\section{Discussion}\label{sec:discussion}
\subsection{Buildup and quenching of massive galaxies}\label{sec:discussion_gal_evol}
\begin{figure}[t]
    \centering
    \includegraphics[width=0.97\linewidth]{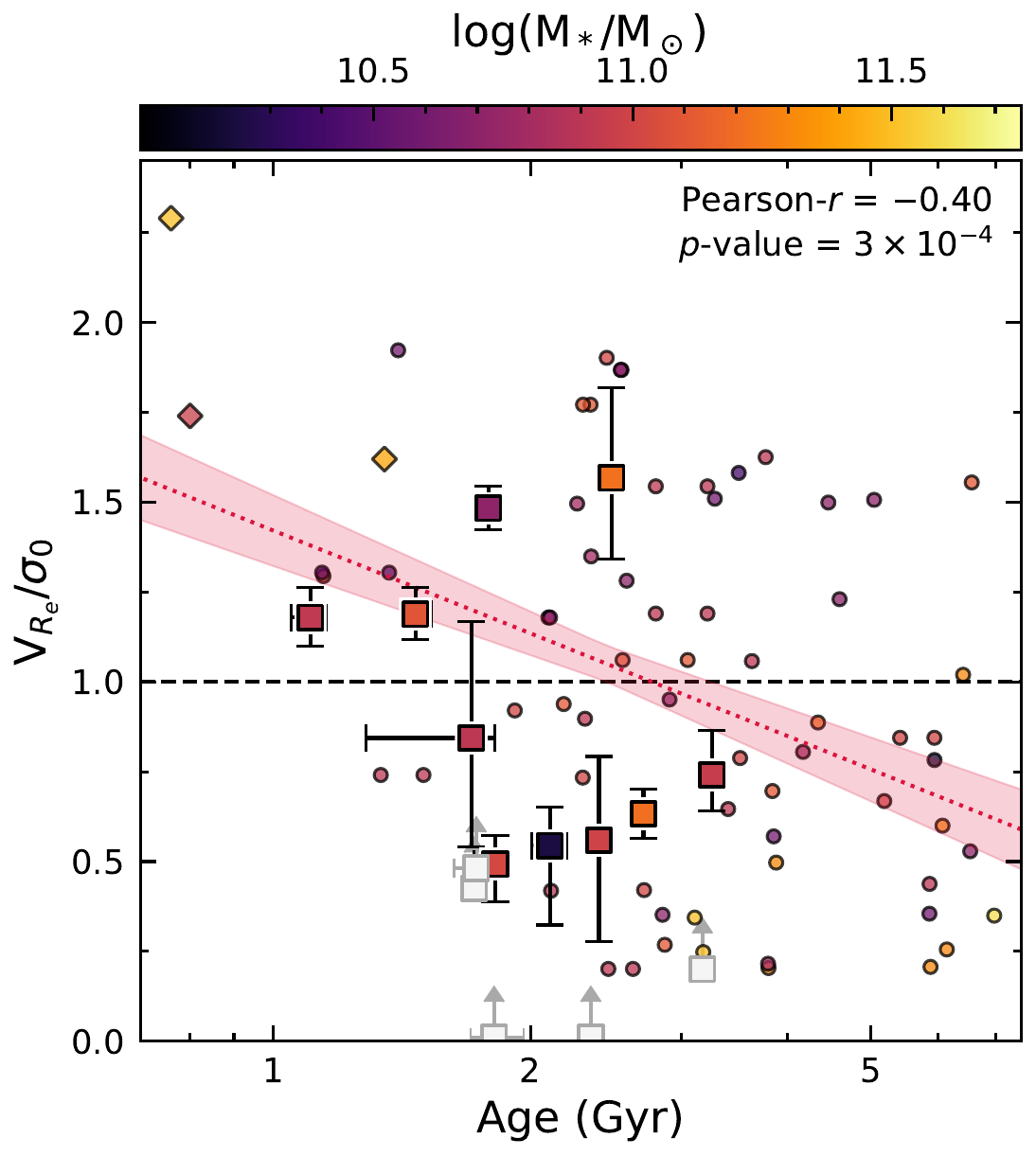}
    \caption{$V_{r_e}/\sigma_0$ as a function of stellar age for the distant quiescent galaxies in our sample (squares). We also include samples from literature using the same symbols as in Figure \ref{fig:v_over_sigma_mass}. The dashed black line represents a ratio of 1, corresponding to the definition of rotational support. The symbols are coloured by stellar mass. The grey squares indicate galaxies for which we can only obtain lower limits to $V_{r_e}/\sigma_0$. We show the Pearson correlation coefficient and $p$-values in the top right. The dotted red line corresponds to a linear fit to the ages and $V_{r_e}/\sigma_0$.}
    \label{fig:V_sigma_overview}
    \vspace{-0.15in}
\end{figure}
In the previous section, we showed that the majority of the distant quiescent galaxies in our sample are rotationally supported. Interestingly, the four distant quiescent galaxies at $z>2$ for which kinematic properties were previously measured all have higher spin parameters. Moreover, our galaxies are, on average, more rotationally supported than massive nearby galaxies (see Figure \ref{fig:lamda_re_vs_ell}). Taken at face value, these results imply that rotational support of quiescent galaxies decreases over cosmic time.

To investigate whether the difference in rotational support between nearby early-type galaxies and distant quiescent galaxies is a genuine redshift effect, we also plot $(V/\sigma)_{r_e}$ as a function of age in Figure \ref{fig:V_sigma_overview}. For our sample, we used the ages derived using {\sc prospector} \citep{BJohnson2021} from \citet{MSlob2024}. For the LEGA-C galaxies, we used ages from spectral fitting with {\sc alf} \citep{CConroy2012b,CConroy2018} presented by \citet{CCheng2025}. \citet{ABeverage2024b} showed that the {\sc alf} ages are consistent with the {\sc prospector} ages, and thus, no systematic biases are expected.

Figure \ref{fig:V_sigma_overview} shows a trend between $(V/\sigma)_{r_e}$ and age, with older galaxies being less rotationally supported than young quiescent galaxies. We calculated the Pearson correlation coefficient for the combined LEGA-C, SUSPENSE, and \citet{ANewman2018b} samples to assess the degree of correlation between \vsigma and age, and find a negative correlation ($r = -0.40$) with a $p$-value $<0.01$. We quantified this relation between age and rotational support by fitting a linear function, which resulted in a best-fit relation of the form
\begin{equation}
    (V/\sigma)_{r_e} = -0.95 \pm 0.13 \times \log(\text{Age}) + 1.4 \pm 0.05.
\end{equation}
Although the trend between rotational support and age has already been identified for low-$z$ galaxies \citep{JVandeSande2018, SCroom2024}, our results imply that it was already in place at higher redshifts. This trend might also explain why our galaxies are less rotationally supported than other quiescent galaxies at $z>2$; they are all relatively old (ages $> 1$~Gyr), while three out of four quiescent rotating discs from \citet{ANewman2018b} and \citet{FdEugenio2024} are post-starburst galaxies with ages of $\sim0.5-0.8$~Gyr.

We found no strong trend between stellar mass and $(V/\sigma)_{r_e}$ for distant quiescent galaxies, with a Pearson-$r$ coefficient of $-0.19$. Furthermore, Figure \ref{fig:mass_size} shows that although the most compact galaxies all rotate significantly, there is no trend between the amount of rotation and the offset from the coeval mass-size relation overall. Figure \ref{fig:lamda_re_vs_ell} also shows no trend between stellar mass and spin parameter for our sample. These results are in contrast with findings for the spin parameter in the nearby universe (see Figure \ref{fig:lamda_re_vs_ell}), where the most massive nearby galaxies are slow rotators, while less massive galaxies are fast rotators. However, we note that the $z>0$ samples for which stellar kinematics have been obtained to date have few galaxies with $\log(M_*/M_{\odot})<10.5$, so that the dynamic range probed by current studies is small. 

The trend between rotational support and age we observe at $z>1$ implies that the difference between rotational properties of nearby and distant massive quiescent galaxies is driven by the fact that more distant galaxies are younger on average, and thus, that their $V/\sigma$ ratios are higher. This trend also implies that significant structural changes must take place after a galaxy has stopped forming stars, instead of a complete destruction of the stellar disc during the quenching event. These structural changes likely happen gradually as galaxies evolve because the rotational support in $z\sim0.7$ LEGA-C galaxies is not significantly lower than the oldest SUSPENSE galaxies.

This gradual decline of rotational support in the quiescent population is consistent with a scenario where dry mergers ``spin down" galaxies down after quenching. Cosmological simulations also showed that multiple consecutive dry mergers are the main mechanism that generates the low-$z$ population of massive slow rotators \citep[e.g.][]{ZPenoyre2017, FSchulze2018, YDubois2016, VRodriguez2017, CLagos2018a, CLagos2018b, ARantala2024}. While both major and minor mergers can be responsible for a decrease in stellar angular momentum, the cumulative effect of a series of dry minor mergers can gradually and more effectively decrease the angular momentum than a single major merger \citep{TNaab2014,CLagos2018a,ARantala2024}. These mergers do not only slow down rotation, but can also explain the size growth \citep[e.g.][]{PvanDokkum2010, RBezanson2009, SPatel2013} and the buildup of the blue and metal-poor outskirts of massive galaxies \citep[e.g.][]{KSuess2020, IMartinNavarro2018, Cheng_2024}. The finding that distant quiescent galaxies have many small companions \citep{ANewman2012, KSuess2023} also directly supports the minor merger scenario. Furthermore, massive relic galaxies in the nearby universe, which formed most of their mass at high redshift and evolved without undergoing mergers after quenching, are predominantly fast-rotating galaxies with discy morphologies \citep[e.g.][]{AFerreMateu2012,AFerreMateu2017,AYilderim2015, CSpiniello2024, CTortora2025}. The kinematic properties of these relic galaxies thus further indicate that mergers may be needed to transform galaxies into slow rotators.

If we assume that the galaxies in our sample are the direct progenitors of massive galaxies at $z\sim0$ and are representative of the general population at cosmic noon, we can estimate how their rotational properties evolve up to $z\sim0$. The galaxies for which we measured rotational support have a median stellar mass of $\log(M_*/M_{\odot})\sim11$ and a median $\lambda_{r_e} = 0.28$. Based on number density arguments, we expect these galaxies to grow $\sim0.3$~dex in stellar mass since cosmic noon \citep[e.g.][]{PvanDokkum2010, AMuzzin2013, AHill2017}. We took all galaxies from the SAMI, ATLAS3D, and MASSIVE surveys with masses $\log(M_*/M_{\odot})= 11.15-11.45$, which are the likely descendants of the SUSPENSE galaxies. The median spin parameter of this population is $\lambda_{r_e} = 0.14$. While the above calculation is only a rough estimate, and the evolution of individual galaxies in our sample will be varied, we can conclude that the spin parameter of massive quiescent galaxies should decrease by a factor of $\sim2$ between cosmic noon and $z\sim0$ on average. Thus, the majority of our population of fast-rotating quiescent galaxies at cosmic noon will likely gradually evolve into the (cores of) old slow-rotating early-type galaxies. We also note, however, that some massive quiescent fast-rotators still exist by $z\sim0$, such that some of our galaxies may still be fast-rotators by $z\sim0$.\\

Our results show that significant structural changes occur after quenching. However, the relatively low $V/\sigma$ ratios ($V/\sigma\sim1$) of our sample compared to cold and ionised gas kinematics of star-forming disc galaxies of similar masses at $z>2$ \citep[$V/\sigma\gtrsim3$; e.g.][]{NForsterSchreiber2009, BEpinat2009,EWisnioski2015,EWisnioski2019,MNeeleman2020,FRizzo2021,FFraternali2021,EParlanti2023,ADanhaive2025} suggest that some loss of rotational support may occur during star-formation quenching. The galaxies in our sample are relatively old (ages $>1$~Gyr), and it is therefore not straightforward to discern whether this decrease in $V/\sigma$ occurred during quenching or in the period between quenching and the time of observation. It is also unclear how the stellar kinematics compare to the gas kinematics in distant star-forming galaxies, although at $z\sim1$, the stellar kinematics lead to significantly lower \vsigma\ ratios than the gas kinematics \citep{HUebler2024}. 

Nonetheless, even the younger quiescent rotating discs from \citet{ANewman2018b} have $V/\sigma$ ratios that are low compared to those of high-redshift star-forming galaxies. \citet{ANewman2018b} argued that while quenching through major mergers might explain this observed decrease in the rotational support, the flattened shapes and lack of significant bulges in their galaxies imply that this scenario is likely too simplistic. Furthermore, we note that based on their rotational velocities, our galaxies fall on the $z\sim1$ stellar mass Tully-Fisher relation for star-forming galaxies \citep{CConselice2005}. This finding implies that the difference in $V/\sigma$ between star-forming and quiescent galaxies is driven by an increase in the velocity dispersion and not by a decrease in the rotational velocity.
In a future study, we plan to test this hypothesis by measuring ionised gas and stellar kinematics of star-forming galaxies beyond $z>1$ using the method described in this work.

\subsection{Implications for the IMF}\label{sec:discussion_mdyn}
In the previous section, we showed that our distant quiescent galaxies likely grow inside-out into low-$z$ early-type galaxies through (mostly) minor mergers. Our results thus imply that the galaxies in our sample are expected to become the cores of massive galaxies today. These cores of the most massive (dispersion-dominated) galaxies are found to have a more bottom-heavy IMF than the Milky Way (e.g. \citealt{TTreu2010, CConroy2012, MCappellari2012,CTortora2013,HLi2017}; see also \citealt{IMartinNavarro2018}). To assess whether our dynamical-to-stellar mass ratios would allow for a bottom-heavy $\sigma_{v}$-dependent IMF, we calculated $\log(M_*)$ assuming the relation from \citet{TTreu2010}. For this relation, the IMF is more bottom heavy than the Salpeter IMF for galaxies with $\sigma_{v}>250$~km/s. We show the resulting stellar masses against the dynamical masses as blue circles in Figure \ref{fig:Mdyn}. The $\sigma_{v}$-dependent IMF assumption results in a median dynamical-to-stellar mass ratio of 2.5 for our sample, implying a dark matter fraction of 59\% within one $r_e$. Furthermore, while for this IMF assumption, the stellar masses of two galaxies in our sample exceed the dynamical mass, the stellar masses of all galaxies are consistent with the dynamical mass within 1$\sigma$. The very low implied DM fraction for some of the galaxies for a bottom-heavy IMF is consistent with the low dark matter fraction found in some relic galaxies (e.g. NGC 1277; \citealt{SComeron2023}).

Previous ground-based studies of distant quiescent galaxies showed that stellar masses calculated using the $\sigma_{v}$-dependent IMF from \citet{TTreu2010} result in median dynamical-to-stellar mass ratios $<1$ \citep{JMendel2020,MKriek2024, BForrest2022}. Thus, these studies claimed that the bottom-heavy IMF observed today could not yet have been in place in distant quiescent galaxies. We found dynamical masses that allowed for a $\sigma_v$-dependent IMF, however, which implies distant quiescent galaxies can become the cores of nearby elliptical galaxies.

The relatively high dynamical-to-stellar mass ratios we derived compared to previous ground-based studies are likely the result of a combination of factors. First, NIRCam provides longer wavelengths, higher spatial resolution, and deeper imaging than HST, enabling more robust structural measurements and resulting dynamical masses. Secondly, previous studies relied on spectra with lower $S/N$, which might have biased their velocity dispersions and the resulting dynamical masses. Our high $S/N$ spectra also allowed more accurate $M/L$ ratio calibrations than ground-based studies, from which we can measure more reliable stellar masses. Moreover, our detailed kinematic modelling, including rotation, likely resulted in more reliable dynamical masses than were inferred from integrated velocity dispersions. Altogether, a range of observational challenges that affect the structural and spectral measurements might cause the lower dynamical-to-stellar mass ratios found in earlier work.

\subsection{Using virial relations for NIRSpec/MSA spectra?}\label{sec:discussion_apertures}
\begin{figure}
    \centering
    \includegraphics[width=0.97\linewidth]{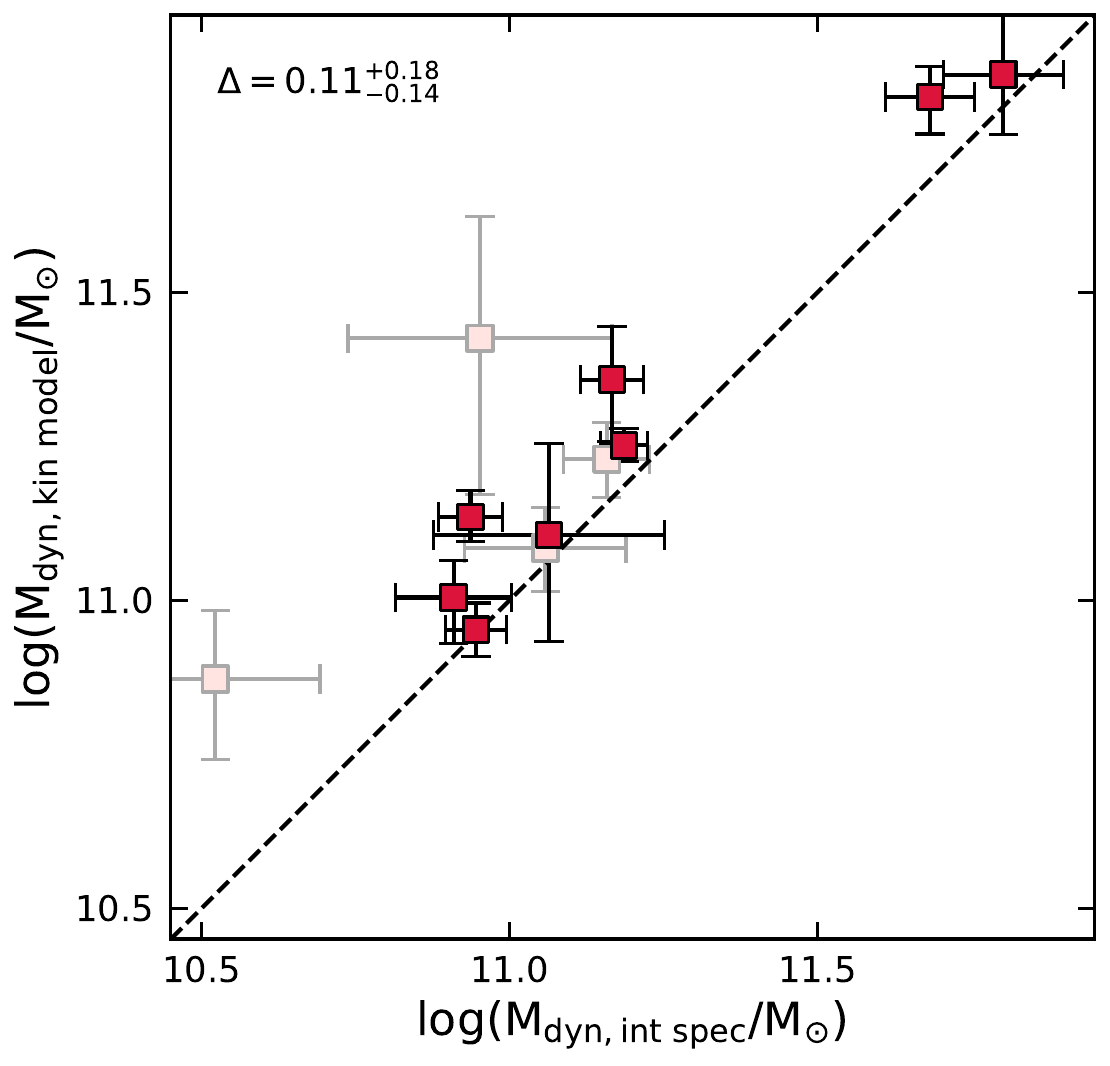}
    \caption{Dynamical masses calculated from the integrated velocity dispersion ($\log(M_{\text{dyn, int spec}})$) against dynamical masses calculated from our kinematic models ($\log(M_{\text{dyn, kin model}})$). The dashed line indicates a one-to-one ratio. The pink squares indicate galaxies for which only lower limits to the rotational velocity could be obtained.}
    \label{fig:mdyn_rot_vs_vir}
    \vspace{-0.15in}
\end{figure}

When no resolved kinematic measurements are available, dynamical masses can be calculated from the observed velocity dispersion of the integrated spectrum, $\sigma_{v,\text{int spec}}$, using virial relations. We assessed how these dynamical masses compare to the dynamical masses from our kinematic models. We used the integrated velocity dispersions from \citet{ABeverage2024b} for our sample to calculate the dynamical masses as 
\begin{equation}
    M_{\text{dyn, int spec}} = \frac{K(n)\ \sigma^2_{v,\text{ int spec}}\ r_e}{G}.
\end{equation}
Here, $K(n)$ is the virial coefficient defined by \citet{MCappellari2006}. In Figure \ref{fig:mdyn_rot_vs_vir} we show these dynamical masses ($\log(M_{\text{dyn, int spec}})$) against the dynamical masses we calculated from our kinematic models ($\log(M_{\text{dyn, kin model}})$) in Section \ref{sec:dyn_mass}. The dynamical masses from our kinematic models are on average $0.11_{-0.14}^{+0.18}$~dex higher than the dynamical masses from the integrated spectra.

The large offset between the two dynamical mass estimates can be attributed to the fact that the integrated velocity dispersions were not corrected for the aperture losses of the NIRSpec/MSA slits. An aperture correction was defined for slit-based observations for both rotation-dominated \citep{SPrice2014} and dispersion-dominated sources \citep{JvandeSande2013}, which can increase the integrated velocity dispersion by a factor up to $\sim1.4$ in rotation-dominated systems. Because most sources that are observed with the NIRSpec/MSA are not centred and aligned within the MSA shutter, however, it is non-trivial to define an aperture correction for NIRSpec/MSA observations. Thus, the dynamical masses derived from the observed integrated velocity dispersion of NIRSpec/MSA spectra are likely underestimated. This is especially true for the extended galaxies in our sample, with effective radii that span across multiple shutters for many sources in our sample.
Using our kinematic models, we assessed how large the offset is between the velocity dispersion of the integrated spectrum ($\sigma_{v,\text{int spec}}$) and the intrinsic line-of-sight integrated velocity dispersion within one effective radius ($\sigma_{v,r_e}$).

To compute $\sigma_{v,r_e}$, we assumed that the observed integrated velocity dispersion is equal to $V^2_{\text{rms}}(r) = \sigma_0^2 + V_{\text{los}}^2(r)$, following \citet{MCappellari2008}. Here, $\sigma_0$ is the best-fit velocity dispersion, and $V_{\text{los}}(r)$ is the best-fit line of sight rotational velocity from our modelling (see the bottom right panels in Figure \ref{fig:v_curves}). From this line-of-sight $V_{\text{rms}}(r)$, we calculated $\sigma_{v,r_e}$ by integrating over the elliptical aperture corresponding to 1~$r_e$, and weighting by the S\'ersic profile of the galaxy, following \citet{JvandeSande2013}.

\begin{figure}[t]
    \centering
    \includegraphics[width=0.97\linewidth]{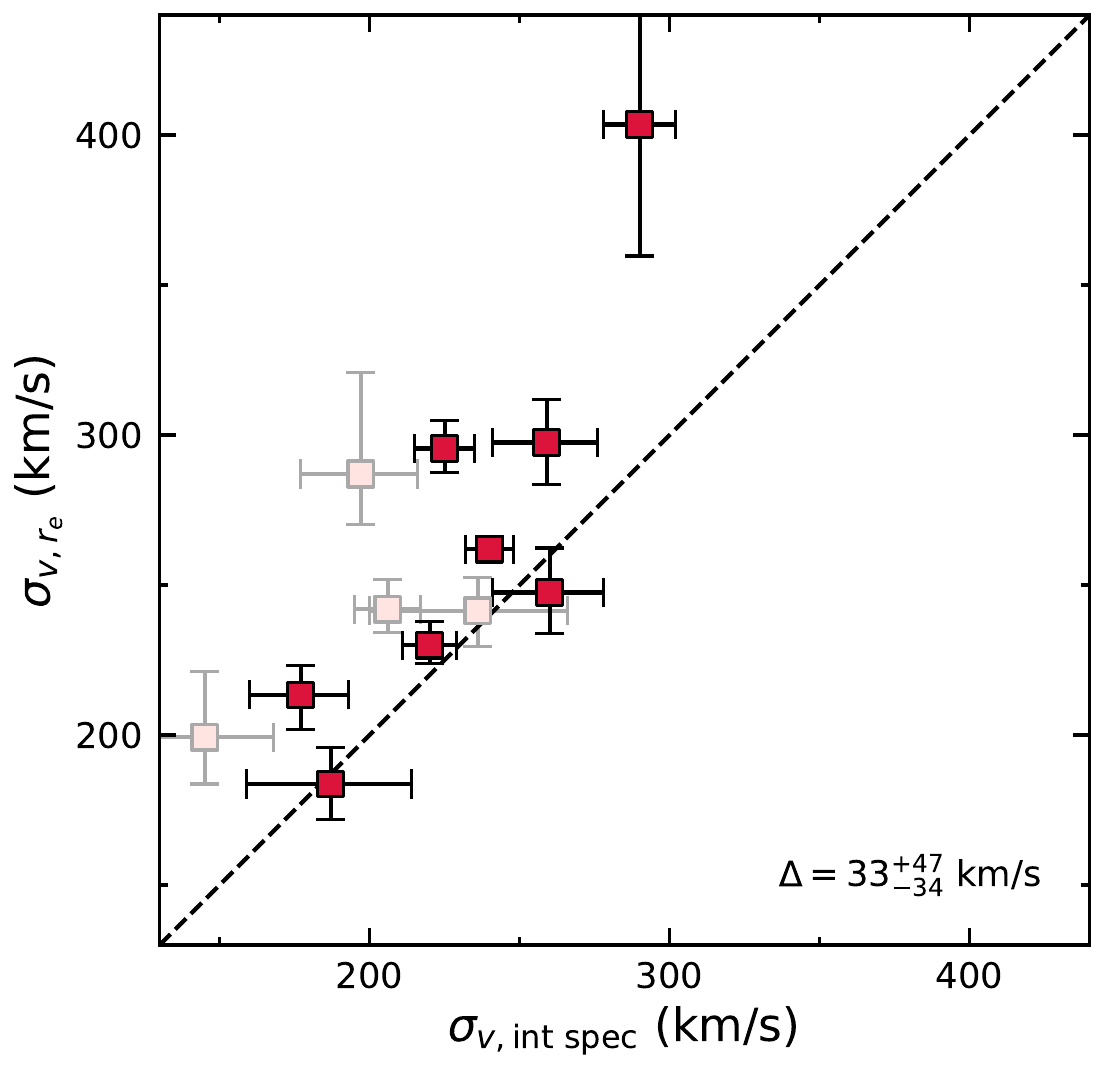}
    \caption{Velocity dispersion of the integrated spectrum ($\sigma_{v,\text{int spec}}$) from \citet{ABeverage2024b} against the inferred integrated line-of-sight velocity dispersion ($\sigma_{v,r_e}$) within 1~$r_e$ for the galaxies in our sample. The dashed line indicates a one-to-one ratio. The pink squares indicate galaxies for which only lower limits to the rotational velocity could be obtained. The median offset between the integrated velocity dispersion and the velocity dispersion within 1~$r_e$ is $33_{-34}^{+47}$~\kms.}
    \label{fig:sigma_re_tot}
    \vspace{-0.15in}
\end{figure}

In Figure \ref{fig:sigma_re_tot} we show the velocity dispersion of the integrated spectrum ($\sigma_{v,\text{int spec}}$), taken from \citet{ABeverage2024b}, against the inferred integrated line-of-sight velocity dispersion within 1~$r_e$ as derived from our kinematic model ($\sigma_{v,r_e}$) for the galaxies in our sample. As expected, the velocity dispersion within 1~$r_e$ is systematically higher than the velocity dispersion of the integrated spectrum, with a median offset of $33_{-34}^{+47}$~\kms. The errors on the median were calculated using Monte Carlo simulations of the data. In Appendix \ref{ap:dyn_masses} we show that the inferred dynamical masses from the aperture-corrected velocity dispersions are consistent with the dynamical masses from the kinematic models.

Our results show that caution is needed when dynamical masses are derived using the integrated velocity dispersion of NIRSpec/MSA spectra without applying an aperture correction, as significant offsets between $\sigma_{v,\text{int spec}}$ and $\sigma_{v,r_e}$ are expected when a source is larger than the size of the micro-shutter. 
We note that this bias from the observed integrated velocity dispersion is also expected for measurements of emission line kinematics of star-forming galaxies observed using NIRSpec/MSA, such that dynamical masses calculated from emission lines are also sensitive to this effect.

\subsection{Caveats}\label{sec:discussion_caveats}
In this work we present the kinematic properties of 15 distant quiescent galaxies from their resolved MSA spectra using a forward-modelling approach that took the source position and alignment with respect to the MSA shutters into account. For all ten galaxies for which we were able to constrain the rotational velocity, we measured strong rotation ($V_{r_e}>114$~\kms). To assess the significance of the rotational support in these galaxies, we computed the intrinsic velocity as a function of radius from Equation \eqref{eq:arctan_v} for 2000 simulations of $v_a$, $r_e$, and $r_t/r_e$, randomly perturbed around their uncertainties. We show these simulated $v(r)$ curves for the ten galaxies in Appendix \ref{ap:v_r}, with the 1$\sigma$ and 2$\sigma$ levels indicated in pink and red, respectively. This test illustrates that the large errors we find for $V_{r_e}$ in part of our sample are not only driven by uncertainties in our kinematic modelling, but is also strongly dependent on the uncertainty of the measurement of $r_e$. Nonetheless, from the 2$\sigma$ confidence intervals, we can conclude that all ten galaxies rotate significantly. 

Interestingly, we find that the galaxies that are most aligned along the major axis with respect to the MSA shutter have the highest $V/\sigma$ measurements. This finding suggests that the degree of rotational support we find for galaxies that are slightly misaligned within their shutter may be underestimated. Detailed IFU observations or duplicate slit-based observations under multiple telescope angles are thus needed to confirm whether our forward-modelling approach systematically underestimates the rotational velocities of misaligned galaxies. However, our overall conclusions would not change qualitatively if the rotational support of the galaxies in our sample were systematically higher.

Finally, the thin-disc assumption that we made in our modelling might bias the dynamical mass and \vsigma\ calculations. For a thin-disc model, the virial coefficient in Equation \eqref{eq:Mdyn_Price22} is $k_{\text{tot}} = 1.8$. Based on the high velocity dispersions ($\sigma_0>250$~km/s) of some of the galaxies in our sample, however, it is more likely that these galaxies are thick discs, with more complicated dynamical profiles. For thick discs, the virial coefficient $k_{\text{tot}}$ is higher \citep{SPrice2022}, which would result in higher inferred dynamical masses. However, our kinematic modelling results and \vsigma\ likely also change when we allow for intrinsically thick discs, as a thin disc assumption maximises $V(r)$. Another limitation to the thin-disc parametrisation is the assumption that $\sigma_v$ is radially invariant, while in reality, the velocity dispersion may be lower in the outskirts of these galaxies. To fully assess how our thin-disc model parametrisation affects the inferred dynamical masses and \vsigma\, we plan to model our sample using JAM methods \citep{MCappellari2008} in a future study. JAM methods assume a variable intrinsic thickness in the models and will allow us to further constrain the dynamical properties. We also note that JAM methods will allow us to calibrate the asymmetric drift parameter for stars we used in our dynamical mass calculations, which will further decrease the uncertainties on the dynamical masses.

\section{Conclusions}\label{sec:summary}
We present the kinematic properties of 15 distant ($z=1.2-2.3$) quiescent galaxies from the JWST-SUSPENSE program, the largest sample of kinematic measurements of quiescent galaxies at $z>1.2$ to date. The galaxies were observed for 16 hours in a single NIRSpec/MSA configuration, and the resulting ultra-deep medium-resolution spectra resolve numerous stellar absorption lines in both the spatial and dispersion direction. From these absorption line spectra, we measured the stellar velocity and velocity dispersion profiles.

As the MSA shutters are misaligned and miscentred with respect to the galaxies, the observed kinematics cannot be directly interpreted. We instead derived the intrinsic kinematics using a forward-modelling approach, making use of the high-resolution images from JWST/NIRCam COSMOS-Web \citep{CCasey2023} and HST/F814W \citep{NScoville2007}. In our forward-modelling approach, we took the source morphologies and positions with respect to the shutters, the NIRSpec PSF, and optical as well as resampling effects from the data reduction into account and assumed a thin-disc model. From the inferred velocity and velocity dispersion profiles, we quantified the degree of rotational support in our sample using $V_{r_e}$ and \vsigma. For ten galaxies, we were able to constrain the rotational velocities, and all have significant rotation. The remaining five galaxies are too misaligned with respect to the MSA shutter to constrain their rotational velocity. Moreover, all ten galaxies with measured stellar rotation are classified as fast rotators from their position in the $\lambda_{r_e}-\epsilon$ plane.

Compared to massive early-type galaxies in the nearby universe,  on average distant quiescent galaxies are more rotationally supported. We find a trend between rotational support and age in the combined LEGA-C, SUSPENSE, and \citet{ANewman2018b} samples at $z\sim0.5-2.5$, with young galaxies having more rotational support. This age trend is thus a likely explanation for the increased rotational support we find at high-$z$, where on average galaxies are younger than in the nearby universe. Surprisingly, we find no trend between stellar mass and rotational support at $z>0.5$. 

Our findings imply that distant quiescent galaxies are still rotating discs just after they stopped forming stars. Thus, the physical process responsible for quenching star formation in massive galaxies likely did not disrupt rotational support. To explain the population of slow-rotating massive quiescent galaxies at low $z$, significant structural changes must instead occur after the galaxies have already become quiescent. Our results support a scenario in which galaxies experience a series of (mostly minor) dry mergers after quenching. These mergers lead to growth in both mass and size, and gradually disturb the ordered rotation of galaxies. We cannot rule out, however, that the galaxies in our sample may have lost part of their specific angular momentum during quenching, as their $V/\sigma$ values are lower than the $V/\sigma$ values (inferred from gas kinematics) of massive star-forming galaxies at $z>2$. On the other hand, the rotational velocities are similar to those of star-forming galaxies of similar masses, implying that the low $V/\sigma$ ratios are mostly driven by an increase in the velocity dispersion.

From the kinematic and structural properties, we also derived dynamical masses. The resulting dynamical masses imply dark matter fractions of $\sim63\%$ within 1$r_e$ for our galaxies, assuming a \citet{GChabrier2003} IMF. The dynamical-to-stellar mass ratios of our sample are higher than in previous studies, and in contrast to these studies, our dynamical-to-stellar mass ratios do allow for a bottom-heavy ($\sigma_v$-dependent) IMF. This finding implies that distant quiescent galaxies can evolve into the cores of nearby massive early-type galaxies, which are found to have more bottom-heavy IMFs. Furthermore, using our kinematic models, we find that the observed integrated velocity dispersions from our NIRSpec/MSA spectra are $\sim10$\% lower than the inferred integrated line-of-sight velocity dispersions from our models. These results show that extreme caution is needed when using integrated velocity dispersions from NIRSpec/MSA spectra to calculate dynamical masses because significant biases can be introduced by not applying the relevant aperture corrections.

Our study demonstrates the power of the NIRSpec/MSA to obtain stellar kinematic properties of large samples of massive quiescent (and possibly also star-forming) galaxies at and beyond cosmic noon. This work is enabled by ultra-deep integration times, our custom observing strategy (large nods and long slits), and the identification of a field of many bright quiescent targets. To further improve the constraints on stellar kinematics beyond $z=1$, larger samples are needed. Although archival data could be used to this end, the shorter integration times and smaller spatial coverage along the MSA shutters of most MSA programs observed to date make this challenging. Thus, future targeted larger and deeper surveys, with a similar observing strategy as the JWST-SUSPENSE programme, may be required to obtain stellar kinematics for larger samples of distant galaxies. Moreover, to assess possible biases in our forward-modelling method, especially for galaxies that are misaligned with respect to the MSA shutter, detailed IFU observations are needed. Finally, the method we presented offers a promising avenue to combine stellar kinematic measurements with direct measurements of the IMF using the NIRSpec/MSA, paving the way for robust probes of dark matter haloes of quiescent galaxies beyond the low-$z$ universe (see the approved JWST programme 5629).

\begin{acknowledgements}
    We thank Gabe Brammer, Ciaran Rogers, and Lucie Rowland for useful conversations about this work.
    We thank the COSMOS-Web team for making their imaging publicly available immediately. 
    This work is based on observations made with the NASA/ESA/CSA James Webb Space Telescope. The data were obtained from the Mikulski Archive for Space Telescopes at the Space Telescope Science Institute, which is operated by the Association of Universities for Research in Astronomy, Inc., under NASA contract NAS 5-03127 for JWST. These observations are associated with program JWST-GO-2110. Support for program JWST-GO-2110 was provided by NASA through a grant from the Space Telescope Science Institute, which is operated by the Association of Universities for Research in Astronomy, Inc., under NASA contract NAS 5-03127. The specific observations analyzed can be accessed via \href{http://dx.doi.org/10.17909/y6rb-fn24}{doi: 10.17909/y6rb-fn24}.
    Some of the data products presented herein were retrieved from the Dawn JWST Archive (DJA). DJA is an initiative of the Cosmic Dawn Center (DAWN), which is funded by the Danish National Research Foundation under grant DNRF140.
    MK acknowledges funding from the Dutch Research Council (NWO) through the award of the Vici grant VI.C.222.047 (project 2010007169).
    D.M. is grateful for the support provided by Leonard and Jane Holmes Bernstein through the Leonard and Jane Holmes Bernstein Professorship in Evolutionary Science.
\end{acknowledgements}
\bibliographystyle{aa}
\bibliography{mybib}

\begin{thebibliography}{123}
\expandafter\ifx\csname natexlab\endcsname\relax\def\natexlab#1{#1}\fi

\bibitem[{{Barnes}(1988)}]{JBarnes1988}
{Barnes}, J.~E. 1988, \apj, 331, 699

\bibitem[{{Belli} {et~al.}(2021){Belli}, {Contursi}, {Genzel}, {Tacconi}, {F{\"o}rster-Schreiber}, {Lutz}, {Combes}, {Neri}, {Garc{\'\i}a-Burillo}, {Schuster}, {Herrera-Camus}, {Tadaki}, {Davies}, {Davies}, {Johnson}, {Lee}, {Leja}, {Nelson}, {Price}, {Shangguan}, {Shimizu}, {Tacchella}, \& {{\"U}bler}}]{SBelli2021}
{Belli}, S., {Contursi}, A., {Genzel}, R., {et~al.} 2021, \apjl, 909, L11

\bibitem[{{Belli} {et~al.}(2017){Belli}, {Newman}, \& {Ellis}}]{SBelli2017}
{Belli}, S., {Newman}, A.~B., \& {Ellis}, R.~S. 2017, \apj, 834, 18

\bibitem[{{Belli} {et~al.}(2024){Belli}, {Park}, {Davies}, {Mendel}, {Johnson}, {Conroy}, {Benton}, {Bugiani}, {Emami}, {Leja}, {Li}, {Maheson}, {Mathews}, {Naidu}, {Nelson}, {Tacchella}, {Terrazas}, \& {Weinberger}}]{SBelli2024}
{Belli}, S., {Park}, M., {Davies}, R.~L., {et~al.} 2024, \nat, 630, 54

\bibitem[{{Beverage} {et~al.}(2025){Beverage}, {Slob}, {Kriek}, {Conroy}, {Barro}, {Bezanson}, {Brammer}, {Cheng}, {de Graaff}, {F{\"o}rster Schreiber}, {Franx}, {Lorenz}, {Mancera Pi{\~n}a}, {Marchesini}, {Muzzin}, {Newman}, {Price}, {Shapley}, {Stefanon}, {Suess}, {van Dokkum}, {Weinberg}, \& {Weisz}}]{ABeverage2024b}
{Beverage}, A.~G., {Slob}, M., {Kriek}, M., {et~al.} 2025, \apj, 979, 249

\bibitem[{{Bezanson} {et~al.}(2018){Bezanson}, {van der Wel}, {Straatman}, {Pacifici}, {Wu}, {Bari{\v{s}}i{\'c}}, {Bell}, {Conroy}, {D'Eugenio}, {Franx}, {Gallazzi}, {van Houdt}, {Maseda}, {Muzzin}, {van de Sande}, {Sobral}, \& {Spilker}}]{RBezanson2018}
{Bezanson}, R., {van der Wel}, A., {Straatman}, C., {et~al.} 2018, \apjl, 868, L36

\bibitem[{{Bezanson} {et~al.}(2009){Bezanson}, {van Dokkum}, {Tal}, {Marchesini}, {Kriek}, {Franx}, \& {Coppi}}]{RBezanson2009}
{Bezanson}, R., {van Dokkum}, P.~G., {Tal}, T., {et~al.} 2009, \apj, 697, 1290

\bibitem[{{Bournaud} {et~al.}(2007){Bournaud}, {Jog}, \& {Combes}}]{FBournaud2007}
{Bournaud}, F., {Jog}, C.~J., \& {Combes}, F. 2007, \aap, 476, 1179

\bibitem[{{Bruce} {et~al.}(2012){Bruce}, {Dunlop}, {Cirasuolo}, {McLure}, {Targett}, {Bell}, {Croton}, {Dekel}, {Faber}, {Ferguson}, {Grogin}, {Kocevski}, {Koekemoer}, {Koo}, {Lai}, {Lotz}, {McGrath}, {Newman}, \& {van der Wel}}]{VBruce2012}
{Bruce}, V.~A., {Dunlop}, J.~S., {Cirasuolo}, M., {et~al.} 2012, \mnras, 427, 1666

\bibitem[{{Bruce} {et~al.}(2014){Bruce}, {Dunlop}, {McLure}, {Cirasuolo}, {Buitrago}, {Bowler}, {Targett}, {Bell}, {McIntosh}, {Dekel}, {Faber}, {Ferguson}, {Grogin}, {Hartley}, {Kocevski}, {Koekemoer}, {Koo}, \& {McGrath}}]{VBruce2014}
{Bruce}, V.~A., {Dunlop}, J.~S., {McLure}, R.~J., {et~al.} 2014, \mnras, 444, 1001

\bibitem[{{Buitrago} {et~al.}(2013){Buitrago}, {Trujillo}, {Conselice}, \& {H{\"a}u{\ss}ler}}]{FBuitrago2013}
{Buitrago}, F., {Trujillo}, I., {Conselice}, C.~J., \& {H{\"a}u{\ss}ler}, B. 2013, \mnras, 428, 1460

\bibitem[{{Burkert} {et~al.}(2010){Burkert}, {Genzel}, {Bouch{\'e}}, {Cresci}, {Khochfar}, {Sommer-Larsen}, {Sternberg}, {Naab}, {F{\"o}rster Schreiber}, {Tacconi}, {Shapiro}, {Hicks}, {Lutz}, {Davies}, {Buschkamp}, \& {Genel}}]{ABurkert2010}
{Burkert}, A., {Genzel}, R., {Bouch{\'e}}, N., {et~al.} 2010, \apj, 725, 2324

\bibitem[{Bushouse {et~al.}(2023)Bushouse, Eisenhamer, Dencheva, Davies, Greenfield, Morrison, Hodge, Simon, Grumm, Droettboom, Slavich, Sosey, Pauly, Miller, Jedrzejewski, Hack, Davis, Crawford, Law, Gordon, Regan, Cara, MacDonald, Bradley, Shanahan, Jamieson, Teodoro, Williams, \& Pena-Guerrero}]{HBushouse2023}
Bushouse, H., Eisenhamer, J., Dencheva, N., {et~al.} 2023, JWST Calibration Pipeline

\bibitem[{{Cappellari}(2008)}]{MCappellari2008}
{Cappellari}, M. 2008, \mnras, 390, 71

\bibitem[{Cappellari(2016)}]{MCappellari2016}
Cappellari, M. 2016, Annual Review of Astronomy and Astrophysics, 54, 597

\bibitem[{{Cappellari}(2017)}]{MCappellari2017}
{Cappellari}, M. 2017, \mnras, 466, 798

\bibitem[{{Cappellari}(2023)}]{MCappellari2023}
{Cappellari}, M. 2023, \mnras, 526, 3273

\bibitem[{{Cappellari} {et~al.}(2006){Cappellari}, {Bacon}, {Bureau}, {Damen}, {Davies}, {de Zeeuw}, {Emsellem}, {Falc{\'o}n-Barroso}, {Krajnovi{\'c}}, {Kuntschner}, {McDermid}, {Peletier}, {Sarzi}, {van den Bosch}, \& {van de Ven}}]{MCappellari2006}
{Cappellari}, M., {Bacon}, R., {Bureau}, M., {et~al.} 2006, \mnras, 366, 1126

\bibitem[{{Cappellari} \& {Emsellem}(2004)}]{MCappellari2004}
{Cappellari}, M. \& {Emsellem}, E. 2004, \pasp, 116, 138

\bibitem[{{Cappellari} {et~al.}(2011){Cappellari}, {Emsellem}, {Krajnovi{\'c}}, {McDermid}, {Scott}, {Verdoes Kleijn}, {Young}, {Alatalo}, {Bacon}, {Blitz}, {Bois}, {Bournaud}, {Bureau}, {Davies}, {Davis}, {de Zeeuw}, {Duc}, {Khochfar}, {Kuntschner}, {Lablanche}, {Morganti}, {Naab}, {Oosterloo}, {Sarzi}, {Serra}, \& {Weijmans}}]{MCappellari2011}
{Cappellari}, M., {Emsellem}, E., {Krajnovi{\'c}}, D., {et~al.} 2011, \mnras, 413, 813

\bibitem[{{Cappellari} {et~al.}(2012){Cappellari}, {McDermid}, {Alatalo}, {Blitz}, {Bois}, {Bournaud}, {Bureau}, {Crocker}, {Davies}, {Davis}, {de Zeeuw}, {Duc}, {Emsellem}, {Khochfar}, {Krajnovi{\'c}}, {Kuntschner}, {Lablanche}, {Morganti}, {Naab}, {Oosterloo}, {Sarzi}, {Scott}, {Serra}, {Weijmans}, \& {Young}}]{MCappellari2012}
{Cappellari}, M., {McDermid}, R.~M., {Alatalo}, K., {et~al.} 2012, \nat, 484, 485

\bibitem[{{Casey} {et~al.}(2023){Casey}, {Kartaltepe}, {Drakos}, {Franco}, {Harish}, {Paquereau}, {Ilbert}, {Rose}, {Cox}, {Nightingale}, {Robertson}, {Silverman}, {Koekemoer}, {Massey}, {McCracken}, {Rhodes}, {Akins}, {Allen}, {Amvrosiadis}, {Arango-Toro}, {Bagley}, {Bongiorno}, {Capak}, {Champagne}, {Chartab}, {Ch{\'a}vez Ortiz}, {Chworowsky}, {Cooke}, {Cooper}, {Darvish}, {Ding}, {Faisst}, {Finkelstein}, {Fujimoto}, {Gentile}, {Gillman}, {Gould}, {Gozaliasl}, {Hayward}, {He}, {Hemmati}, {Hirschmann}, {Jahnke}, {Jin}, {Khostovan}, {Kokorev}, {Lambrides}, {Laigle}, {Larson}, {Leung}, {Liu}, {Liaudat}, {Long}, {Magdis}, {Mahler}, {Mainieri}, {Manning}, {Maraston}, {Martin}, {McCleary}, {McKinney}, {McPartland}, {Mobasher}, {Pattnaik}, {Renzini}, {Rich}, {Sanders}, {Sattari}, {Scognamiglio}, {Scoville}, {Sheth}, {Shuntov}, {Sparre}, {Suzuki}, {Talia}, {Toft}, {Trakhtenbrot}, {Urry}, {Valentino}, {Vanderhoof}, {Vardoulaki}, {Weaver}, {Whitaker}, {Wilkins}, {Yang}, \& {Zavala}}]{CCasey2023}
{Casey}, C.~M., {Kartaltepe}, J.~S., {Drakos}, N.~E., {et~al.} 2023, \apj, 954, 31

\bibitem[{{Chabrier}(2003)}]{GChabrier2003}
{Chabrier}, G. 2003, \pasp, 115, 763

\bibitem[{{Chang} {et~al.}(2013){Chang}, {van der Wel}, {Rix}, {Holden}, {Bell}, {McGrath}, {Wuyts}, {H{\"a}ussler}, {Barden}, {Faber}, {Mozena}, {Ferguson}, {Guo}, {Galametz}, {Grogin}, {Kocevski}, {Koekemoer}, {Dekel}, {Huang}, {Hathi}, \& {Donley}}]{YChang2013b}
{Chang}, Y.-Y., {van der Wel}, A., {Rix}, H.-W., {et~al.} 2013, \apj, 773, 149

\bibitem[{Chang {et~al.}(2012)Chang, van~der Wel, Rix, Wuyts, Zibetti, Ramkumar, \& Holden}]{YChang2013a}
Chang, Y.-Y., van~der Wel, A., Rix, H.-W., {et~al.} 2012, The Astrophysical Journal, 762, 83

\bibitem[{{Cheng} {et~al.}(2025){Cheng}, {Kriek}, {Beverage}, {Slob}, {Bezanson}, {Franx}, {Leja}, {Mancera Pi{\~n}a}, {Suess}, {van der Wel}, {van de Sande}, \& {van Dokkum}}]{CCheng2025}
{Cheng}, C.~M., {Kriek}, M., {Beverage}, A.~G., {et~al.} 2025, \mnras, 540, 1527

\bibitem[{{Cheng} {et~al.}(2024){Cheng}, {Kriek}, {Beverage}, {van der Wel}, {Bezanson}, {D'Eugenio}, {Franx}, {Mancera Pi{\~n}a}, {Nersesian}, {Slob}, {Suess}, {van Dokkum}, {Wu}, {Gallazzi}, \& {Zibetti}}]{Cheng_2024}
{Cheng}, C.~M., {Kriek}, M., {Beverage}, A.~G., {et~al.} 2024, \mnras, 532, 3604

\bibitem[{{Comer{\'o}n} {et~al.}(2023){Comer{\'o}n}, {Trujillo}, {Cappellari}, {Buitrago}, {Gardu{\~n}o}, {Zaragoza-Cardiel}, {Zinchenko}, {Lara-L{\'o}pez}, {Ferr{\'e}-Mateu}, \& {Dib}}]{SComeron2023}
{Comer{\'o}n}, S., {Trujillo}, I., {Cappellari}, M., {et~al.} 2023, \aap, 675, A143

\bibitem[{{Conroy} \& {van Dokkum}(2012{\natexlab{a}})}]{CConroy2012b}
{Conroy}, C. \& {van Dokkum}, P. 2012{\natexlab{a}}, \apj, 747, 69

\bibitem[{{Conroy} \& {van Dokkum}(2012{\natexlab{b}})}]{CConroy2012}
{Conroy}, C. \& {van Dokkum}, P.~G. 2012{\natexlab{b}}, \apj, 760, 71

\bibitem[{{Conroy} {et~al.}(2018){Conroy}, {Villaume}, {van Dokkum}, \& {Lind}}]{CConroy2018}
{Conroy}, C., {Villaume}, A., {van Dokkum}, P.~G., \& {Lind}, K. 2018, \apj, 854, 139

\bibitem[{Conselice {et~al.}(2005)Conselice, Bundy, Ellis, Brichmann, Vogt, \& Phillips}]{CConselice2005}
Conselice, C.~J., Bundy, K., Ellis, R.~S., {et~al.} 2005, The Astrophysical Journal, 628, 160

\bibitem[{{Courteau}(1997)}]{SCourteau1997}
{Courteau}, S. 1997, \aj, 114, 2402

\bibitem[{{Croom} {et~al.}(2012){Croom}, {Lawrence}, {Bland-Hawthorn}, {Bryant}, {Fogarty}, {Richards}, {Goodwin}, {Farrell}, {Miziarski}, {Heald}, {Jones}, {Lee}, {Colless}, {Brough}, {Hopkins}, {Bauer}, {Birchall}, {Ellis}, {Horton}, {Leon-Saval}, {Lewis}, {L{\'o}pez-S{\'a}nchez}, {Min}, {Trinh}, \& {Trowland}}]{SCroom2012}
{Croom}, S.~M., {Lawrence}, J.~S., {Bland-Hawthorn}, J., {et~al.} 2012, \mnras, 421, 872

\bibitem[{{Croom} {et~al.}(2021){Croom}, {Owers}, {Scott}, {Poetrodjojo}, {Groves}, {van de Sande}, {Barone}, {Cortese}, {D'Eugenio}, {Bland-Hawthorn}, {Bryant}, {Oh}, {Brough}, {Agostino}, {Casura}, {Catinella}, {Colless}, {Cecil}, {Davies}, {Drinkwater}, {Driver}, {Ferreras}, {Foster}, {Fraser-McKelvie}, {Lawrence}, {Leslie}, {Liske}, {L{\'o}pez-S{\'a}nchez}, {Lorente}, {McElroy}, {Medling}, {Obreschkow}, {Richards}, {Sharp}, {Sweet}, {Taranu}, {Taylor}, {Tescari}, {Thomas}, {Tocknell}, \& {Vaughan}}]{SCroom2021}
{Croom}, S.~M., {Owers}, M.~S., {Scott}, N., {et~al.} 2021, \mnras, 505, 991

\bibitem[{{Croom} {et~al.}(2024){Croom}, {van de Sande}, {Vaughan}, {Rutherford}, {Lagos}, {Barsanti}, {Bland-Hawthorn}, {Brough}, {Bryant}, {Colless}, {Cortese}, {D'Eugenio}, {Fraser-McKelvie}, {Goodwin}, {Lorente}, {Richards}, {Ristea}, {Sweet}, {Yi}, \& {Zafar}}]{SCroom2024}
{Croom}, S.~M., {van de Sande}, J., {Vaughan}, S.~P., {et~al.} 2024, \mnras, 529, 3446

\bibitem[{{Daddi} {et~al.}(2005){Daddi}, {Renzini}, {Pirzkal}, {Cimatti}, {Malhotra}, {Stiavelli}, {Xu}, {Pasquali}, {Rhoads}, {Brusa}, {di Serego Alighieri}, {Ferguson}, {Koekemoer}, {Moustakas}, {Panagia}, \& {Windhorst}}]{EDaddi2005}
{Daddi}, E., {Renzini}, A., {Pirzkal}, N., {et~al.} 2005, \apj, 626, 680

\bibitem[{{Danhaive} {et~al.}(2025){Danhaive}, {Tacchella}, {\textbackslash''Ubler}, {de Graaff}, {Egami}, {Johnson}, {Sun}, {Arribas}, {Bunker}, {Carniani}, {Jones}, {Maiolino}, {McClymont}, {Parlanti}, {Simmonds}, {Villanueva}, {Baker}, {Jaffe}, {Eisenstein}, {Hainline}, {Helton}, {Ji}, {Lin}, {Pusk\textbackslash'as}, {Rieke}, {Rinaldi}, {Robertson}, {Scholz}, {Williams}, \& {Willmer}}]{ADanhaive2025}
{Danhaive}, A.~L., {Tacchella}, S., {\textbackslash''Ubler}, H., {et~al.} 2025, arXiv e-prints, arXiv:2503.21863

\bibitem[{{Davies} {et~al.}(2024){Davies}, {Belli}, {Park}, {Mendel}, {Johnson}, {Conroy}, {Benton}, {Bugiani}, {Emami}, {Leja}, {Li}, {Maheson}, {Mathews}, {Naidu}, {Nelson}, {Tacchella}, {Terrazas}, \& {Weinberger}}]{RDavies2024}
{Davies}, R.~L., {Belli}, S., {Park}, M., {et~al.} 2024, \mnras, 528, 4976

\bibitem[{{de Graaff} {et~al.}(2024){de Graaff}, {Rix}, {Carniani}, {Suess}, {Charlot}, {Curtis-Lake}, {Arribas}, {Baker}, {Boyett}, {Bunker}, {Cameron}, {Chevallard}, {Curti}, {Eisenstein}, {Franx}, {Hainline}, {Hausen}, {Ji}, {Johnson}, {Jones}, {Maiolino}, {Maseda}, {Nelson}, {Parlanti}, {Rawle}, {Robertson}, {Tacchella}, {{\"U}bler}, {Williams}, {Willmer}, \& {Willott}}]{AdeGraaff2024_kinematics}
{de Graaff}, A., {Rix}, H.-W., {Carniani}, S., {et~al.} 2024, \aap, 684, A87

\bibitem[{{D'Eugenio} {et~al.}(2024){D'Eugenio}, {P{\'e}rez-Gonz{\'a}lez}, {Maiolino}, {Scholtz}, {Perna}, {Circosta}, {{\"U}bler}, {Arribas}, {B{\"o}ker}, {Bunker}, {Carniani}, {Charlot}, {Chevallard}, {Cresci}, {Curtis-Lake}, {Jones}, {Kumari}, {Lamperti}, {Looser}, {Parlanti}, {Rix}, {Robertson}, {Rodr{\'\i}guez Del Pino}, {Tacchella}, {Venturi}, \& {Willott}}]{FdEugenio2024}
{D'Eugenio}, F., {P{\'e}rez-Gonz{\'a}lez}, P.~G., {Maiolino}, R., {et~al.} 2024, Nature Astronomy, 8, 1443

\bibitem[{{Dubois} {et~al.}(2016){Dubois}, {Peirani}, {Pichon}, {Devriendt}, {Gavazzi}, {Welker}, \& {Volonteri}}]{YDubois2016}
{Dubois}, Y., {Peirani}, S., {Pichon}, C., {et~al.} 2016, \mnras, 463, 3948

\bibitem[{{Emsellem} {et~al.}(2011){Emsellem}, {Cappellari}, {Krajnovi{\'c}}, {Alatalo}, {Blitz}, {Bois}, {Bournaud}, {Bureau}, {Davies}, {Davis}, {de Zeeuw}, {Khochfar}, {Kuntschner}, {Lablanche}, {McDermid}, {Morganti}, {Naab}, {Oosterloo}, {Sarzi}, {Scott}, {Serra}, {van de Ven}, {Weijmans}, \& {Young}}]{EEmsellem2011}
{Emsellem}, E., {Cappellari}, M., {Krajnovi{\'c}}, D., {et~al.} 2011, \mnras, 414, 888

\bibitem[{{Emsellem} {et~al.}(2007){Emsellem}, {Cappellari}, {Krajnovi{\'c}}, {van de Ven}, {Bacon}, {Bureau}, {Davies}, {de Zeeuw}, {Falc{\'o}n-Barroso}, {Kuntschner}, {McDermid}, {Peletier}, \& {Sarzi}}]{EEmsellem2007}
{Emsellem}, E., {Cappellari}, M., {Krajnovi{\'c}}, D., {et~al.} 2007, \mnras, 379, 401

\bibitem[{{Epinat} {et~al.}(2009){Epinat}, {Contini}, {Le F{\`e}vre}, {Vergani}, {Garilli}, {Amram}, {Queyrel}, {Tasca}, \& {Tresse}}]{BEpinat2009}
{Epinat}, B., {Contini}, T., {Le F{\`e}vre}, O., {et~al.} 2009, \aap, 504, 789

\bibitem[{{Ferr{\'e}-Mateu} {et~al.}(2017){Ferr{\'e}-Mateu}, {Trujillo}, {Mart{\'\i}n-Navarro}, {Vazdekis}, {Mezcua}, {Balcells}, \& {Dom{\'\i}nguez}}]{AFerreMateu2017}
{Ferr{\'e}-Mateu}, A., {Trujillo}, I., {Mart{\'\i}n-Navarro}, I., {et~al.} 2017, \mnras, 467, 1929

\bibitem[{{Ferr{\'e}-Mateu} {et~al.}(2012){Ferr{\'e}-Mateu}, {Vazdekis}, {Trujillo}, {S{\'a}nchez-Bl{\'a}zquez}, {Ricciardelli}, \& {de la Rosa}}]{AFerreMateu2012}
{Ferr{\'e}-Mateu}, A., {Vazdekis}, A., {Trujillo}, I., {et~al.} 2012, \mnras, 423, 632

\bibitem[{{Ferruit} {et~al.}(2022){Ferruit}, {Jakobsen}, {Giardino}, {Rawle}, {Alves de Oliveira}, {Arribas}, {Beck}, {Birkmann}, {B{\"o}ker}, {Bunker}, {Charlot}, {de Marchi}, {Franx}, {Henry}, {Karakla}, {Kassin}, {Kumari}, {L{\'o}pez-Caniego}, {L{\"u}tzgendorf}, {Maiolino}, {Manjavacas}, {Marston}, {Moseley}, {Muzerolle}, {Pirzkal}, {Rauscher}, {Rix}, {Sabbi}, {Sirianni}, {te Plate}, {Valenti}, {Willott}, \& {Zeidler}}]{PFerruit2022}
{Ferruit}, P., {Jakobsen}, P., {Giardino}, G., {et~al.} 2022, \aap, 661, A81

\bibitem[{{Forrest} {et~al.}(2022){Forrest}, {Wilson}, {Muzzin}, {Marchesini}, {Cooper}, {Marsan}, {Annunziatella}, {McConachie}, {Zaidi}, {Gomez}, {Urbano Stawinski}, {Chang}, {de Lucia}, {La Barbera}, {Lubin}, {Nantais}, {Pe{\~n}a}, {Saracco}, {Surace}, \& {Stefanon}}]{BForrest2022}
{Forrest}, B., {Wilson}, G., {Muzzin}, A., {et~al.} 2022, \apj, 938, 109

\bibitem[{{F{\"o}rster Schreiber} {et~al.}(2009){F{\"o}rster Schreiber}, {Genzel}, {Bouch{\'e}}, {Cresci}, {Davies}, {Buschkamp}, {Shapiro}, {Tacconi}, {Hicks}, {Genel}, {Shapley}, {Erb}, {Steidel}, {Lutz}, {Eisenhauer}, {Gillessen}, {Sternberg}, {Renzini}, {Cimatti}, {Daddi}, {Kurk}, {Lilly}, {Kong}, {Lehnert}, {Nesvadba}, {Verma}, {McCracken}, {Arimoto}, {Mignoli}, \& {Onodera}}]{NForsterSchreiber2009}
{F{\"o}rster Schreiber}, N.~M., {Genzel}, R., {Bouch{\'e}}, N., {et~al.} 2009, \apj, 706, 1364

\bibitem[{{Fraternali} {et~al.}(2021){Fraternali}, {Karim}, {Magnelli}, {G{\'o}mez-Guijarro}, {Jim{\'e}nez-Andrade}, \& {Posses}}]{FFraternali2021}
{Fraternali}, F., {Karim}, A., {Magnelli}, B., {et~al.} 2021, \aap, 647, A194

\bibitem[{{Greene} {et~al.}(2015){Greene}, {Janish}, {Ma}, {McConnell}, {Blakeslee}, {Thomas}, \& {Murphy}}]{JGreene2015}
{Greene}, J.~E., {Janish}, R., {Ma}, C.-P., {et~al.} 2015, \apj, 807, 11

\bibitem[{{Griffith} {et~al.}(2012){Griffith}, {Cooper}, {Newman}, {Moustakas}, {Stern}, {Comerford}, {Davis}, {Lotz}, {Barden}, {Conselice}, {Capak}, {Faber}, {Kirkpatrick}, {Koekemoer}, {Koo}, {Noeske}, {Scoville}, {Sheth}, {Shopbell}, {Willmer}, \& {Weiner}}]{RGriffith2012}
{Griffith}, R.~L., {Cooper}, M.~C., {Newman}, J.~A., {et~al.} 2012, \apjs, 200, 9

\bibitem[{{Hernquist}(1993)}]{LHernquist1993}
{Hernquist}, L. 1993, \apj, 409, 548

\bibitem[{{Hill} {et~al.}(2017){Hill}, {Muzzin}, {Franx}, {Clauwens}, {Schreiber}, {Marchesini}, {Stefanon}, {Labbe}, {Brammer}, {Caputi}, {Fynbo}, {Milvang-Jensen}, {Skelton}, {van Dokkum}, \& {Whitaker}}]{AHill2017}
{Hill}, A.~R., {Muzzin}, A., {Franx}, M., {et~al.} 2017, \apj, 837, 147

\bibitem[{Hopkins {et~al.}(2008)Hopkins, Cox, Kereš, \& Hernquist}]{PHopkins2008}
Hopkins, P.~F., Cox, T.~J., Kereš, D., \& Hernquist, L. 2008, The Astrophysical Journal Supplement Series, 175, 390

\bibitem[{{Jafariyazani} {et~al.}(2020){Jafariyazani}, {Newman}, {Mobasher}, {Belli}, {Ellis}, \& {Patel}}]{Jafariyazani_2020}
{Jafariyazani}, M., {Newman}, A.~B., {Mobasher}, B., {et~al.} 2020, \apjl, 897, L42

\bibitem[{{Johnson} {et~al.}(2021){Johnson}, {Leja}, {Conroy}, \& {Speagle}}]{BJohnson2021}
{Johnson}, B.~D., {Leja}, J., {Conroy}, C., \& {Speagle}, J.~S. 2021, \apjs, 254, 22

\bibitem[{{Kriek} {et~al.}(2024){Kriek}, {Beverage}, {Price}, {Suess}, {Barro}, {Bezanson}, {Conroy}, {Cutler}, {Franx}, {Lin}, {Lorenz}, {Ma}, {Momcheva}, {Mowla}, {Pasha}, {van Dokkum}, \& {Whitaker}}]{MKriek2024}
{Kriek}, M., {Beverage}, A.~G., {Price}, S.~H., {et~al.} 2024, \apj, 966, 36

\bibitem[{{Lagos} {et~al.}(2022){Lagos}, {Emsellem}, {van de Sande}, {Harborne}, {Cortese}, {Davison}, {Foster}, \& {Wright}}]{CLagos2022}
{Lagos}, C. d.~P., {Emsellem}, E., {van de Sande}, J., {et~al.} 2022, \mnras, 509, 4372

\bibitem[{{Lagos} {et~al.}(2018){Lagos}, {Schaye}, {Bah{\'e}}, {van de Sande}, {Kay}, {Barnes}, {Davis}, \& {Dalla Vecchia}}]{CLagos2018a}
{Lagos}, C. d.~P., {Schaye}, J., {Bah{\'e}}, Y., {et~al.} 2018, \mnras, 476, 4327

\bibitem[{Lagos {et~al.}(2018)Lagos, Stevens, Bower, Davis, Contreras, Padilla, Obreschkow, Croton, Trayford, Welker, \& Theuns}]{CLagos2018b}
Lagos, C. d.~P., Stevens, A. R.~H., Bower, R.~G., {et~al.} 2018, Monthly Notices of the Royal Astronomical Society, 473, 4956–4974

\bibitem[{{Li} {et~al.}(2017){Li}, {Ge}, {Mao}, {Cappellari}, {Long}, {Li}, {Emsellem}, {Dutton}, {Li}, {Bundy}, {Thomas}, {Drory}, \& {Lopes}}]{HLi2017}
{Li}, H., {Ge}, J., {Mao}, S., {et~al.} 2017, \apj, 838, 77

\bibitem[{{Ma} {et~al.}(2014){Ma}, {Greene}, {McConnell}, {Janish}, {Blakeslee}, {Thomas}, \& {Murphy}}]{CMa2014}
{Ma}, C.-P., {Greene}, J.~E., {McConnell}, N., {et~al.} 2014, \apj, 795, 158

\bibitem[{{Mart{\'\i}n-Navarro} {et~al.}(2018){Mart{\'\i}n-Navarro}, {Vazdekis}, {Falc{\'o}n-Barroso}, {La Barbera}, {Y{\i}ld{\i}r{\i}m}, \& {van de Ven}}]{IMartinNavarro2018}
{Mart{\'\i}n-Navarro}, I., {Vazdekis}, A., {Falc{\'o}n-Barroso}, J., {et~al.} 2018, \mnras, 475, 3700

\bibitem[{{McGrath} {et~al.}(2008){McGrath}, {Stockton}, {Canalizo}, {Iye}, \& {Maihara}}]{EMcGrath2008}
{McGrath}, E.~J., {Stockton}, A., {Canalizo}, G., {Iye}, M., \& {Maihara}, T. 2008, \apj, 682, 303

\bibitem[{{McLure} {et~al.}(2013){McLure}, {Pearce}, {Dunlop}, {Cirasuolo}, {Curtis-Lake}, {Bruce}, {Caputi}, {Almaini}, {Bonfield}, {Bradshaw}, {Buitrago}, {Chuter}, {Foucaud}, {Hartley}, \& {Jarvis}}]{RMcClure2013}
{McLure}, R.~J., {Pearce}, H.~J., {Dunlop}, J.~S., {et~al.} 2013, \mnras, 428, 1088

\bibitem[{{Mendel} {et~al.}(2020){Mendel}, {Beifiori}, {Saglia}, {Bender}, {Brammer}, {Chan}, {F{\"o}rster Schreiber}, {Fossati}, {Galametz}, {Momcheva}, {Nelson}, {Wilman}, \& {Wuyts}}]{JMendel2020}
{Mendel}, J.~T., {Beifiori}, A., {Saglia}, R.~P., {et~al.} 2020, \apj, 899, 87

\bibitem[{{Muzzin} {et~al.}(2013){Muzzin}, {Marchesini}, {Stefanon}, {Franx}, {McCracken}, {Milvang-Jensen}, {Dunlop}, {Fynbo}, {Brammer}, {Labb{\'e}}, \& {van Dokkum}}]{AMuzzin2013}
{Muzzin}, A., {Marchesini}, D., {Stefanon}, M., {et~al.} 2013, \apj, 777, 18

\bibitem[{Naab {et~al.}(2009)Naab, Johansson, \& Ostriker}]{TNaab2009}
Naab, T., Johansson, P.~H., \& Ostriker, J.~P. 2009, The Astrophysical Journal, 699, L178

\bibitem[{{Naab} {et~al.}(2014){Naab}, {Oser}, {Emsellem}, {Cappellari}, {Krajnovi{\'c}}, {McDermid}, {Alatalo}, {Bayet}, {Blitz}, {Bois}, {Bournaud}, {Bureau}, {Crocker}, {Davies}, {Davis}, {de Zeeuw}, {Duc}, {Hirschmann}, {Johansson}, {Khochfar}, {Kuntschner}, {Morganti}, {Oosterloo}, {Sarzi}, {Scott}, {Serra}, {van de Ven}, {Weijmans}, \& {Young}}]{TNaab2014}
{Naab}, T., {Oser}, L., {Emsellem}, E., {et~al.} 2014, \mnras, 444, 3357

\bibitem[{{Neeleman} {et~al.}(2020){Neeleman}, {Prochaska}, {Kanekar}, \& {Rafelski}}]{MNeeleman2020}
{Neeleman}, M., {Prochaska}, J.~X., {Kanekar}, N., \& {Rafelski}, M. 2020, \nat, 581, 269

\bibitem[{{Newman} {et~al.}(2015){Newman}, {Belli}, \& {Ellis}}]{ANewman2015}
{Newman}, A.~B., {Belli}, S., \& {Ellis}, R.~S. 2015, \apjl, 813, L7

\bibitem[{{Newman} {et~al.}(2018){Newman}, {Belli}, {Ellis}, \& {Patel}}]{ANewman2018b}
{Newman}, A.~B., {Belli}, S., {Ellis}, R.~S., \& {Patel}, S.~G. 2018, \apj, 862, 126

\bibitem[{{Newman} {et~al.}(2012){Newman}, {Ellis}, {Bundy}, \& {Treu}}]{ANewman2012}
{Newman}, A.~B., {Ellis}, R.~S., {Bundy}, K., \& {Treu}, T. 2012, \apj, 746, 162

\bibitem[{{Newman} {et~al.}(2025){Newman}, {Gu}, {Belli}, {Ellis}, {Gangula}, {Greene}, {Walsh}, {Suyu}, {Ertl}, {Caminha}, {Granata}, {Grillo}, {Schuldt}, {Barone}, {Bird}, {Glazebrook}, {Jafariyazani}, {Kriek}, {Matthews}, {Morishita}, {Nanayakkara}, {Pierel}, {Acebr\textbackslash'on}, {Bergamini}, {Cha}, {Diego}, {Foo}, {Frye}, {Fudamoto}, {Jee}, {Kamieneski}, {Koekemoer}, {Meena}, {Nishida}, {Oguri}, {Rosati}, \& {Zitrin}}]{ANewman2025}
{Newman}, A.~B., {Gu}, M., {Belli}, S., {et~al.} 2025, arXiv e-prints, submitted to Nature, arXiv:2503.17478

\bibitem[{{Parlanti} {et~al.}(2023){Parlanti}, {Carniani}, {Pallottini}, {Cignoni}, {Cresci}, {Kohandel}, {Mannucci}, \& {Marconi}}]{EParlanti2023}
{Parlanti}, E., {Carniani}, S., {Pallottini}, A., {et~al.} 2023, \aap, 673, A153

\bibitem[{{Pascalau} {et~al.}(2025){Pascalau}, {D'Eugenio}, {Tacchella}, {Maiolino}, {Cappellari}, {Lagos}, {Bunker}, {Jones}, {Scholtz}, {{\"U}bler}, {Cresci}, {Arribas}, {Perna}, {van der Wel}, {Danhaive}, {McClymont}, {Vani}, {Maseda}, {Carnall}, {Charlot}, {Carniani}, {Duan}, {Goh}, {de Graaff}, {Ji}, \& {P{\'e}rez-Gonz{\'a}lez}}]{RPascalau2025}
{Pascalau}, R.~G., {D'Eugenio}, F., {Tacchella}, S., {et~al.} 2025, arXiv e-prints, submitted to MNRAS, arXiv:2505.06349

\bibitem[{{Patel} {et~al.}(2013){Patel}, {van Dokkum}, {Franx}, {Quadri}, {Muzzin}, {Marchesini}, {Williams}, {Holden}, \& {Stefanon}}]{SPatel2013}
{Patel}, S.~G., {van Dokkum}, P.~G., {Franx}, M., {et~al.} 2013, \apj, 766, 15

\bibitem[{{Peng} {et~al.}(2002){Peng}, {Ho}, {Impey}, \& {Rix}}]{CPeng2002}
{Peng}, C.~Y., {Ho}, L.~C., {Impey}, C.~D., \& {Rix}, H.-W. 2002, \aj, 124, 266

\bibitem[{{Peng} {et~al.}(2010){Peng}, {Ho}, {Impey}, \& {Rix}}]{CPeng2010}
{Peng}, C.~Y., {Ho}, L.~C., {Impey}, C.~D., \& {Rix}, H.-W. 2010, \aj, 139, 2097

\bibitem[{{Penoyre} {et~al.}(2017){Penoyre}, {Moster}, {Sijacki}, \& {Genel}}]{ZPenoyre2017}
{Penoyre}, Z., {Moster}, B.~P., {Sijacki}, D., \& {Genel}, S. 2017, \mnras, 468, 3883

\bibitem[{{Perna} {et~al.}(2023){Perna}, {Arribas}, {Marshall}, {D'Eugenio}, {{\"U}bler}, {Bunker}, {Charlot}, {Carniani}, {Jakobsen}, {Maiolino}, {Rodr{\'\i}guez Del Pino}, {Willott}, {B{\"o}ker}, {Circosta}, {Cresci}, {Curti}, {Husemann}, {Kumari}, {Lamperti}, {P{\'e}rez-Gonz{\'a}lez}, \& {Scholtz}}]{MPerna2023}
{Perna}, M., {Arribas}, S., {Marshall}, M., {et~al.} 2023, \aap, 679, A89

\bibitem[{{Perrin} {et~al.}(2014){Perrin}, {Sivaramakrishnan}, {Lajoie}, {Elliott}, {Pueyo}, {Ravindranath}, \& {Albert}}]{MPerrin2014}
{Perrin}, M.~D., {Sivaramakrishnan}, A., {Lajoie}, C.-P., {et~al.} 2014, in Society of Photo-Optical Instrumentation Engineers (SPIE) Conference Series, Vol. 9143, Space Telescopes and Instrumentation 2014: Optical, Infrared, and Millimeter Wave, ed. J.~M. {Oschmann}, Jr., M.~{Clampin}, G.~G. {Fazio}, \& H.~A. {MacEwen}, 91433X

\bibitem[{{Price} {et~al.}(2020){Price}, {Kriek}, {Barro}, {Shapley}, {Reddy}, {Freeman}, {Coil}, {Shivaei}, {Azadi}, {de Groot}, {Siana}, {Mobasher}, {Sanders}, {Leung}, {Fetherolf}, {Zick}, {{\"U}bler}, \& {F{\"o}rster Schreiber}}]{SPrice2020}
{Price}, S.~H., {Kriek}, M., {Barro}, G., {et~al.} 2020, \apj, 894, 91

\bibitem[{{Price} {et~al.}(2014){Price}, {Kriek}, {Brammer}, {Conroy}, {F{\"o}rster Schreiber}, {Franx}, {Fumagalli}, {Lundgren}, {Momcheva}, {Nelson}, {Skelton}, {van Dokkum}, {Whitaker}, \& {Wuyts}}]{SPrice2014}
{Price}, S.~H., {Kriek}, M., {Brammer}, G.~B., {et~al.} 2014, \apj, 788, 86

\bibitem[{{Price} {et~al.}(2016){Price}, {Kriek}, {Shapley}, {Reddy}, {Freeman}, {Coil}, {de Groot}, {Shivaei}, {Siana}, {Azadi}, {Barro}, {Mobasher}, {Sanders}, \& {Zick}}]{SPrice2016}
{Price}, S.~H., {Kriek}, M., {Shapley}, A.~E., {et~al.} 2016, \apj, 819, 80

\bibitem[{{Price} {et~al.}(2022){Price}, {{\"U}bler}, {F{\"o}rster Schreiber}, {de Zeeuw}, {Burkert}, {Genzel}, {Tacconi}, {Davies}, \& {Price}}]{SPrice2022}
{Price}, S.~H., {{\"U}bler}, H., {F{\"o}rster Schreiber}, N.~M., {et~al.} 2022, \aap, 665, A159

\bibitem[{{Rantala} {et~al.}(2024){Rantala}, {Rawlings}, {Naab}, {Thomas}, \& {Johansson}}]{ARantala2024}
{Rantala}, A., {Rawlings}, A., {Naab}, T., {Thomas}, J., \& {Johansson}, P.~H. 2024, \mnras, 535, 1202

\bibitem[{{Rizzo} {et~al.}(2021){Rizzo}, {Vegetti}, {Fraternali}, {Stacey}, \& {Powell}}]{FRizzo2021}
{Rizzo}, F., {Vegetti}, S., {Fraternali}, F., {Stacey}, H.~R., \& {Powell}, D. 2021, \mnras, 507, 3952

\bibitem[{{Rodriguez-Gomez} {et~al.}(2017){Rodriguez-Gomez}, {Sales}, {Genel}, {Pillepich}, {Zjupa}, {Nelson}, {Griffen}, {Torrey}, {Snyder}, {Vogelsberger}, {Springel}, {Ma}, \& {Hernquist}}]{VRodriguez2017}
{Rodriguez-Gomez}, V., {Sales}, L.~V., {Genel}, S., {et~al.} 2017, \mnras, 467, 3083

\bibitem[{{Schulze} {et~al.}(2018){Schulze}, {Remus}, {Dolag}, {Burkert}, {Emsellem}, \& {van de Ven}}]{FSchulze2018}
{Schulze}, F., {Remus}, R.-S., {Dolag}, K., {et~al.} 2018, \mnras, 480, 4636

\bibitem[{Scoville {et~al.}(2007)Scoville, Abraham, Aussel, Barnes, Benson, Blain, Calzetti, Comastri, Capak, Carilli, Carlstrom, Carollo, Colbert, Daddi, Ellis, Elvis, Ewald, Fall, Franceschini, Giavalisco, Green, Griffiths, Guzzo, Hasinger, Impey, Kneib, Koda, Koekemoer, Lefevre, Lilly, Liu, McCracken, Massey, Mellier, Miyazaki, Mobasher, Mould, Norman, Refregier, Renzini, Rhodes, Rich, Sanders, Schiminovich, Schinnerer, Scodeggio, Sheth, Shopbell, Taniguchi, Tyson, Urry, Van~Waerbeke, Vettolani, White, \& Yan}]{NScoville2007}
Scoville, N., Abraham, R.~G., Aussel, H., {et~al.} 2007, The Astrophysical Journal Supplement Series, 172, 38

\bibitem[{{Sersic}(1968)}]{JSersic1968}
{Sersic}, J.~L. 1968, {Atlas de Galaxias Australes}

\bibitem[{{Slob} {et~al.}(2024){Slob}, {Kriek}, {Beverage}, {Suess}, {Barro}, {Bezanson}, {Brammer}, {Cheng}, {Conroy}, {de Graaff}, {F{\"o}rster Schreiber}, {Franx}, {Lorenz}, {Mancera Pi{\~n}a}, {Marchesini}, {Muzzin}, {Newman}, {Price}, {Shapley}, {Stefanon}, {van Dokkum}, \& {Weisz}}]{MSlob2024}
{Slob}, M., {Kriek}, M., {Beverage}, A.~G., {et~al.} 2024, \apj, 973, 131

\bibitem[{{Spiniello} {et~al.}(2024){Spiniello}, {D'Ago}, {Coccato}, {Hartke}, {Tortora}, {Ferr{\'e}-Mateu}, {Pulsoni}, {Cappellari}, {Maksymowicz-Maciata}, {Arnaboldi}, {Bevacqua}, {Gallazzi}, {Hunt}, {La Barbera}, {Mart{\'\i}n-Navarro}, {Napolitano}, {Radovich}, {Saracco}, {Scognamiglio}, {Spavone}, \& {Zibetti}}]{CSpiniello2024}
{Spiniello}, C., {D'Ago}, G., {Coccato}, L., {et~al.} 2024, \mnras, 527, 8793

\bibitem[{{Straatman} {et~al.}(2022){Straatman}, {van der Wel}, {van Houdt}, {Bezanson}, {Bell}, {van Dokkum}, {D'Eugenio}, {Franx}, {Gallazzi}, {de Graaff}, {Maseda}, {Meidt}, {Muzzin}, {Sobral}, \& {Wu}}]{CStraatman2022}
{Straatman}, C. M.~S., {van der Wel}, A., {van Houdt}, J., {et~al.} 2022, \apj, 928, 126

\bibitem[{{Struck}(1999)}]{CStruck1999}
{Struck}, C. 1999, \physrep, 321, 1

\bibitem[{{Suess} {et~al.}(2019){Suess}, {Kriek}, {Price}, \& {Barro}}]{KSuess2019}
{Suess}, K.~A., {Kriek}, M., {Price}, S.~H., \& {Barro}, G. 2019, \apjl, 885, L22

\bibitem[{{Suess} {et~al.}(2020){Suess}, {Kriek}, {Price}, \& {Barro}}]{KSuess2020}
{Suess}, K.~A., {Kriek}, M., {Price}, S.~H., \& {Barro}, G. 2020, \apjl, 899, L26

\bibitem[{{Suess} {et~al.}(2021){Suess}, {Kriek}, {Price}, \& {Barro}}]{KSuess2021}
{Suess}, K.~A., {Kriek}, M., {Price}, S.~H., \& {Barro}, G. 2021, \apj, 915, 87

\bibitem[{{Suess} {et~al.}(2023){Suess}, {Williams}, {Robertson}, {Ji}, {Johnson}, {Nelson}, {Alberts}, {Hainline}, {D'Eugenio}, {{\"U}bler}, {Rieke}, {Rieke}, {Bunker}, {Carniani}, {Charlot}, {Eisenstein}, {Maiolino}, {Stark}, {Tacchella}, \& {Willott}}]{KSuess2023}
{Suess}, K.~A., {Williams}, C.~C., {Robertson}, B., {et~al.} 2023, \apjl, 956, L42

\bibitem[{{Toft} {et~al.}(2017){Toft}, {Zabl}, {Richard}, {Gallazzi}, {Zibetti}, {Prescott}, {Grillo}, {Man}, {Lee}, {G{\'o}mez-Guijarro}, {Stockmann}, {Magdis}, \& {Steinhardt}}]{SToft2017}
{Toft}, S., {Zabl}, J., {Richard}, J., {et~al.} 2017, \nat, 546, 510

\bibitem[{{Tortora} {et~al.}(2013){Tortora}, {Romanowsky}, \& {Napolitano}}]{CTortora2013}
{Tortora}, C., {Romanowsky}, A.~J., \& {Napolitano}, N.~R. 2013, \apj, 765, 8

\bibitem[{{Tortora} {et~al.}(2025){Tortora}, {Tozzi}, {Agapito}, {Barbera}, {Spiniello}, {Li}, {Carl{\`a}}, {D'Ago}, {Ghose}, {Mannucci}, {Napolitano}, {Pinna}, {Arnaboldi}, {Bevacqua}, {Ferr{\'e}-Mateu}, {Gallazzi}, {Hartke}, {Hunt}, {Maksymowicz-Maciata}, {Pulsoni}, {Saracco}, {Scognamiglio}, \& {Spavone}}]{CTortora2025}
{Tortora}, C., {Tozzi}, G., {Agapito}, G., {et~al.} 2025, \mnras, 540, 2555

\bibitem[{{Treu} {et~al.}(2010){Treu}, {Auger}, {Koopmans}, {Gavazzi}, {Marshall}, \& {Bolton}}]{TTreu2010}
{Treu}, T., {Auger}, M.~W., {Koopmans}, L. V.~E., {et~al.} 2010, \apj, 709, 1195

\bibitem[{{{\"U}bler} {et~al.}(2024){{\"U}bler}, {F{\"o}rster Schreiber}, {van der Wel}, {Bezanson}, {Price}, {D'Eugenio}, {Wisnioski}, {Genzel}, {Tacconi}, {Wuyts}, {Naab}, {Lutz}, {Straatman}, {Shimizu}, {Davies}, {Liu}, \& {Mendel}}]{HUebler2024}
{{\"U}bler}, H., {F{\"o}rster Schreiber}, N.~M., {van der Wel}, A., {et~al.} 2024, \mnras, 527, 9206

\bibitem[{{van de Sande} {et~al.}(2017){van de Sande}, {Bland-Hawthorn}, {Fogarty}, {Cortese}, {d'Eugenio}, {Croom}, {Scott}, {Allen}, {Brough}, {Bryant}, {Cecil}, {Colless}, {Couch}, {Davies}, {Elahi}, {Foster}, {Goldstein}, {Goodwin}, {Groves}, {Ho}, {Jeong}, {Jones}, {Konstantopoulos}, {Lawrence}, {Leslie}, {L{\'o}pez-S{\'a}nchez}, {McDermid}, {McElroy}, {Medling}, {Oh}, {Owers}, {Richards}, {Schaefer}, {Sharp}, {Sweet}, {Taranu}, {Tonini}, {Walcher}, \& {Yi}}]{JvandeSande2017}
{van de Sande}, J., {Bland-Hawthorn}, J., {Fogarty}, L. M.~R., {et~al.} 2017, \apj, 835, 104

\bibitem[{{van de Sande} {et~al.}(2013){van de Sande}, {Kriek}, {Franx}, {van Dokkum}, {Bezanson}, {Bouwens}, {Quadri}, {Rix}, \& {Skelton}}]{JvandeSande2013}
{van de Sande}, J., {Kriek}, M., {Franx}, M., {et~al.} 2013, \apj, 771, 85

\bibitem[{{van de Sande} {et~al.}(2019){van de Sande}, {Lagos}, {Welker}, {Bland-Hawthorn}, {Schulze}, {Remus}, {Bah{\'e}}, {Brough}, {Bryant}, {Cortese}, {Croom}, {Devriendt}, {Dubois}, {Goodwin}, {Konstantopoulos}, {Lawrence}, {Medling}, {Pichon}, {Richards}, {Sanchez}, {Scott}, \& {Sweet}}]{Jvandesande2019}
{van de Sande}, J., {Lagos}, C. D.~P., {Welker}, C., {et~al.} 2019, \mnras, 484, 869

\bibitem[{{van de Sande} {et~al.}(2018){van de Sande}, {Scott}, {Bland-Hawthorn}, {Brough}, {Bryant}, {Colless}, {Cortese}, {Croom}, {d'Eugenio}, {Foster}, {Goodwin}, {Konstantopoulos}, {Lawrence}, {McDermid}, {Medling}, {Owers}, {Richards}, \& {Sharp}}]{JVandeSande2018}
{van de Sande}, J., {Scott}, N., {Bland-Hawthorn}, J., {et~al.} 2018, Nature Astronomy, 2, 483

\bibitem[{{van der Wel} {et~al.}(2021){van der Wel}, {Bezanson}, {D'Eugenio}, {Straatman}, {Franx}, {van Houdt}, {Maseda}, {Gallazzi}, {Wu}, {Pacifici}, {Barisic}, {Brammer}, {Munoz-Mateos}, {Vervalcke}, {Zibetti}, {Sobral}, {de Graaff}, {Calhau}, {Kaushal}, {Muzzin}, {Bell}, \& {van Dokkum}}]{AvanderWel2021}
{van der Wel}, A., {Bezanson}, R., {D'Eugenio}, F., {et~al.} 2021, \apjs, 256, 44

\bibitem[{{van der Wel} {et~al.}(2014){van der Wel}, {Franx}, {van Dokkum}, {Skelton}, {Momcheva}, {Whitaker}, {Brammer}, {Bell}, {Rix}, {Wuyts}, {Ferguson}, {Holden}, {Barro}, {Koekemoer}, {Chang}, {McGrath}, {H{\"a}ussler}, {Dekel}, {Behroozi}, {Fumagalli}, {Leja}, {Lundgren}, {Maseda}, {Nelson}, {Wake}, {Patel}, {Labb{\'e}}, {Faber}, {Grogin}, \& {Kocevski}}]{AvanderWel2014}
{van der Wel}, A., {Franx}, M., {van Dokkum}, P.~G., {et~al.} 2014, \apj, 788, 28

\bibitem[{{van der Wel} {et~al.}(2011){van der Wel}, {Rix}, {Wuyts}, {McGrath}, {Koekemoer}, {Bell}, {Holden}, {Robaina}, \& {McIntosh}}]{AvanderWel2011}
{van der Wel}, A., {Rix}, H.-W., {Wuyts}, S., {et~al.} 2011, \apj, 730, 38

\bibitem[{{van der Wel} {et~al.}(2022){van der Wel}, {van Houdt}, {Bezanson}, {Franx}, {D'Eugenio}, {Straatman}, {Bell}, {Muzzin}, {Sobral}, {Maseda}, {de Graaff}, \& {Holden}}]{AvanderWel2022}
{van der Wel}, A., {van Houdt}, J., {Bezanson}, R., {et~al.} 2022, \apj, 936, 9

\bibitem[{van Dokkum {et~al.}(2008)van Dokkum, Franx, Kriek, Holden, Illingworth, Magee, Bouwens, Marchesini, Quadri, Rudnick, Taylor, \& Toft}]{PvanDokkum2008}
van Dokkum, P.~G., Franx, M., Kriek, M., {et~al.} 2008, The Astrophysical Journal, 677, L5

\bibitem[{van Dokkum {et~al.}(2010)van Dokkum, Whitaker, Brammer, Franx, Kriek, Labbé, Marchesini, Quadri, Bezanson, Illingworth, Muzzin, Rudnick, Tal, \& Wake}]{PvanDokkum2010}
van Dokkum, P.~G., Whitaker, K.~E., Brammer, G., {et~al.} 2010, The Astrophysical Journal, 709, 1018

\bibitem[{{van Houdt} {et~al.}(2021){van Houdt}, {van der Wel}, {Bezanson}, {Franx}, {d'Eugenio}, {Barisic}, {Bell}, {Gallazzi}, {de Graaff}, {Maseda}, {Pacifici}, {van de Sande}, {Sobral}, {Straatman}, \& {Wu}}]{JvanHoudt2021}
{van Houdt}, J., {van der Wel}, A., {Bezanson}, R., {et~al.} 2021, \apj, 923, 11

\bibitem[{{Veale} {et~al.}(2017){Veale}, {Ma}, {Thomas}, {Greene}, {McConnell}, {Walsh}, {Ito}, {Blakeslee}, \& {Janish}}]{MVeale2017}
{Veale}, M., {Ma}, C.-P., {Thomas}, J., {et~al.} 2017, \mnras, 464, 356

\bibitem[{{Williams} {et~al.}(2021){Williams}, {Spilker}, {Whitaker}, {Dav{\'e}}, {Woodrum}, {Brammer}, {Bezanson}, {Narayanan}, \& {Weiner}}]{CWilliams2021}
{Williams}, C.~C., {Spilker}, J.~S., {Whitaker}, K.~E., {et~al.} 2021, \apj, 908, 54

\bibitem[{{Wisnioski} {et~al.}(2015){Wisnioski}, {F{\"o}rster Schreiber}, {Wuyts}, {Wuyts}, {Bandara}, {Wilman}, {Genzel}, {Bender}, {Davies}, {Fossati}, {Lang}, {Mendel}, {Beifiori}, {Brammer}, {Chan}, {Fabricius}, {Fudamoto}, {Kulkarni}, {Kurk}, {Lutz}, {Nelson}, {Momcheva}, {Rosario}, {Saglia}, {Seitz}, {Tacconi}, \& {van Dokkum}}]{EWisnioski2015}
{Wisnioski}, E., {F{\"o}rster Schreiber}, N.~M., {Wuyts}, S., {et~al.} 2015, \apj, 799, 209

\bibitem[{Wisnioski {et~al.}(2019)Wisnioski, Schreiber, Fossati, Mendel, Wilman, Genzel, Bender, Wuyts, Davies, Übler, Bandara, Beifiori, Belli, Brammer, Chan, Davies, Fabricius, Galametz, Lang, Lutz, Nelson, Momcheva, Price, Rosario, Saglia, Seitz, Shimizu, Tacconi, Tadaki, van Dokkum, \& Wuyts}]{EWisnioski2019}
Wisnioski, E., Schreiber, N. M.~F., Fossati, M., {et~al.} 2019, The Astrophysical Journal, 886, 124

\bibitem[{{Y{\i}ld{\i}r{\i}m} {et~al.}(2015){Y{\i}ld{\i}r{\i}m}, {van den Bosch}, {van de Ven}, {Husemann}, {Lyubenova}, {Walsh}, {Gebhardt}, \& {G{\"u}ltekin}}]{AYilderim2015}
{Y{\i}ld{\i}r{\i}m}, A., {van den Bosch}, R. C.~E., {van de Ven}, G., {et~al.} 2015, \mnras, 452, 1792

\end{thebibliography}

\begin{appendix}
\onecolumn
\section{Structural parameters from GALFIT}\label{ap:galfit}
As described in Section \ref{sec:galfit}, we used GALFIT \citep{CPeng2002} to measure the structural parameters of the galaxies in our sample for which COSMOS-Web imaging was available. We describe our fitting method below. 

First, we construct a synthetic PSF for each of the four NIRCam filters that we use to fit the structural parameters (F115W, F150W, F277W, and F444W) using the WebbPSF software \citep{MPerrin2014}. In our fitting, we use the distorted PSF, sampled to the NIRCam pixel scale of each filter. As the image input for GALFIT we use the COSMOS-Web images that are available in MAST\footnote{\url{https://mast.stsci.edu/}}, and cut out images of $200\times200$ pixels around each galaxy in our sample. We then construct segmentation maps for each cutout using {\sc photutils}, and mask out any additional galaxies in the cutout. We subtract a local background estimate from each masked cutout using {\sc photutils}. Finally, we fit the background-subtracted cutout for each galaxy and each filter using GALFIT.

\begin{figure}[h!]
    \flushleft   
    \includegraphics[width=\linewidth]{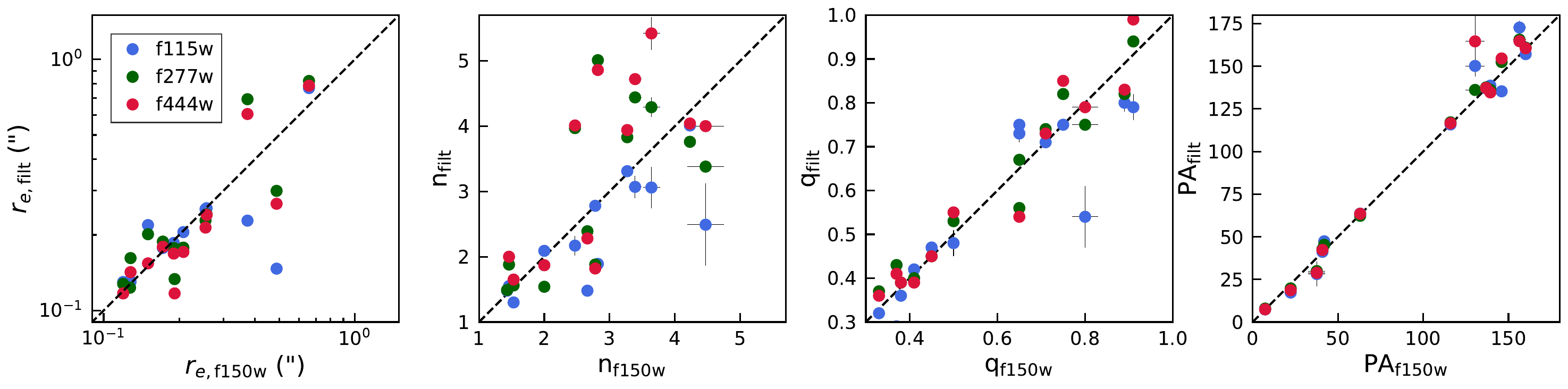}
    \caption{Measured effective radius ($r_e$; first panel), S\'ersic index ($n$; second panel), axis ratio ($q$; third panel), and position angle (PA; last panel) for the galaxies in our sample in the F150W filter against the F115W (blue points), F277W (green points), and F444W (red points) filters. The dashed line in each panel represents a one-to-one ratio.}
    \label{fig:galfit_comparison}
\end{figure}

In Figure \ref{fig:galfit_comparison} we show the effective radius ($r_e$), S\'ersic index ($n$), axis ratio ($q$), and position angle (PA) for each galaxy in the F150W filter versus the F115W (blue points), F277W (green points), and F444W (red points) filters. For $r_e$, $q$, and PA the measurements from the four filters are in good agreement, with little scatter and no systematic offsets. For the S\'ersic index the scatter between the four different filters is more prominent, but there is no structural offset between the different filters. In our modelling we use the median value of each structural parameter in the NIRCam filters which cover rest-frame optical wavelengths for our galaxies. For galaxies at $z>1.3$ this means we exclude the F115W filter, and for galaxies at $z>2$ we exclude the F115W and F150W filters.

Although this might lead to larger uncertainties in the S\'ersic index, we note that the choice of $n$ does not affect our modelling significantly. On the other hand, $r_e$, $q$ and PA have a more direct impact on our modelling results. The small scatter between the different filters for these parameters implies that no additional uncertainties are introduced by using the median of the different filters.
\newpage
\section{Kinematic modelling results for misaligned galaxies}\label{ap:unconstrained_fits}
\begin{figure}[h!]
    \flushleft
    \vspace{-0.15in}
    \subfigure{\includegraphics[width=0.473\linewidth]{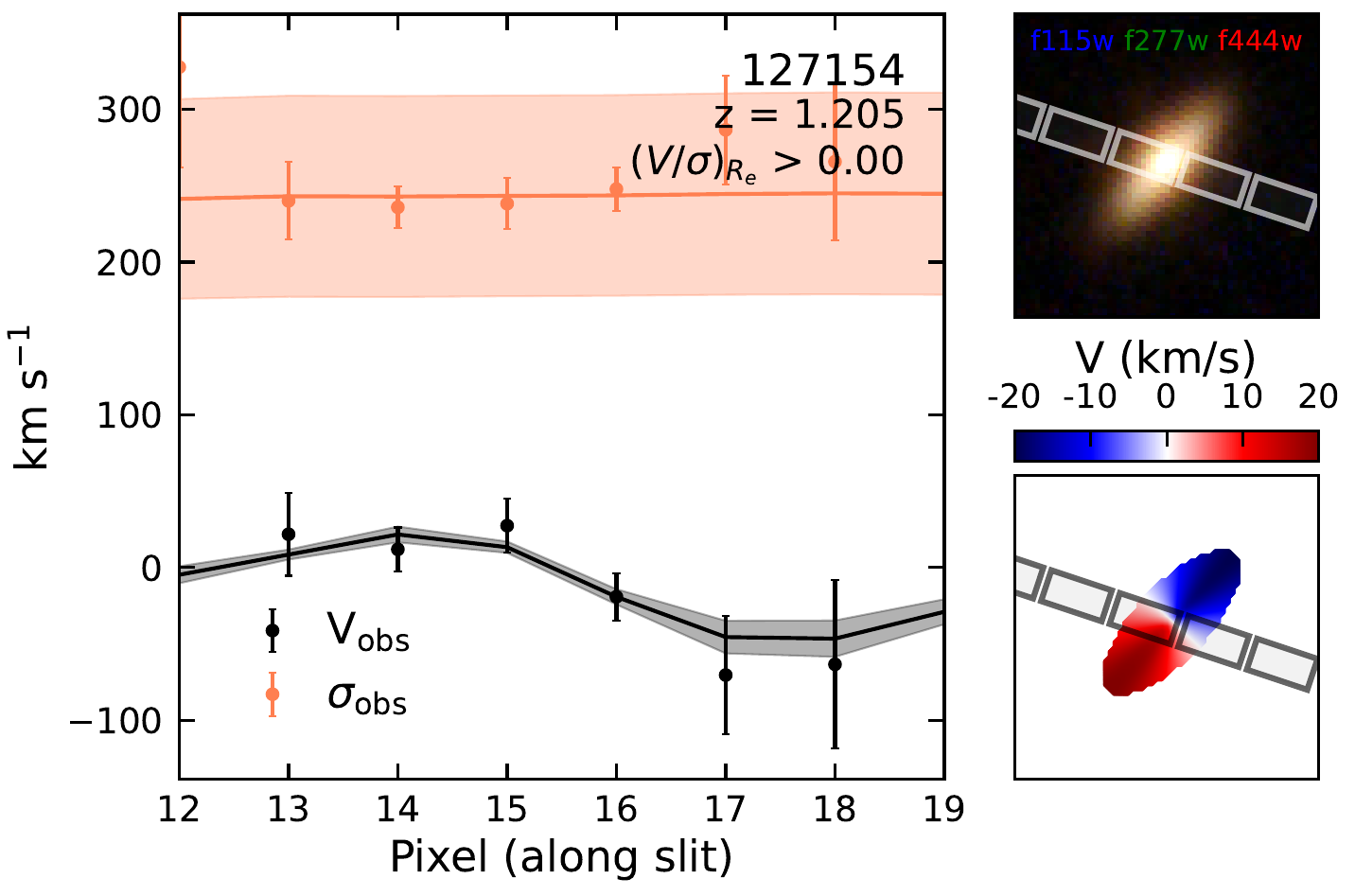}}%
    \subfigure{\includegraphics[width=0.473\linewidth]{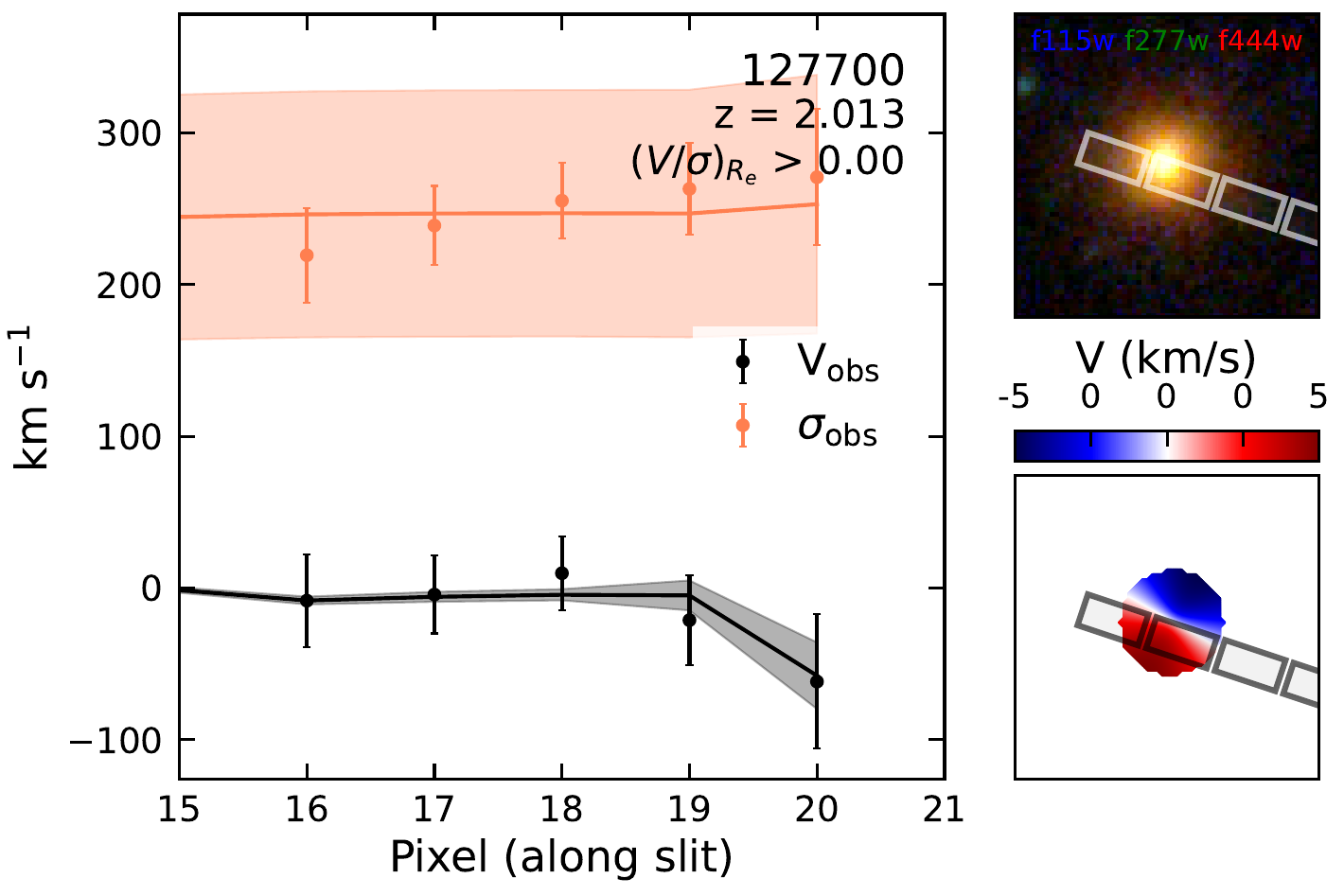}}
    \subfigure{\includegraphics[width=0.473\linewidth]{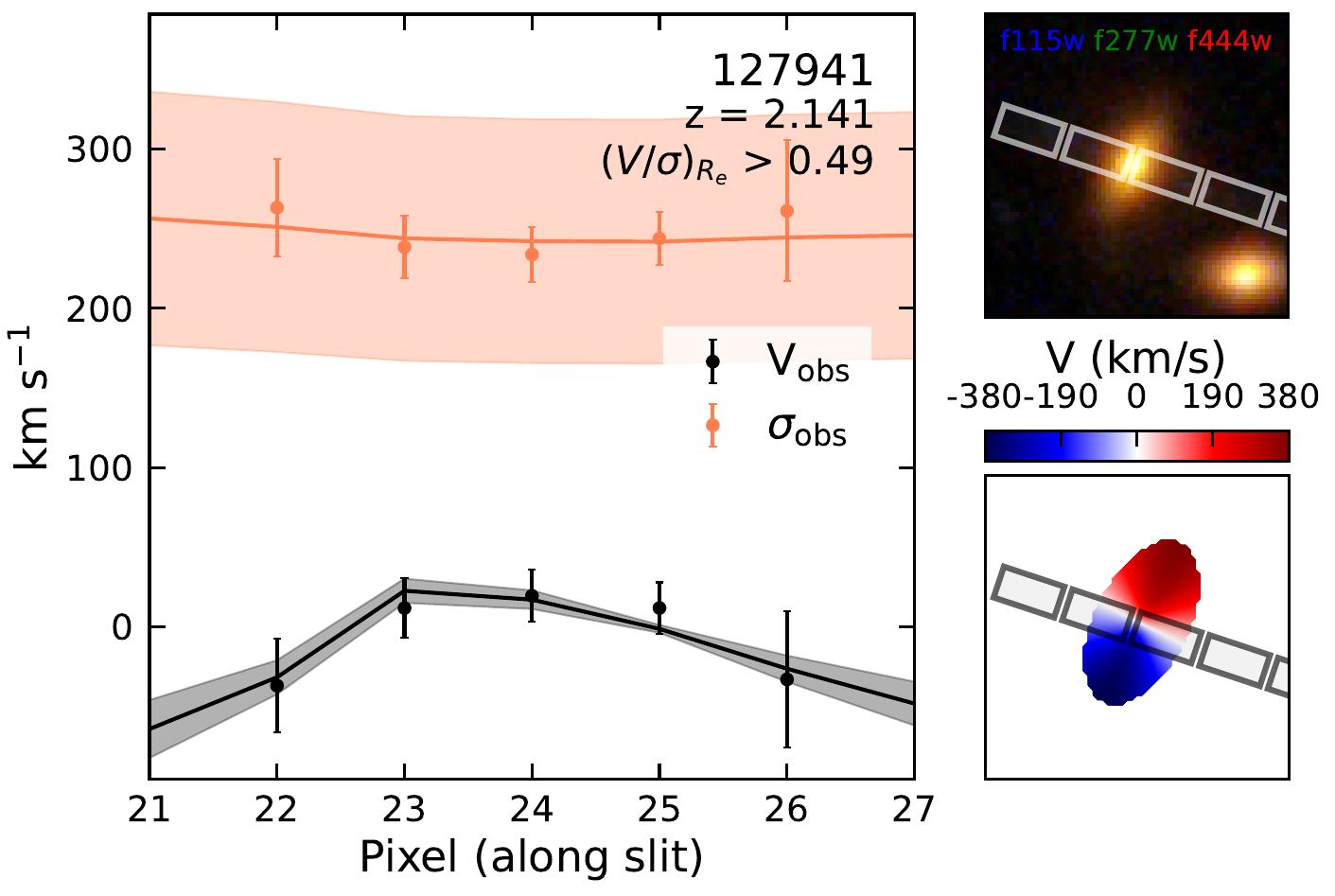}}%
    \subfigure{\includegraphics[width=0.473\linewidth]{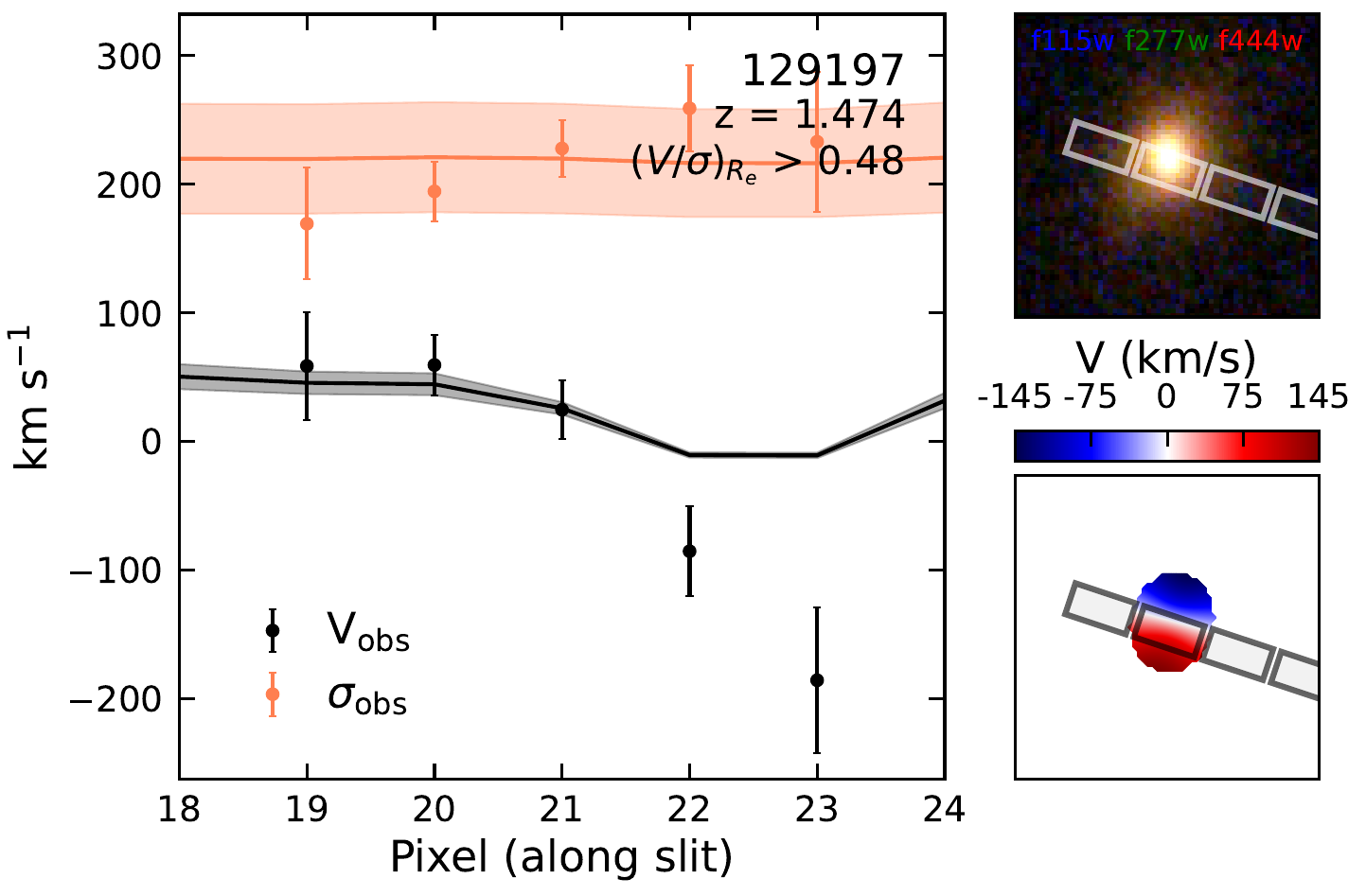}}
    \subfigure{\includegraphics[width=0.473\linewidth]{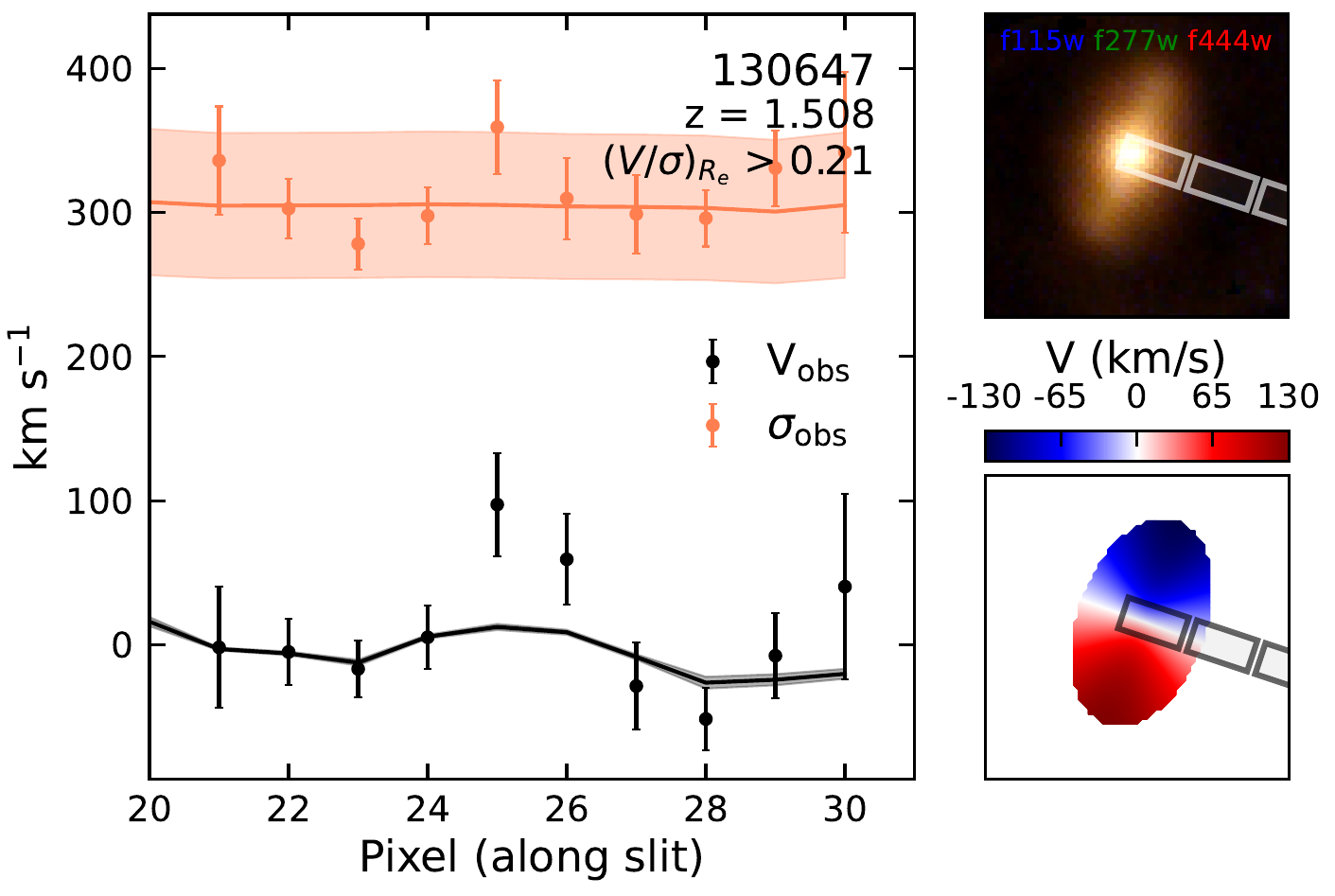}}%
    \label{fig:v_models_unconstrained}
    \vspace{-0.1in}
    \caption{Same as Figure \ref{fig:v_curves} but for the five galaxies in our sample for which the rotational velocity could not be constrained.}
\end{figure}

For five out of 15 galaxies we could not constrain the rotational velocities due to the fact that they were (mostly) aligned along the minor axis with respect to the MSA shutter. We show the best-fit kinematic models for these galaxies in Figure \ref{fig:v_models_unconstrained}.

\section{Aperture-corrected dynamical masses}\label{ap:dyn_masses}
\begin{figure}[h!]
    \centering
    \includegraphics[width=0.425\linewidth]{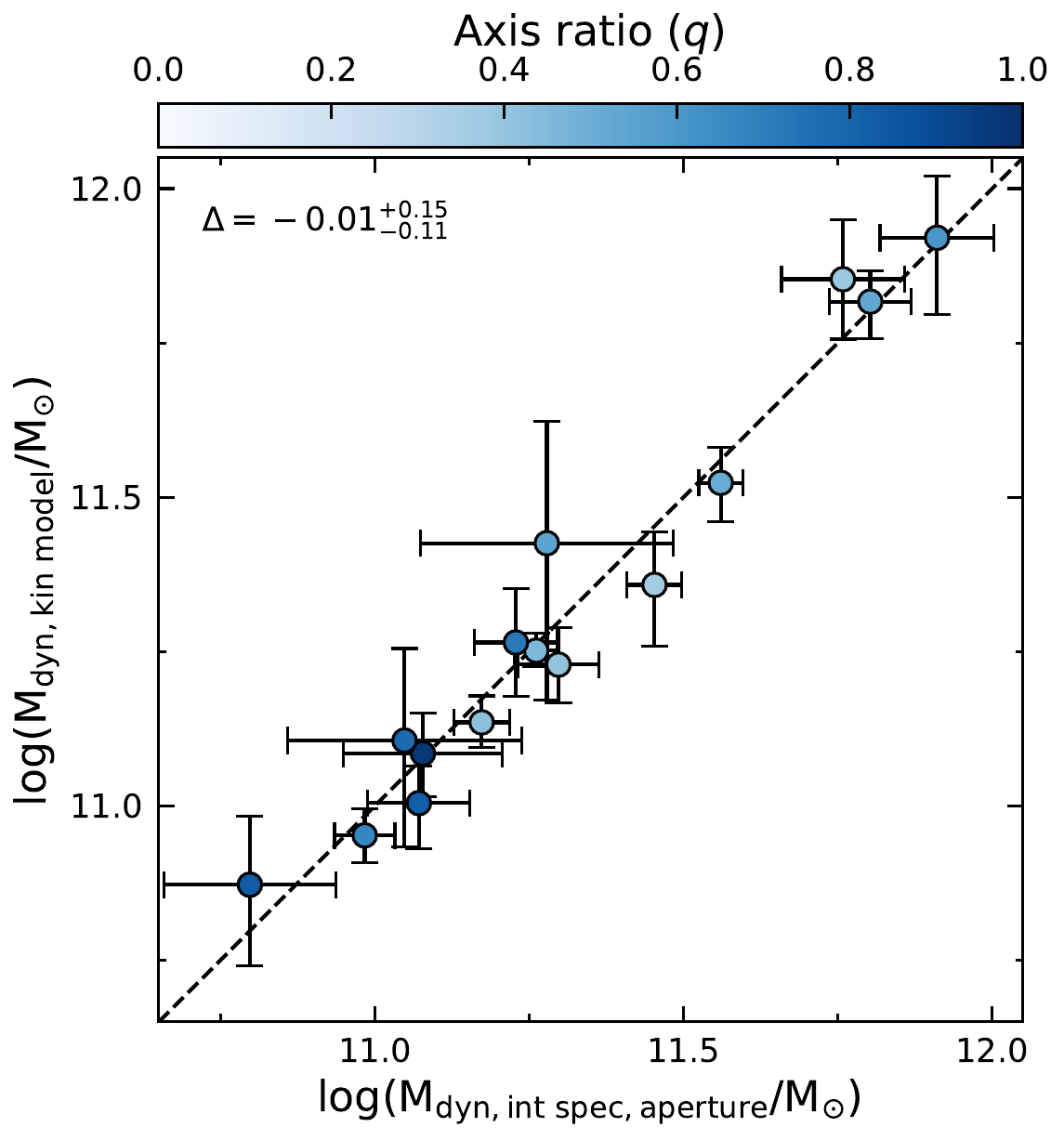}
    \caption{Dynamical masses calculated from the aperture corrected inferred integrated velocity dispersion using the virial theorem against dynamical masses calculated from our kinematic models. We colour the points by the axis ratio, $q$. The median offset between the two dynamical masses is shown in the top left corner.}
    \label{fig:mdyn_apertures}
\end{figure}

In Section \ref{sec:discussion_apertures} we derive integrated line-of-sight velocity dispersions ($\sigma_{v,r_e}$) within one $r_e$ from our best-fit models. We now use these $\sigma_{v,r_e}$ values to calculate the virial dynamical masses to assess the influence of the aperture correction on our inferred dynamical masses. In Figure \ref{fig:mdyn_apertures} we show the virial dynamical masses against the dynamical masses from our kinematic modelling (Section \ref{sec:dyn_mass}). As expected, we find that these dynamical masses are higher compared to the virial masses without an aperture correction from Section \ref{sec:discussion_apertures}, and are in good agreement with the dynamical masses presented in Section \ref{sec:dyn_mass} ($\Delta=-0.01_{-0.11}^{+0.15}$).

Interestingly, we find no trend between the observed axis ratio of galaxies and the offset between the virial dynamic mass and the dynamical mass from our kinematic modelling. This is in contrast to findings for the LEGA-C sample from \citet{AvanderWel2022}, who find that the ratio between the Jeans mass and virial mass show a significant trend with the axis ratio. \citet{AvanderWel2022} attribute this trend to projection effects and galaxy geometry, which are especially relevant for rotating systems. They correct for this variation empirically by defining a second order homology correction $K(q)$ which removes any residual trends. For the rotationally supported galaxies in our sample we expect that projection effects would have a similar effect as for the LEGA-C sample, and it is thus puzzling that we see no trend with the observed axis ratio. However, we also note that our dynamical masses may be biased due to the thin disc assumption made in our kinematic modelling, and Jeans masses are needed to further investigate this effect. Furthermore, the smaller number of sources in our sample compared to the LEGA-C sample may bias our results.

\section{Distributions of intrinsic velocity curves}\label{ap:v_r}
\begin{figure}[h!]
    \centering
    \includegraphics[width=0.935\linewidth]{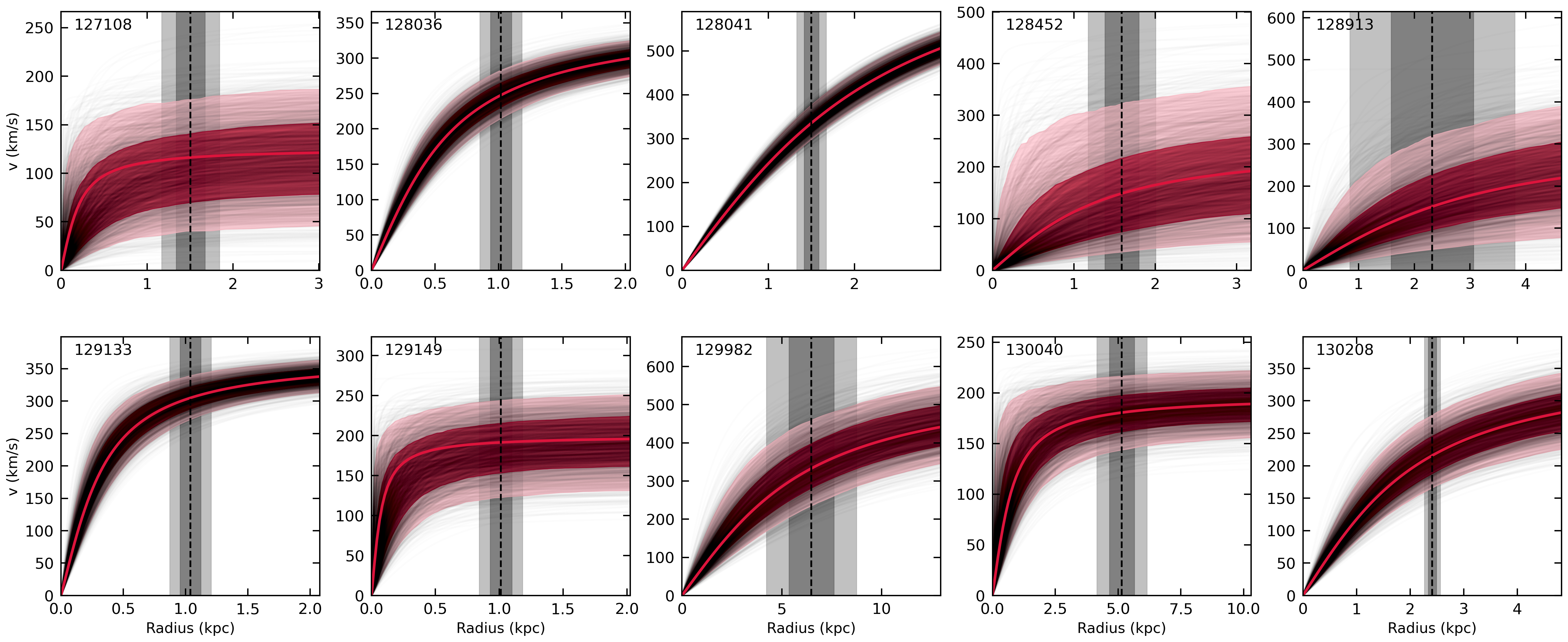}
    \caption{Intrinsic velocity as a function of radius from Equation \eqref{eq:arctan_v}, for 2000 simulations of $v_a$, $r_e$, and $r_t/r_e$, randomly perturbed around their uncertainties (black lines). The solid red line shows the best-fit model, and the 1$\sigma$ and 2$\sigma$ levels of $v(r)$ are indicated in pink and red, respectively. The dashed line indicates the measured $r_e$, with the 1$\sigma$ and 2$\sigma$ levels indicated with grey bands. All 10 galaxies have significant rotation from the 2$\sigma$ rotation curve limits.}
    \label{fig:vr_distributions}
\end{figure}

As described in Section \ref{sec:discussion_caveats}, we assess the significance of the rotational support in our galaxies by creating 2000 simulations of the intrinsic rotational velocity $v(r)$. In Figure \ref{fig:vr_distributions} we show these rotation curves, with their 1$\sigma$ and 2$\sigma$ distributions. For all 10 galaxies the rotational support is significant within 2$\sigma$.

\end{appendix}

\end{document}